\documentclass[onecolumn,preprintnumbers,amssymb,amsmath,superscriptaddress,letterpaper]{revtex4}[10pt]

\usepackage{graphicx}
\usepackage{dcolumn}
\usepackage{bm}
\usepackage{pstricks}
\usepackage{color}
\usepackage{epsfig}
\usepackage{subfigure}

\newcommand{\be}{\begin{equation}}
\newcommand{\ee}{\end{equation}}
\newcommand{\bea}{\begin{eqnarray}}
\newcommand{\eea}{\end{eqnarray}}

\newcommand{\gsim}{ \mathop{}_{\textstyle \sim}^{\textstyle >} }
\newcommand{\lsim}{ \mathop{}_{\textstyle \sim}^{\textstyle <}}

\newcommand{\cm}{{\rm ~cm}}

\newcommand{\GeV}{{\rm ~GeV}}
\newcommand{\gev}{{\rm ~GeV}}

\begin{document}

\title{High Energy Positrons and the WMAP Haze from Exciting Dark Matter}

\author{Ilias Cholis}
\email{ijc219@nyu.edu}
\affiliation{Center for Cosmology and Particle Physics, Department of Physics, New York University, 
New York, NY 10003}

\author{Lisa Goodenough}
\email{lcg261@nyu.edu}
\affiliation{Center for Cosmology and Particle Physics, Department of Physics, New York University, 
New York, NY 10003}

\author{Neal Weiner}
\email{neal.weiner@nyu.edu}
\affiliation{Center for Cosmology and Particle Physics, Department of Physics, New York University, 
New York, NY 10003}

\date{\today}

\begin{abstract}
We consider the signals of positrons and electrons from ``exciting'' dark matter (XDM) annihilation. Because of the light ($m_\phi \lsim 1\gev$) force carrier $\phi$ into which the dark matter states can annihilate, the electrons and positrons are generally very boosted, yielding a hard spectrum, in addition to the low energy positrons needed for INTEGRAL observations of the galactic center. We consider the relevance of this scenario for HEAT, PAMELA and the WMAP ``haze,'' focusing on light ($m_\phi \lsim 2 m_{\pi}$) $\phi$ bosons, and find that significant signals can be found for all three, although significant signals generally require high dark matter densities. We find that measurements of the positron fraction are generally insensitive to the halo model, but do suffer significant astrophysical uncertainties. We discuss the implications for upcoming PAMELA results.

\end{abstract}


\maketitle

\section{Introduction}
An overwhelming amount of evidence has established cold dark matter (CDM) as the standard paradigm for the missing matter of the universe. Beginning with early observations of velocity dispersions of galaxies in clusters \cite{Zwicky:1933gu}, and later measurements of galactic rotation curves \cite{Rubin:1980zd,Rubin:1985ze}, CDM has been supported by numerous additional observations. These include strong lensing of background galaxies \cite{Tyson:1998vp}, x-ray emission from galaxy clusters \cite{Cen:1993az}, the combination of CMB and type Ia supernovae data \cite{Spergel:2006hy}, measurements of the distributions of galaxies \cite{Cole:2005sx,Tegmark:2006az}, as well as the highly remarkable recent study of the bullet cluster \cite{Clowe:2006eq}.

The nature of this dark matter remains an open question. Measurements of the CMB, the need for early structure growth, and the success of big-bang nucleosynthesis rule out baryonic matter as being the dark matter, necessitating a new particle beyond the standard model. One of the most appealing of these is the thermal WIMP, whose relic abundance has a simple relation with the annihilation cross section. Specifically, for thermal s-wave freezeout, one has \cite{kolbandturner}
\be
\Omega h^2 = 0.1 \times \left( \frac{2.5 \times 10^{-26}\; \rm cm^3 s^{-1}}{\left\langle\sigma_{ann}\left|v\right|\right\rangle} \right) f(m)
\ee
where $f(m)$ is a logarithmically varying function of mass, with $f(500\gev)=1$. Because $\left\langle\sigma_{ann}\left|v\right|\right\rangle = 2.5 \times 10^{-26}\; \rm cm^3 s^{-1}$ is a cross section appropriate for a weak scale particle, a scale already suggested by the hierarchy problem, we have strong motivation to consider a WIMP with a mass in the range of a few hundred GeV.

\subsection{Indirect Detection of Dark Matter}
Indirect detection involves observing either the annihilation products of dark matter or the decay products of unstable states.  These products include positrons, photons, anti-protons, neutrinos, and anti-nuclei.  Searches are often made for the anti-particles, because their astrophysical backgrounds tend to be smaller than those of particles.  Though the DM annihilation cross-sections are generally small, if the density of dark matter particles is large, as is believed to be the case in the center of the Galaxy, then the annihilation rate can be large enough to produce observable effects.  Currently experiments are underway to detect neutrinos (AMANDA \cite{Ahrens:2002eb}, IceCube \cite{Ahrens:2003ix}) from annihilation of dark matter captured in the Earth and Sun, and to detect anti-particles (PAMELA \cite{Picozza:2006nm}) and photons (GLAST \cite{Gehrels:1999ri}, HESS \cite{Hinton:2004eu}, INTEGRAL \cite{Weidenspointner:2006nua}) from dark matter annihilation in the Galactic halo.

In many theories of SUSY dark matter, the dark matter particles are Majorana fermions that annihilate through processes such as $\chi  \chi \longrightarrow f \bar{f}$.
Annihilations to heavy fermions are favored over those to the light fermions, because the cross-section is proportional to $m_f^{2}$.  Therefore, the $e^{+}$ and $e^{-}$ spectra are generally soft.  This can make the DM signal difficult to detect, as the Galactic background spectra for electrons and positrons are largest at low energies and are currently not precisely understood.  For this reason, a theory of dark matter that produces a hard spectrum is perhaps the best chance for an observable indirect detection signal from electron and positrons.

Recently, a number of different experiments have made measurements which are consistent with excesses in positrons and/or electrons that may be indicative of a dark matter signal. Two of these (HEAT and the WMAP haze) are indicative of high energy particles, whereas the results of INTEGRAL/SPI are indicative of low energy particles.

In this paper, we will explore the possibility that a single source can produce all of these signals, specifically the ``exciting dark matter'' (XDM) proposal. We will focus on the cases most favorable for achieving high energy positron signals, in which the annihilation products of the WIMP are lighter than about 250 MeV, and decay either into $\mu^+ \mu^-$ or $e^+ e^-$ pairs. In the next section, we will review the evidence for excess electronic activity which may arise from dark matter. We then will describe the XDM scenario, and how it produces large quantities of high energy electrons and positrons. We describe the calculation of observable positron fraction and synchrotron signals in section \ref{sec:calc}. In section \ref{sec:results} we present our results. We find the XDM can easily explain both the HEAT excess as well as the WMAP haze, although significant signals tend to require large densities of dark matter. We find a wide range of spectra that may be found at PAMELA, including very hard positron spectra arising from the highly boosted annihilation products of XDM. Finally, in section \ref{sec:conc}, we conclude.

\section{Signals from Positrons and Electrons}
Positrons and electrons are the products of dark matter annihilation in numerous theories of WIMP dark matter, so their careful study could prove fruitful in the quest to understand the nature of dark matter.  The existence of electrons and positrons can be observed directly through a measurement of their particle fluxes, or indirectly through the measurement of their associated radiation, including $\gamma$-rays resulting from their annihilation and synchrotron radiation.

\subsubsection{Direct Measurement of $e^{+}$ and $e^{-}$ Spectra}
Direct measurements of $e^{+}$ and $e^{-}$ spectra have been done in several balloon-borne cosmic ray experiments, most recently in HEAT2000 and CAPRICE98 \cite{Barbiellini:1996ba}.  PAMELA, a satellite experiment, is currently taking data.  (Note: Many experiments state their results for the ratio $e^{+}/(e^{+}+e^{-})$, the positron fraction, to eliminate some systematic errors in the measurements of the individual fluxes.)  Cosmic ray electrons and positrons are of two varieties: ``primary'' particles created by astrophysical sources, and ``secondary particles'' produced through the interactions of primaries with gas in the interstellar medium (ISM).  Above 100 MeV the primary positron flux due to standard astrophysical sources is thought to be negligible \cite{Protheroe:1982pp}, while the secondary flux behaves like an inverse power law.   Therefore, the positron background signal is small at large energies, which makes the positron signal a promising candidate for observing new physics.

For instance, the balloon-borne HEAT (High-Energy Antimatter Telescope) experiment flew twice in the mid-1990's, and measured the individual and combined energy spectra of electrons and positrons \cite{Barwick:1997ig}.  From these measurements the positron fraction was determined for the energy range from 1 to 50 GeV. The HEAT results suggest that for energies larger than 10 GeV an excess in the positron flux above that expected from pure secondary production cannot be ruled out \cite{Barwick:1997ig}.  There may exist sources that give rise to primary positrons with energies above 10 GeV.  Additionally, there appears to be some structure in the positron fraction above 7 GeV that a pure secondary spectrum cannot explain \cite{Coutu:1999ws}.

An important experiment which will clarify the results of HEAT is PAMELA. The Payload for Antimatter Matter Exploration and Light-nuclei Astrophysics experiment is a satellite-borne apparatus launched in 2006 that is currently measuring the energy spectra of many cosmic ray particle species, including electrons, positrons, protons, anti-protons, and light nuclei.  One of the stated primary objectives of the experiment is to search for evidence of annihilations of dark matter particles by measuring the electron and positron energy spectra in the energy range from 50 MeV to 270 GeV \cite{Picozza:2006nm}.  PAMELA will extend the measurement of the positron fraction out to an energy almost an order of magnitude larger than that achieved by the HEAT experiment.  Additionally, the PAMELA results will be based on very high statistics, approximately $ 10^{5}\; e^{+}$ per year \cite{Picozza:2006nm}.  Initial high energy results from the PAMELA experiment are expected soon.

\subsubsection{High Energy Positron and Electron Signals in Synchrotron Radiation}
In addition to directly detecting the cosmic ray positrons, one can detect $\gamma$-rays coming from electron-positron annihilation and synchrotron radiation coming from electrons and positrons propagating through the galactic magnetic field.  An excess of either of these radiation signals could indicate the existence of dark matter annihilations. The former is generally a measure of stopped or low energy positrons, while the latter is a measure of higher energy (multi-GeV) positrons and electrons. 

Let us begin by considering the possibility of synchrotron radiation.  The Wilkinson Microwave Anisotropy Probe (WMAP), launched in 2001, measured the microwave emission at five frequencies, 22.5, 32.7, 40.6, 60.7, and 93.1 GHz, over the full sky.  Multiple frequency measurements were necessary to separate Galactic foreground signals, each with a unique spectrum and spatial distribution, from the Cosmic Microwave Background (CMB) signal.

Using the one-year data from WMAP released in 2003, Finkbeiner identified the microwave emission due to the three well-understood Galactic signals, thermal radiation from dust, free-free (bremsstrahlung) radiation, and synchrotron radiation from high-energy electrons (accelerated by supernovae shocks) spiraling in the Galactic magnetic field, and additional emission due to spinning dust \cite{Finkbeiner:2003im}.  His analysis indicated that there is excess emission, not correlated with any of the known Galactic foregrounds, that is distributed with approximate radial symmetry within $\sim 20^{\circ}$ of the Galactic center and that decreases rapidly with projected distance from the Galactic center.  He called this excess microwave emission the ``haze'' and argued that it is consistent with synchrotron emission from high-energy electrons and positrons created in the center of the Galaxy \cite{Finkbeiner:2004us}.  One explanation for the presence of these electrons and positrons in the inner Galaxy is that they are produced by annihilating dark matter \cite{Finkbeiner:2004us,Hooper:2007kb}. 

\subsubsection{INTEGRAL/SPI}
The strongest gamma-ray line signal from our galaxy comes from electron-positron annihilation \cite{Weidenspointner:2007ru}.  The SPI imaging spectrometer aboard ESA's INTernational Gamma-Ray Astrophysics Laboratory (INTEGRAL) gamma-ray satellite observatory has measured the 511 keV line coming from direct annihilation of $e^{+}e^{-}$ pairs and decay of para-positronium, as well as the continuum spectrum from the 3-photon decay of ortho-positronium.  Both signals are strongest within a few degrees of the Galactic center, indicating that this region has the largest concentration of electron-positron annihilation \cite{Weidenspointner:2007ru}.  The 511 keV emission from the bulge has ellipsoidal symmetry about the Galactic center with an angular extension of $6.5^{\circ +1.1}_{\,\, \, -0.9}$ (FWHM) in longitude and $5.1^{\circ +0.8}_{\,\,\, -0.8}$ (FWHM) in latitude \cite{Weidenspointner:2007ru}.  The flux from the bulge at 511 keV is $(7.04 \pm 0.32)\times 10^{-4} \rm \;ph \;cm^{-2} \;s^{-1}$, while the emission from the disk is $(1.41 \pm 0.17)\times 10^{-3} \rm \;ph \;cm^{-2} \;s^{-1}$ \cite{Weidenspointner:2007ru}.

It is presently a great challenge for conventional astrophysical sources, such as cosmic ray interactions with the ISM, neutron stars, black holes, supernovae, low mass x-ray binaries, and pulsars, to explain the INTEGRAL signal, although, with myriad uncertainties, they may prove to be responsible. Recently, it was noted that the disk component of positron emission, in particular, was likely due in large part to LMXBs \cite{Weidenspointner:2008zz}, although it is still uncertain whether such objects can provide a significant piece of the bulge emission.

A very appealing possibility is that the excess arises in some fashion from dark matter. However, such explanations are challenging, because the INTEGRAL signal is inconsistent with injected positrons at energies much higher than a few MeV \cite{Beacom:2004pe,Jean:2005af,Beacom:2005qv}. As such, we cannot simply identify these positrons with annihilation products of weak-scale dark matter.

\section{Exciting Dark Matter (XDM)}

Theories of ``light'' dark matter, for example scalar dark matter \cite{Boehm:2003hm,Boehm:2003bt,Hooper:2008im}, with masses in the MeV range well below the weak scale, as well as decaying dark matter scenarios \cite{Hooper:2004qf,Pospelov:2007xh}, have been proposed to explain the INTEGRAL signal. 
However, these theories are ad hoc to some degree from a particle physics perspective.  An alternative explanation for the INTEGRAL signal with a weak scale dark matter particle arises in the scenario of ``exciting'' dark matter (XDM) \cite{Finkbeiner:2007kk}.  XDM proposes that the electron-positron pairs needed to explain the INTEGRAL signal are created in the decay of an excited state of WIMP dark matter. The theory contains a weak scale dark matter particle $\chi$ with an excited state $\chi^{*}$ that has an energy at least $2m_{e}$ above that of the ground state.  The coupling of the pseudo-Dirac dark matter particle to standard model particles is through a light scalar particle $\phi$, $m_{\phi} \sim .1 - 1$ GeV, which couples to the Higgs. The decay of the excited state into the ground state can result in the emission of an $e^{+}e^{-}$ pair.

\begin{figure}[htpb]
\begin{center}
\includegraphics[width=0.4\textwidth]{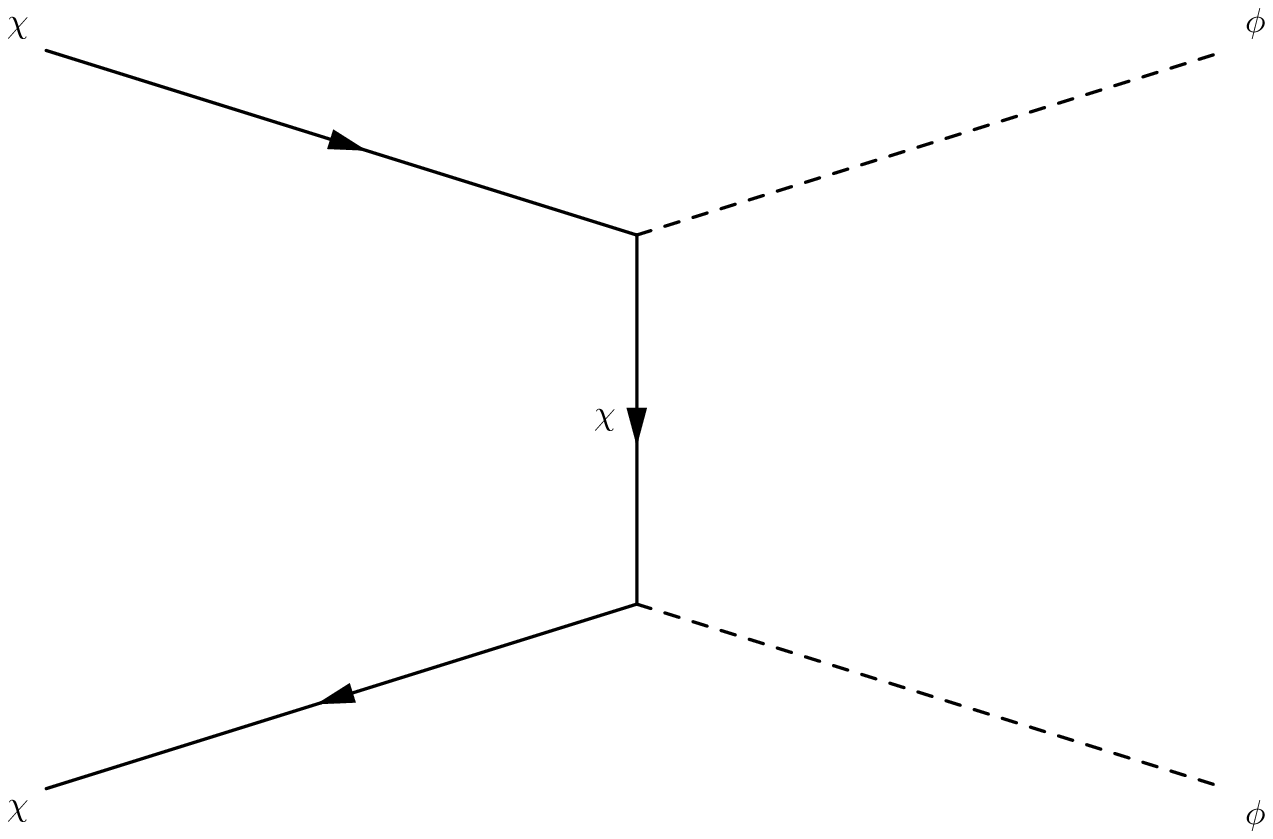}
\hskip 0.15in
\includegraphics[width=0.4\textwidth]{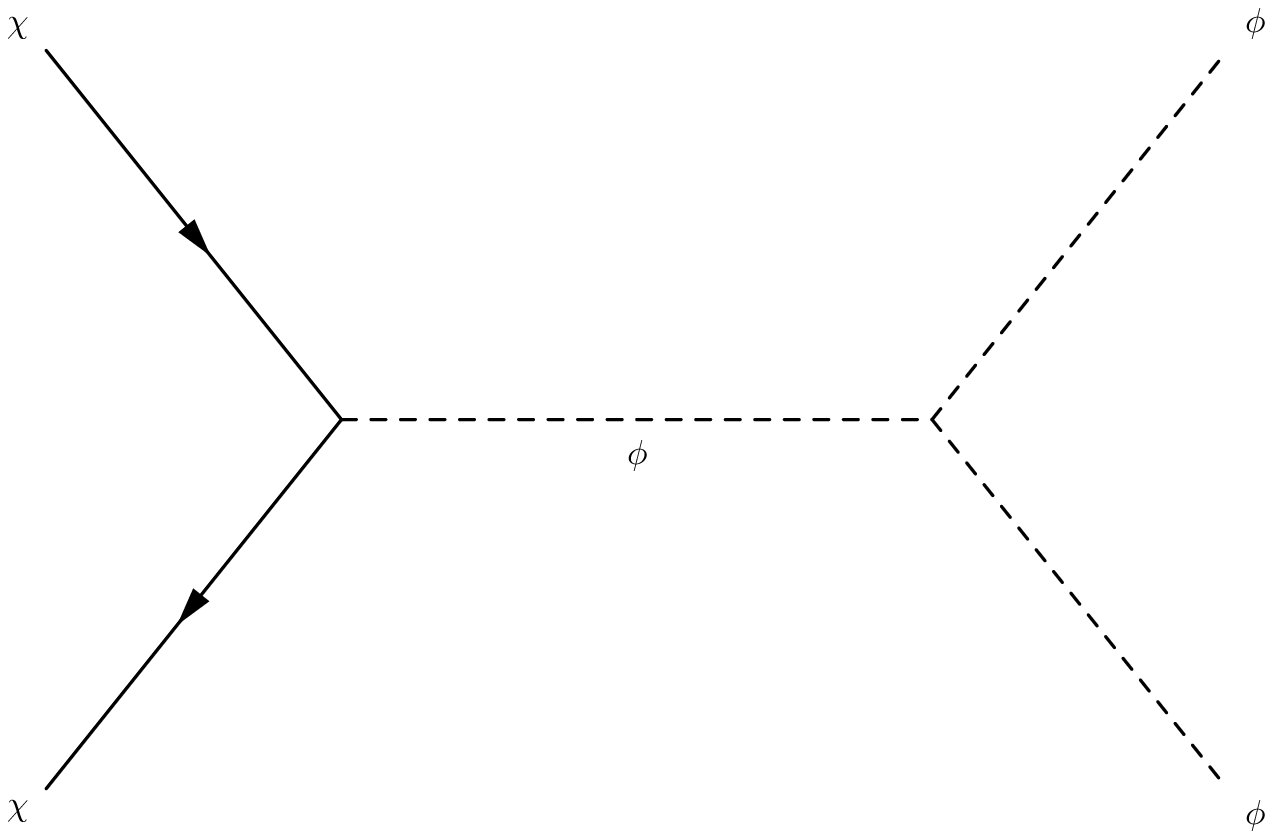}\\
\end{center}
\caption{Diagrams contributing to annihilation of $\chi$.}
\label{fig:annih1}
\end{figure}

The energy needed for the excitation process comes from the kinetic energy of the WIMPs; a 500 GeV particle with a speed of 500 km/s has a kinetic energy greater than $m_{e}$.  XDM can account for the creation of $3 \times 10^{42}$ $e^{+}e^{-}$ pairs per second in the Galactic bulge, enough to explain the INTEGRAL $\gamma$-ray signal.  Although the model requires large scattering cross sections, and cuspy halos, it is an intriguing suggestion that allows one to generate low energy positrons from weak-scale dark matter particles.  Beyond particles produced from excitation and subsequent decay of WIMPs, additional signals of XDM arise through annihilation of $\chi$ into $\phi$, which proceeds through the processes shown in Fig.~\ref{fig:annih1}.

Since present-day dark matter is moving very slowly ($\beta \sim 10^{-3}$), only s-wave processes are relevant for indirect detection in the halo (at least for thermal dark matter). Since we are motivated by possible {\em existing} signals, it is important to consider whether XDM can produce models with significant s-wave annihilations. This is not a trivial point. Although XDM functions as a Dirac fermion at freezeout, where the splitting between $\chi$ and $\chi^*$ is small, in the present universe, by assumption, only $\chi$ is present. This is a Majorana fermion, and thus to annihilate via s-wave, must be put into a state with CP = -1 (see the discussion in \cite{Drees:1992am}). As a consequence, Majorana XDM cannot annihilate into two identical real $\phi$ particles. However, relatively minor modifications can allow this annihilation to proceed via s-wave. In particular, we can allow $O(1)$ CP violation in the dark matter sector, or we can promote $\phi$ to a complex scalar. In models of XDM where the dark matter is a Dirac fermion with an excited state (in analogy with hadrons), no such modifications are necessary. In any event, the modifications to the model in \cite{Finkbeiner:2007kk} needed to generate s-wave annihilations are small, so the resulting differences in the signals in question are small.

Since $m_{\chi} \gg m_{\phi}$, the $\phi$ particles come out boosted by a factor of $\gamma = \frac{m_{\chi}}{m_{\phi}}$.  The $\phi$s can then decay into SM particles through their mixing with the Higgs. For $m_{\phi}<2m_{e}$, $\phi$ can only decay into two photons.  For $2m_{e}< m_{\phi}<2m_{\mu}$, the decay can proceed through two channels, $\phi\longrightarrow \gamma \gamma$ and $\phi\longrightarrow e^{-}  e^{+}$. The decay of $\phi$ to two $\gamma$s is due to one loop spin-0, spin-1/2 and spin-1 contributions.  These almost cancel giving a branching ratio (B.R.) for $\phi\longrightarrow \gamma \gamma$ of at most 14$\%$ in that energy region \cite{HiggsHG}. In the energy range $2m_{\mu}< m_{\phi}< 2m_{\pi^{0}}$, the channel $\phi\longrightarrow \mu^{-}  \mu^{+}$ becomes relevant, and since the branching ratio of $\phi$ to $f\bar{f}$ goes as $m_{f}^{2}$, the decay $\phi\longrightarrow \mu^{-} \mu^{+}$ completely dominates over $\phi\longrightarrow e^{-}  e^{+}$.  In this energy range the decay of $\phi$ into 2 photons is negligible \cite{HiggsHG}.  Above the threshold energy $2m_{\pi^{0}}$, the channel to neutral pions is opened.  For  $\phi$ masses above $2m_{\pi^{\pm}}$, it has been shown that the B.R.s for the processes $\phi\longrightarrow \pi^{-}  \pi^{+}$ and $\phi\longrightarrow 2\pi^{0}$ cannot be ignored \cite{Voloshin:1980zf,Voloshin:1985tc,Grinstein:1988yu}.  For a 1 GeV $\phi$ the B.R. for $\phi\longrightarrow \mu^{-}  \mu^{+}$ is $\approx 20\%$ \cite{HiggsHG} while for the decays to $\pi$s it is $\approx 40\%$. For $m_{\phi}> 1$ GeV decay channels to heavier mesons appear.  The work presented in this paper investigates the mass range $2m_{e}<m_{\phi}<2m_{\pi}$.  We leave other mass ranges for future work.

We considered two distinct cases for the mass of $m_{\phi}$: $m_{\phi}=0.1$ GeV, for which the B.R. for $\phi\longrightarrow e^{-} e^{+}$ is $\cong 90\%$, and $m_{\phi}=0.25$ GeV, for which the B.R. for decay to $\mu^{+}$ and $\mu^{-}$ is $\cong100\%$.
In both scenarios, the resulting spectra for the electrons and positrons are much harder than typical $e^{+}e^{-}$ spectra coming from weak scale WIMP annihilation.  The positron fraction has a characteristic bump that starts around 6 GeV. (The precise location of the peak depends on the masses of $\chi$ and $\phi$.)  There is evidence for additional structure around this energy in the HEAT measurements of the positron fraction; XDM provides an explanation for this.  The presence of additional high energy $e^{+}e^{-}$ pairs from XDM annihilation in the center of the Galaxy would give rise to additional synchrotron radiation as they spiral through the Galactic magnetic field.  In this way XDM provides an explanation for the excess microwave emission, the ``haze'' \cite{Finkbeiner:2004us}, observed by WMAP.

\section{Calculations}
\label{sec:calc}
We have calculated the local positron fraction $e^{+}/(e^{+}+e^{-})$ up to energies of 1 TeV and the local synchrotron radiation due to dark matter for the haze frequencies, 22.5, 32.7, 40.6, 60.7, and 93.1 GHz, using GALPROP  \cite{Moskalenko:1999sb} version 50p \footnote{GALPROP and the resource files we employed are available at  http://galprop.stanford.edu .}.  The ratio calculation requires knowledge of the local primary and secondary positron and electron spectra, as well as the contributions to both local spectra from dark matter particle annihilations.  The contribution to the positron fraction from XDM annihilation is significant only for energies above $\sim$5 GeV, so our starting assumption is that the background fraction should fit well the HEAT data at energies below $\sim$5 GeV.  We calculated our background ratio in agreement with the first five HEAT data points and required the total positron fraction to fit well all nine HEAT data points.  In light of prelimary PAMELA data on positrons below 10 GeV, we also fit to those, as well. We predict the results of the PAMELA experiment for the positron fraction in the XDM scenario.  Additionally, we compare our calculation of the XDM synchrotron radiation to the ``haze''.

\subsection{The Injection Spectra for $e^{+}$ $e^{-}$ Arising from DM Annihilation }
\begin{figure}[htpb]
\begin{center}
\subfigure[Direct decay channel]{\label{fig:directspec}\epsfig{figure=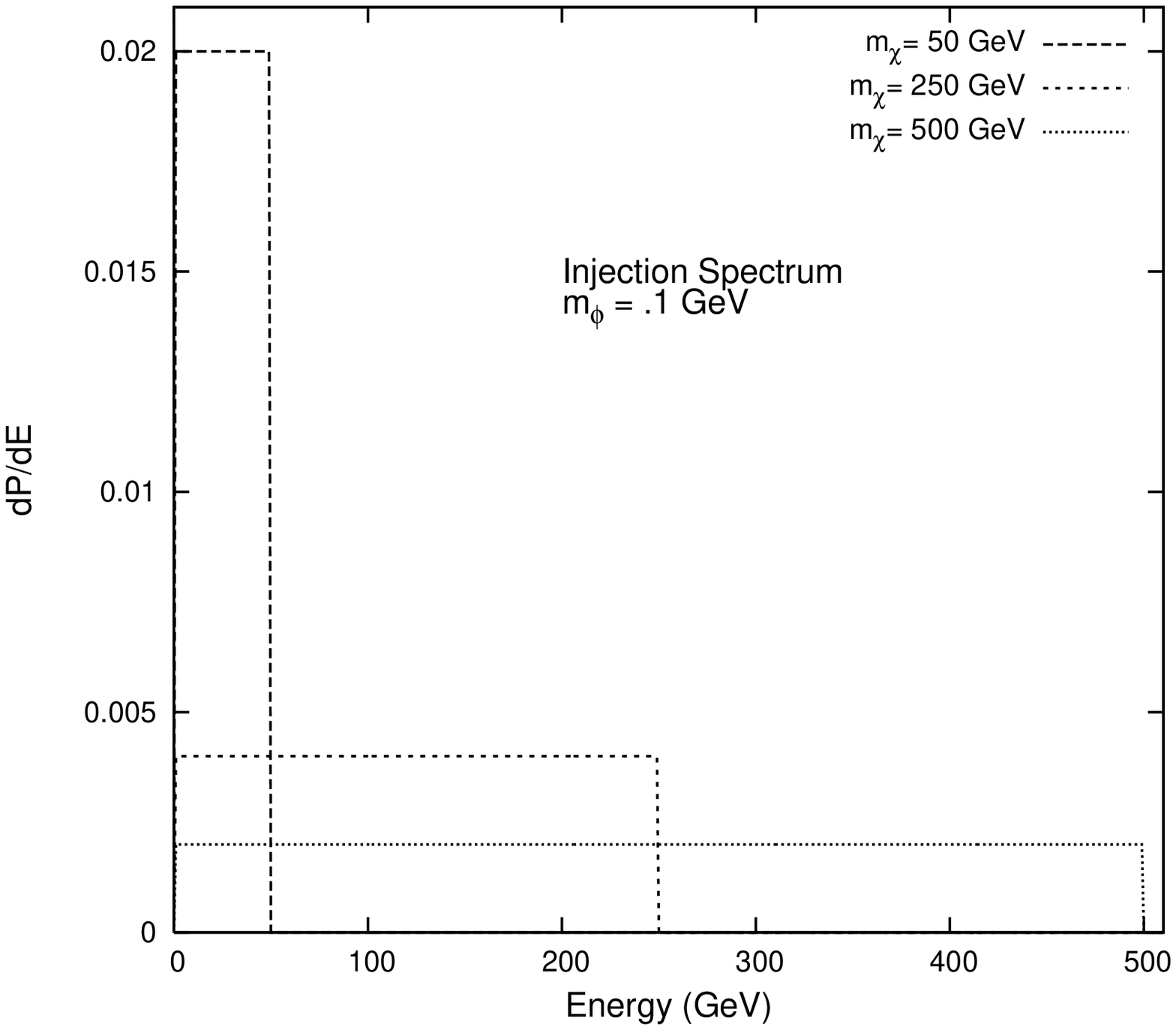,scale=.4}}
\hskip 0.15in 
\subfigure[Muon decay channel]{\label{fig:muonspec}\epsfig{figure=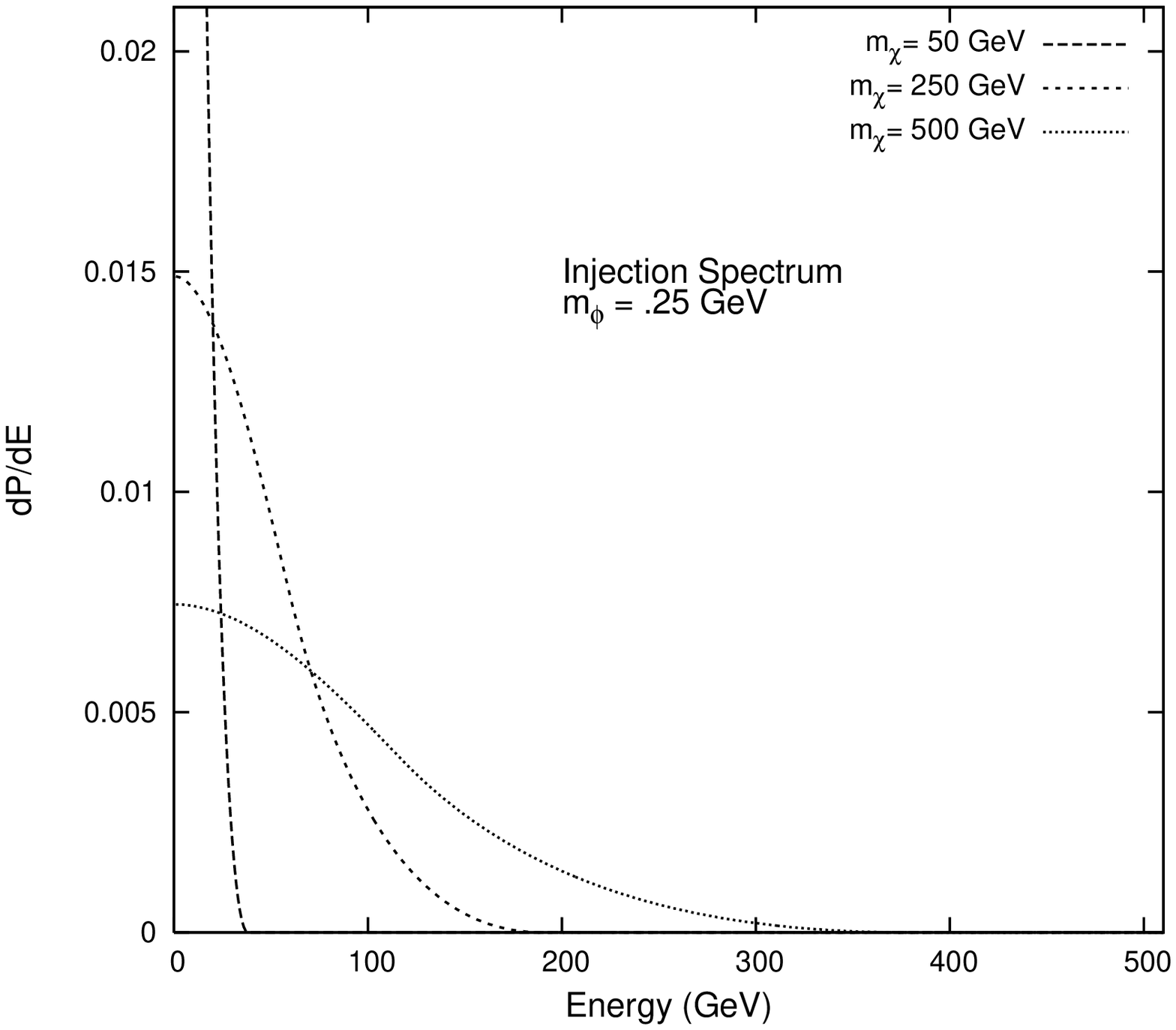,scale=.4}}
\end{center}
\caption{Injection spectra of positrons.}
\label{fig:decaychannels}
\end{figure}

The annihilation process for XDM is through the intermediate $\phi$ particle.  The $\phi$ mixes with the Higgs allowing it to decay into the final state fermions.  We consider two scenarios.  The first is the ``direct'' decay channel, in which the annihilation proceeds directly into $e^{+}$ $e^{-}$ pairs, which occurs for $m_{\phi} < 2 m_{\mu}$.  For $2m_\mu \lsim m_\phi \lsim 2 m_{\pi^{0}}$ we have the ``muon'' decay channel, where $\phi$ decays to muons, which in turn produce electrons and positrons.  Higher mass $\phi$ can decay into pions, kaons, and other hadrons. We defer the analysis of these scenarios to a future paper.  For the direct channel, the injection spectrum is flat (see Fig.~\ref{fig:directspec}), while for the muon channel it is somewhat softer (see Fig.~\ref{fig:muonspec}). We give the details and formulae in Appendix \ref{ap:muonspec}.  

It is important to note that these light particles produce few $\gamma$-rays directly, and are kinematically incapable of producing anti-protons, which are a strong constraint on these models. The absence of indications of anti-proton excesses may be indicative of such a light mediator producing a positron signal.

The GALPROP code of Moskalenko and Strong calculates cosmic ray propagation through the Galaxy by solving the propagation equation numerically on a grid \cite{Moskalenko:1999sb}.  The spatial dynamics include diffusion resulting from cosmic rays scattering on MHD waves, and convection on Galactic winds \cite{Strong:2007nh}.  In momentum space, diffusive re-acceleration, the result of stochastic acceleration due to scattering on MHD waves, and energy loss from ionization, bremsstrahlung, inverse Compton scattering and synchrotron radiation are included \cite{Strong:2007nh,Moskalenko:1999sb}.  Using the most recent measurements of the source abundances for the primary species (nuclei, electrons and $\gamma$-rays), GALPROP propagates the primaries through the Galaxy, then iteratively computes the resulting spallation source functions for all species, and propagates the equivalent full, primary + secondary, source function for each species until a converging result is obtained \cite{Strong:1999sv}.  GALPROP assumes free escape of the particles as the spatial boundary condition \cite{Galprop1}.  For our calculations, we used a 2-dimensional spatial grid (and a one dimensional energy grid), and assumed cylindrical symmetry as well as mirror symmetry with respect to the Galactic plane.

Through the galdef file, the parameter input file, the code allows for considerable freedom in the choice of many astrophysical parameters, including the primary electron injection spectrum, the diffusion coefficient, and the strength of re-acceleration.  We used the ``conventional'' model of CR production and propagation as used by Abdo et al. \cite{Abdo:2006fq} as the starting point for our choice of parameters.  The conventional model calculates the local proton and electron spectra in agreement with the locally measured values \cite{Strong:2004de}.  For a given choice of Alfv\'{e}n velocity, we varied only four physical parameters to fit the low energy HEAT data, specifically,  the energy of the break in the primary electron spectrum (Break Rigidity), the injection index below and above the break, and the overall flux normalization (see Table \ref{tab:elecspecparams}). We also vary one calculational parameter, the energy grid spacing, in our calculations of the positron fraction.

\begin{table}
\begin{center}
\caption{Parameters of the primary electron spectrum for fits to HEAT}
\label{tab:elecspecparams}
\begin{tabular}{c|c|c|c}
\hline\hline
$v_A$ & Injection Index & Break Rigidity & Electron Flux Norm.\\
(km/s) & Below Break Rigidity & (MV) & $\rm (cm^{-2} sr^{-1} s^{-1} MeV^{-1})$\\
\hline
0 & 1.60 & $4.0\times 10^3$ & $0.2488\times10^{-9}$ \\
20 & 2.10 & $6.0\times 10^3$ & $0.2612\times10^{-9}$ \\
35 & 2.15 & $4.0\times 10^3$ & $0.2887\times10^{-9}$ \\
\hline
\end{tabular}
\end{center}
\end{table}

\subsection{Dark Matter Density Functions (Halo Profiles)}
In order to understand the dependence of the signals upon halo models, we performed our calculations using three different halo profiles:

\[\rho(r)=\rho_{0}\; \frac{r_{c}}{r}\frac{1}{(1+\frac{r}{r_{c}})^2}\hspace{1cm}\mbox{N.F.W. Profile \cite{Navarro:1995iw}}\]

\[\rho(r)=\rho_{0}\; \frac{r_{c}^{2}+R_{\odot}^{2}}{r_{c}^{2}+r^{2}}\hspace{1cm}\mbox{Isothermal Profile \cite{Moskalenko:1999sb}}\]

\[\rho(r)=\rho_{0}\; exp[-\frac{2}{\alpha}(\frac{r^{\alpha}-R_{\odot}^{\alpha}}{r_{-2}^{\alpha}})]\hspace{1cm}\mbox{Merritt Profile \cite{Merritt:2005xc}.}\]

\begin{figure}[h!]
\begin{center}
\includegraphics[scale=.4]{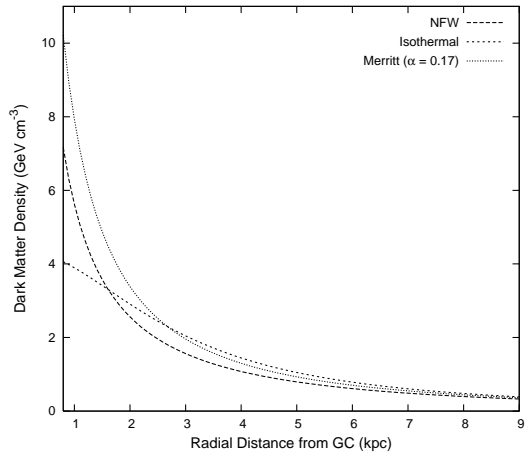}
\caption{Dark Matter density profiles.}
\label{fig:profiles}
\end{center}
\end{figure}

Here $R_{\odot}=8.5$ kpc is the solar distance from the Galactic center, $r^2=R^2+z^2$ is the spherical radial coordinate, $.13\leq\alpha\leq.22$ \footnote{The range of values $0.13<\alpha<0.22$ corresponds to that allowed by simulations done by \cite{Merritt:2005xc}.} is the Merritt parameter which defines the cuspiness of the profile, and $r_{-2}=25$ kpc is the radius at which the logarithmic slope of the Merritt profile is -2. $r_c$ is the core radius and $\rho_{0}$ is the local value of the dark matter mass density.  The values of these parameters for the three profiles are listed in Table~\ref{tab:profileparams}.  See Fig.~\ref{fig:profiles} for a comparison of these three DM density profiles.  We limited ourselves to these three halo models, which arise under various assumptions and circumstances, and did not consider theoretically unmotivated profiles, such as a flat halo profile. The Merritt profile, being the cuspiest, is generally the best for generating the INTEGRAL signal from XDM.  The isothermal profile provides a theoretical simplification, which is often employed for analyzing, for example, direct detection limits.

\begin{table}[h]
\begin{center}
\caption{Parameters of the dark matter density profiles}
\label{tab:profileparams}
\begin{tabular}{lcr}
\hline\hline
Model & $r_c$ (kpc) \\
\hline
N.F.W. & 20.0 \\
Isothermal & 2.8 \\
Merritt & 25.0 \\
\hline
\end{tabular}
\end{center}
\end{table}

\subsection{The Electron and Positron Backgrounds}
The background positron fraction is given by
\[\frac{e^{+}_{secondary}}{e^{+}_{secondary}+e^{-}_{primary}+e^{-}_{secondary}}\; .\]
The primary positron spectrum is taken to be zero, as it is negligible in comparison with the secondary spectrum.  The sources of the primary $e^{-}$ spectrum are believed to include supernovae remnants and pulsars, though their exact nature is unclear.  Additionally, there is uncertainty in the primary $e^{-}$ spectrum itself, particularly at energies below 10 GeV, where the effect of solar modulation is to attenuate the spectrum at lower energies.  However, it is generally agreed that for the energy range of interest to us, 1 GeV to 1 TeV, the spectrum is well described by a power law $E^{-\alpha}$ with an index $\alpha$ that increases at higher energies.  The GALPROP code provides the freedom to modify the primary electron injection spectrum.  Inputs into GALPROP include constant values of $\alpha$ in three energy regions, the two energies defining these regions, the normalization of the electron flux, and the energy at which this normalization holds. 

Secondary positrons and electrons are created in the interactions of primary CR protons and Helium as they propagate through the Galaxy.  p-p and p-He collisions result in charged pions and kaons, which give rise to electrons and positrons through the decays $K^{\pm} \rightarrow \pi^{\pm} + \pi^0$,
$K^{\pm} \rightarrow \mu^{\pm} + \nu_{\mu}$,
$\pi^{\pm} \rightarrow \mu^{\pm} + \nu_{\mu}$, and
$\mu^{\pm} \rightarrow e^{\pm} + \overline{\nu_{\mu}} + \nu_{e}$.  The secondary positron and electron spectra are equal to one another by CP invariance of weak decays \cite{Kelner:2006tc}.

\begin{figure}[htpb]
\begin{center}
\includegraphics[scale=.4]{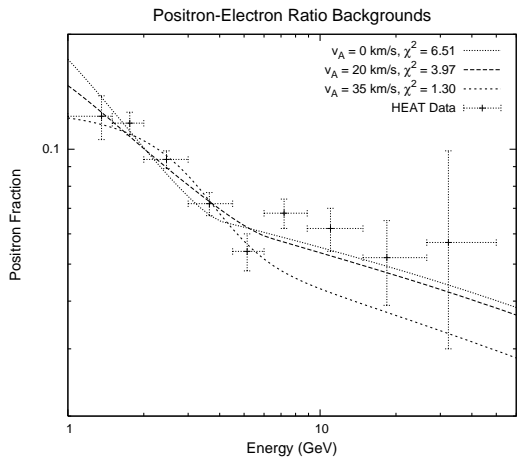}
\end{center}
\caption{Positron fraction backgrounds for $v_A = 0$, $v_A = 20$, and $v_A = 35$ km/s. The $\chi^2$ quoted is for the fit to the five lowest energy points.}
\label{fig:backgrounds}
\end{figure}

We used a standard least-squares fit of the calculated background ratio to the first five HEAT data points.  The backgrounds are shown in Fig.~\ref{fig:backgrounds}.  It is clear from the figure that the backgrounds are very sensitive to the choice of parameters for the primary electron spectrum.  In spite of the differences in the backgrounds, as we show later, there are common features in the total positron fraction.  As a consistency check on our backgrounds, we calculated the resulting local H and He fluxes and compared them to the measured fluxes at the top of the atmosphere \cite{Sanuki:2000wh,Menn:2000,Alcaraz:2000vp}.  We had good agreement for both spectra for energies above a few hundred MeV.  It is important to note that differences in the low energy background carry over to high energies, and thus lead to differences in the positron fraction at high energies.

\subsection{The Positron Fraction}
We calculated the total positron fraction \footnote{We calculate the positron fraction, instead of the absolute particle fluxes, for comparison with experimental results, which are often presented as the fraction to reduce systematic errors.} for each of the three backgrounds, $v_A$ = 0, 20, and 35 km/sec, varying the following parameters as described: five dark matter halo profiles, N.F.W., Isothermal, Merritt($\alpha = 0.13$), Merritt($\alpha = 0.17$), and Merritt($\alpha = 0.22$); five DM masses, $m_{\chi}$=  50, 100, 250, 500, and 800 GeV; and two decay channels, direct and muon, corresponding to $m_{\phi} =$ .1 and .25 GeV, respectively.  We fit the total fraction to all nine points of the HEAT data using a least squares fitting routine with the annihilation cross-section as the fit parameter.  As stated previously, a value of the thermally averaged annihilation cross-section $\left\langle\sigma_{ann}\left|v\right|\right\rangle$ that gives the correct DM relic abundance is roughly $2.5\times 10^{-26}\; \rm cm^{2} s^{-1}$. However, because the physical signal, in particular for high energy positrons, is dominated by the local density, we quote our results in terms of the physically relevant parameter $\left\langle \rho^2 \sigma v\right\rangle$.

Inhomogeneities in the dark matter density can lead to additional ``boost'' factors, $b= \left\langle\rho^2\right\rangle/\left\langle\rho\right\rangle^2$.
Based on present halo simulations, a reasonable value for a boost factor for gamma ray production in the halo is $b \lsim 3$, although extrapolations to smaller scales maybe allow boosts as large as 13 \cite{Diemand:2006ik}. Here, we are interested in two separate signals, both from the inner region of the halo, and it is not yet clear what a reasonable boost factor is. However, it is more likely that clumps are tidally destroyed in the region closer to the galactic center, thus we expect, for reasonable scenarios, that a larger boost factor would arise for the positrons produced away from the GC, i.e. those relevant for HEAT and PAMELA, than for positrons produced near the GC, i.e. those relevant for the haze.  That said, uncertainties in the galactic magnetic field and halo profile can also play an important role in the haze signal, so the boost factor that we report here is still uncertain.

\subsection{DM Synchrotron Radiation}
\label{sec:DMsynchrad} 
The calculation of the synchrotron radiation due to $e^{+} e^{-}$ pairs arising from dark matter annihilation is unencumbered by the complications of the background fluxes, since the dark matter contribution is completely independent of the contributions from standard astrophysical sources.  We fit the calculated intensity for 22.5 GHz synchrotron radiation to the 29 haze data points extending over 6$^{\circ}$ - 34$^{\circ}$ from the Galactic Center using the annihilation cross-section and a constant offset as fit parameters.  (The inner  5$^{\circ}$ are masked due to the bright dust emission.) Because we only expect dark matter to dominate roughly the inner $15^{\circ}$, and because the overall normalization of the haze is somewhat uncertain, we fit the existing data to a dark matter component, plus a constant offset. This, then, prevents the $15^{\circ}-35^{\circ}$ region from dominating the fit.

\subsection{Uncertainties}

A given model generally makes a very robust prediction for the primary positron injection spectra once all kinematical elements are specified (specifically, dark matter masses and the masses of the annihilation products). However, when converting these injection spectra into observed rates and spectra at an experiment such as PAMELA, there are many uncertainties that must be considered.

\subsubsection{DM Halo Uncertainties}
The amplitude of the signal is first and foremost determined by the cross section and the density of the dark matter $\rho_0$, entering in the form $\rho_{0}^{2}\left\langle\sigma\left|v\right|\right\rangle$. While a thermal WIMP will generally have a cross section $2-3 \times 10^{-26} \rm cm^3 s^{-1}$, non-thermal WIMPs can have significantly higher cross sections giving higher rates at HEAT and PAMELA and for the WMAP haze \cite{Hooper:2004xn,Grajek:2008jb,Nagai:2008se}.

 The local dark-matter-produced positron signal at the highest energies is simply determined by the local density of dark matter. This is because high energy positrons propagate a very short distance before losing much of their energy. A 300 GeV positron, for instance, typically drops to $e^{-1} \times 300$ GeV in a distance of $\sim$ 300 pc. Consequently, the signal for HEAT and PAMELA is relatively insensitive to the broad shape of the halo, but {\em is} very sensitive to the local value of $\rho_0$.

Conventionally, $\rho_0=0.3\ \rm GeV cm^{-3}$ is adopted for many analyses. While providing a useful baseline, it is important to recognize that there is significant uncertainty in this value. The standard reference for a determination of $\rho_0$ is \cite{Gates:1995dw}. There, the expected range is given as $0.2\ \rm GeV cm^{-3} \lsim \rho_0 \lsim 0.7 \ \rm GeV cm^{-3}$, while noting simultaneously that additional flattening of the halo could increase this by a factor of two. \cite{Bergstrom:1997fj} considers halo models with densities up to $0.8\ \rm GeV cm^{-3}$, while \cite{Bottino:2008mf} considers local densities of $1.07\ \rm GeV cm^{-3}$.   

The clumpiness of the halo (commonly referred to as a ``boost factor'', $b= <\rho^2>/<\rho>^2$), is simply the observation that local overdensities can increase the annihilation rate. Studies of this for gamma rays have indicated that boosts $b\approx 3$ are quite reasonable, and boosts $b\approx 13$ might occur, if the boosts found in the simulation are extrapolated to the subhalos \cite{Diemand:2006ik}, although it is not clear that such boosts survive to interior regions of the Galaxy as are relevant for HEAT and PAMELA. Moreover, such overdensities can impact the spectrum if a particularly significant overdensity is located near us \cite{Hooper:2004xn}. 

Ultimately, it is reasonable to consider values as large as $\rho_{0}^{2}\left\langle\sigma\left|v\right|\right\rangle \sim 3 \times 10^{-26} \rm GeV^2 cm^{-3}s^{-1}$ without invoking boost factors, and possibly significantly higher, $\rho_{0}^{2}\left\langle\sigma\left|v\right|\right\rangle \sim 4 \times 10^{-25} \rm GeV^2 cm^{-3}s^{-1}$, if one considers boost factors as well. Non-thermal candidates can invoke yet higher values. We shall find that most regions of parameter space that produce significant signals at PAMELA are on the high end of this range. However, in the event of a significant signal at PAMELA, it will be unclear whether a local overdensity or boost, a high value of $\rho_0$ (perhaps from a flattened halo), or large non-thermal cross sections are responsible for it.

\subsubsection{Uncertainties in the Propagation Parameters}
\noindent{\it{Relevance to the Positron Fraction}}

\vspace{\baselineskip}
We require our results to be consistent with local CR measurements, since it is the local positron fraction we are interested in.  Local cosmic ray measurements allow for uncertainties in the primary electron and secondary electron and positron fluxes, and these in turn give rise to different shapes for the positron fraction.  The uncertainties in the CR spectra can be associated with uncertainties in the production mechanisms for the spectra, i.e. the shapes of the spectra themselves at production, or with uncertainties in the CR propagation parameters.

The total (primary+secondary+dark matter) electron spectrum is constrained up to $\sim100$ GeV by measurements of the BETS \cite{Torii:2001aw}, HEAT \cite{DuVernois:2001bb}, CAPRICE \cite{Boezio:2000}, and MASS \cite{Grimani:2002yz} experiments, among others.  Since the secondary and dark matter contributions to the spectrum are at least an order of magnitude smaller than the primary contribution at all energies, the fit of the total electron spectrum to the data is determined almost exclusively by the shape of the primary spectrum.  For a given set a propagation parameters, the data determines the following electron flux parameters (though it does provide for some freedom in these parameters): the primary electron spectrum normalization, power law indices, and power law break rigidity.  The values of the injection power law index for energies greater than a few GeV that are consistent with local measurements include the values of 2.54 and 2.6, which we have used in our background models. Our results deviate from the local electron spectrum at energies below $E\cong 5$ GeV where solar modulation plays an important role.  This is because our calculations using GALPROP don't take into account the effects of solar modulation.  This disagreement is expected to show up as in \cite{Strong:2007nh}.  Changes in the propagation parameters may give rise to changes in the electron flux.  As an example, electrons lose energy relatively quickly as they propagate, so the high energy contribution to the local flux must come from electrons produced in nearby regions of the Galaxy.  Therefore, changes in the propagation parameters that affect propagation on short distance scales can affect the flux at high energies.  In this way, the uncertainties in the propagation parameters can lead to uncertainties in the local particle fluxes.

Secondary positrons and electrons are created as primary CR protons and Helium propagate through the Galaxy.  The local proton flux is constrained from a few tenths of a GeV up to $\sim180$ GeV by the measurements of BESS \cite{Sanuki:2000wh}, IMAX \cite{Menn:2000}, and AMS \cite{Alcaraz:2000vp}.  The primary nuclei spectra are described by a power law $E^{-\alpha}$, with an increasing value of $\alpha$ at higher energies. GALPROP allows for defining the constant values of $\alpha$ in two rigidity regions, the value of the rigidity that defines those regions (break rigidity), the normalization of the proton differential flux, and the proton kinetic energy at which the normalization is being carried out.  The proton data allows for some, though not much, uncertainty in these proton flux parameters.  As a result, there is some corresponding uncertainty in the secondary positron spectrum.  For energies higher than $E\cong10$ GeV, values of $\alpha$ in the range $\cong2.6 - 2.7$ are consistent with the locally observed spectra. Lower values, as in \cite{Abdo:2006fq}, where an index value of 2.42 was used, are also within the current uncertainties. Higher values of $\alpha$ for energies $E\gsim 10 \ \rm GeV$ result in a decrease in the flux of the secondary electrons and positrons. This in turn allows for a higher (up to a factor of $\cong 2$) $e^{+}_{DM}/e^{+}_{secondary}$ for a given local DM density $\rho_{0}$ and annihilation cross section.  Additionally, since positrons and nuclei propagate differently, the choice of propagation parameters can affect the positron spectrum for a particular choice of proton spectrum parameters.

The propagation of cosmic rays is governed by the transport equation, which includes the effects of diffusion and energy losses, among other things.  We chose to hold fixed the energy loss mechanisms provided in the GALPROP code.  However, in addition to varying the parameters associated with re-acceleration, we varied the following diffusion parameters: the diffusion zone length $2z_{max}$ (the perpendicular distance from the Galactic plane at which free escape of cosmic rays is assumed), and the energy-dependent diffusion coefficient, specifically through its normalization, break rigidity, and power law index above and below the break.  
 
Mechanisms for re-acceleration of charged particles within the halo are well motivated, but still poorly understood. In order to provide some sense of the effects of this uncertainty, we employed three different values of Alfv\'{e}n velocity, a parameter describing the strength of diffusive re-acceleration.  (The Alfv\'{e}n velocity relates the spatial diffusion coefficient to the momentum-space diffusion coefficient.)  In our calculations we take $v_{A}=0$ km/s corresponding to no re-acceleration, and $v_{A}=20$ km/s and 35 km/s corresponding to the inclusion of re-acceleration in CR propagation.  These values have been used extensively in the literature \cite{Strong:1999su} and are within the acceptable range of values \cite{Moskemail,Schlickeiser}.  The effect of re-acceleration is to shift the flux of electrons and positrons to higher energies.  In our calculations, significant contributions to the fluxes occurred only for energies well below 20 GeV, nonetheless, we often see differences in the high energy positron fraction when varying $v_A$. This arises as follows: our primary and secondary fluxes in the low energy range of the HEAT data are highly dependent on the value of the Alfv\'{e}n velocity.  In order to fit our three backgrounds to the first five HEAT data points, as previously described, we varied the primary electron spectrum through the low energy power law index, the break energy of this index, and the normalization. (The values of these parameters for the three backgrounds are listed in Table~\ref{tab:elecspecparams}.) The changes in the low energy background then carry over to high energies, and thus lead to variations in the positron fraction at $E\gsim 20~{\rm GeV}$.

The GALPROP code provides for an energy-dependent spatial diffusion coefficient $D_{xx}$ defined as:
\be
\label{eq:spatial diff. coef. eq.} D_{xx} = \beta D_{0xx}\left(\frac{R}{D_{rigid \; br}}\right)^{D_{g}}
\ee
where $R$ is the rigidity of the particles, $\beta=v/c$, $D_{0xx}$ is the diffusion coefficient divided by $\beta$ at rigidity $D_{rigid \; br}$, and the index $D_{g}$ can take the value $D_{g1}$ for $R<D_{rigid \; br}$ and a different different value $D_{g2}$ for $R>D_{rigid \; br}$ \cite{Galprop1}.
For a given $z_{max}$ and re-accleration strength, the diffusion coefficient is determined by fitting the Boron-to-Carbon (B/C) ratio to local measurements.  In general, an overall decrease in the diffusion coefficient gives an enhancement in the flux of low energy electrons and positrons, while having little effect on the proton spectrum.  This effect provides for additional uncertainty in the secondary positron spectrum; the fits to the proton data can be maintained for many variations on the secondary positron spectrum.

In the models presented here we use $D_{0xx}=5.8\times 10^{28} \rm cm^{2} s^{-1}$.  Increasing the value of $D_{0xx}$ results in an increase in the ratio $e^{+}_{DM}/e^{+}_{secondary}$ at the energies of interest to us, $E\gsim100$ GeV, but for values of $D_{0xx}\gsim6-7\times 10^{28} cm^2 s^{-1}$ with $D_{rigid \; br}\cong$ few GeV, serious inconsistencies arise between the calculated electron flux and the local electron flux, unless we choose a smaller power law index for the primary electron spectrum for energies above a few GeV.

For energies where the XDM positron contribution becomes comparable to the secondary positron flux, the power law index $D_{g}$ is significant. Values between $D_{g}=1/3$ (corresponding to Kolmogorov turbulence) and $D_{g}=1/2$ (corresponding to Kraichnan turbulence) are considered reasonable \cite{Hooper:2004bq,Galprop1}. Increasing $D_{g}$ from $1/3$ to $1/2$ can increase $e^{+}_{DM}/e^{+}_{secondary}$ up to a factor of 2.  However, to be consistent with the local proton and CR nuclei measurements, the primary proton power las index must be decreased in the region of energies higher than $E\cong10 \rm GeV$. This cancels the gain in $e^{+}_{DM}/e^{+}_{secondary}$ by about the same factor.

The diffusion zone width $2z_{max}$ can take on a range of values.  The allowed values for propagation models that include re-acceleration and for those models with both no re-acceleration and no convection are in the range $4<z_{max}<12$ kpc \cite{Strong:1998pw}.  Increasing the diffusion zone width allows fewer particles to escape from the confines of the Galaxy, so particles from farther away are able to make their way to our location in the Galaxy.  As mentioned earlier, electrons and positrons lose energy rapidly as they propagate.  The result is that the flux of low energy electrons and positrons is enhanced.  The positron flux at energies $E\gsim 100\rm GeV$ is affected negligibly, as the highly energetic positrons we measure come from a sphere with a much smaller radius than the diffusion zone width.

\vspace{\baselineskip} 
\noindent{\it{Relevance to Synchrotron Radiation (The Haze)}}

\vspace{\baselineskip}
Since the emissivity of synchrotron radiation at a specific frequency is proportional to the magnetic field, the most dominant parameter determining the shape of the synchrotron spectrum, apart from the DM halo profile, is the magnetic field.  A reasonable value for the local magnetic field is $B\cong 3\ \rm \mu G$ with $5\  \rm \mu G$ being within the limits.  For the center of the Galaxy, values within the range of $5-20\  \mu G$ are acceptable. Little can be said about the parametric dependence of the intergalactic magnetic field on $z$, the vertical distance from the galactic plane, or on the radial distance $R$ from the center of the Galaxy at $z=0$. We considered, as in \cite{Abdo:2006fq}, a magnetic field of the form $B=B_{l}e^{-\frac{R-R_{\odot}}{R_{c}}}e^{-\frac{z}{z_{c}}}$, where $B_{l}$ is the local value of the B-field, and $R_{c}$ and $z_{c}$ are the characteristic scales of the B-field, taken to be $10~{\rm kpc}$ and $2~{\rm kpc}$, respectively. We took the local value of the magnetic field to be $5\  \rm \mu G$. More homogeneous magnetic fields have been used in the literature \cite{Hooper:2007kb}.

The synchrotron spectrum also depends on the CR diffusion parameters.  A smaller diffusion coefficient suppresses the diffusion of $e^{+}$ and $e^{-}$, thus they remain inside the diffusion zone for longer, yielding larger synchrotron fluxes. Because the synchrotron radiation at the Haze frequencies gets it's dominant contribution from $e^{+}$ and $e^{-}$ with energies above a few GeV, the dependence of the diffusion coefficient on the rigidity at high energies can have a small effect on the synchrotron flux. Since the Haze is in the region of $\cong6-15$ degrees in latitude or $r\cong0.9-2.3$ kpc from the center of the Galaxy, varying the diffusion zone width above $z\cong4$ kpc doesn't have significant effects on the shape of the synchrotron radiation as a function of latitude. 

Additionally, varying the ratio of the energy density in the magnetic field to the energy density in the radiation field $U_{mag}/U_{rad}$, which is equal to the ratio of energy loss of high energy $e^{+} e^{-}$ through synchrotron radiation to that through inverse Compton scattering, has an affect on the synchroton radiation produced \cite{Grajek:2008jb}.  A larger value of $U_{rad}$ results in faster depletion of the number of high energy $e^{+}e^{-}$ pairs, while lower values can yield a stronger haze signal.

\section{Results}
\label{sec:results}

\subsection{Baseline Models}

\begin{figure}[htpb]
\begin{center}
\subfigure[Positron fraction, direct decay channel]{\epsfig{figure=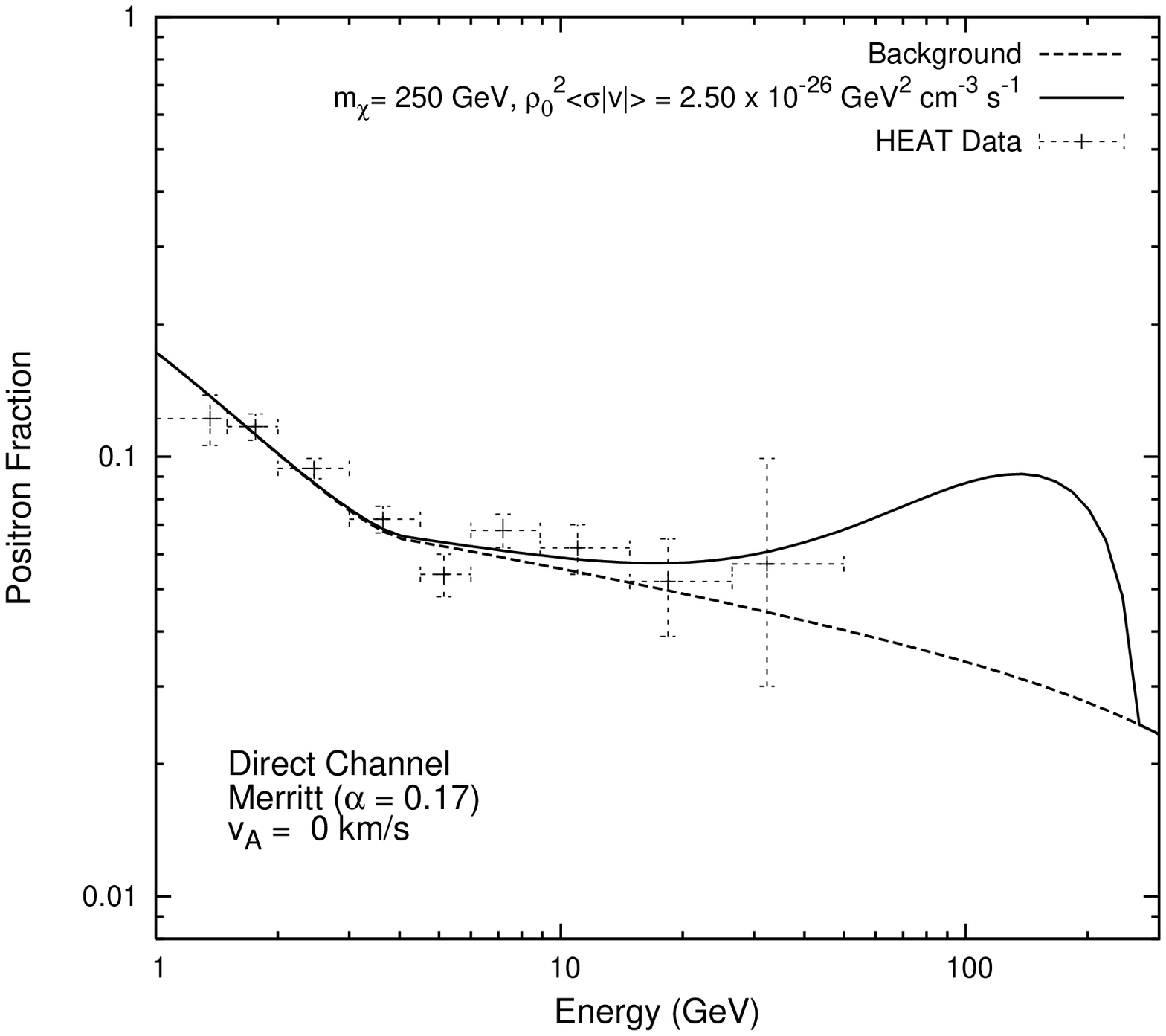,scale=.4}}
\hskip 0.15in
\subfigure[Synchrotron radiation, direct decay channel]{\epsfig{figure=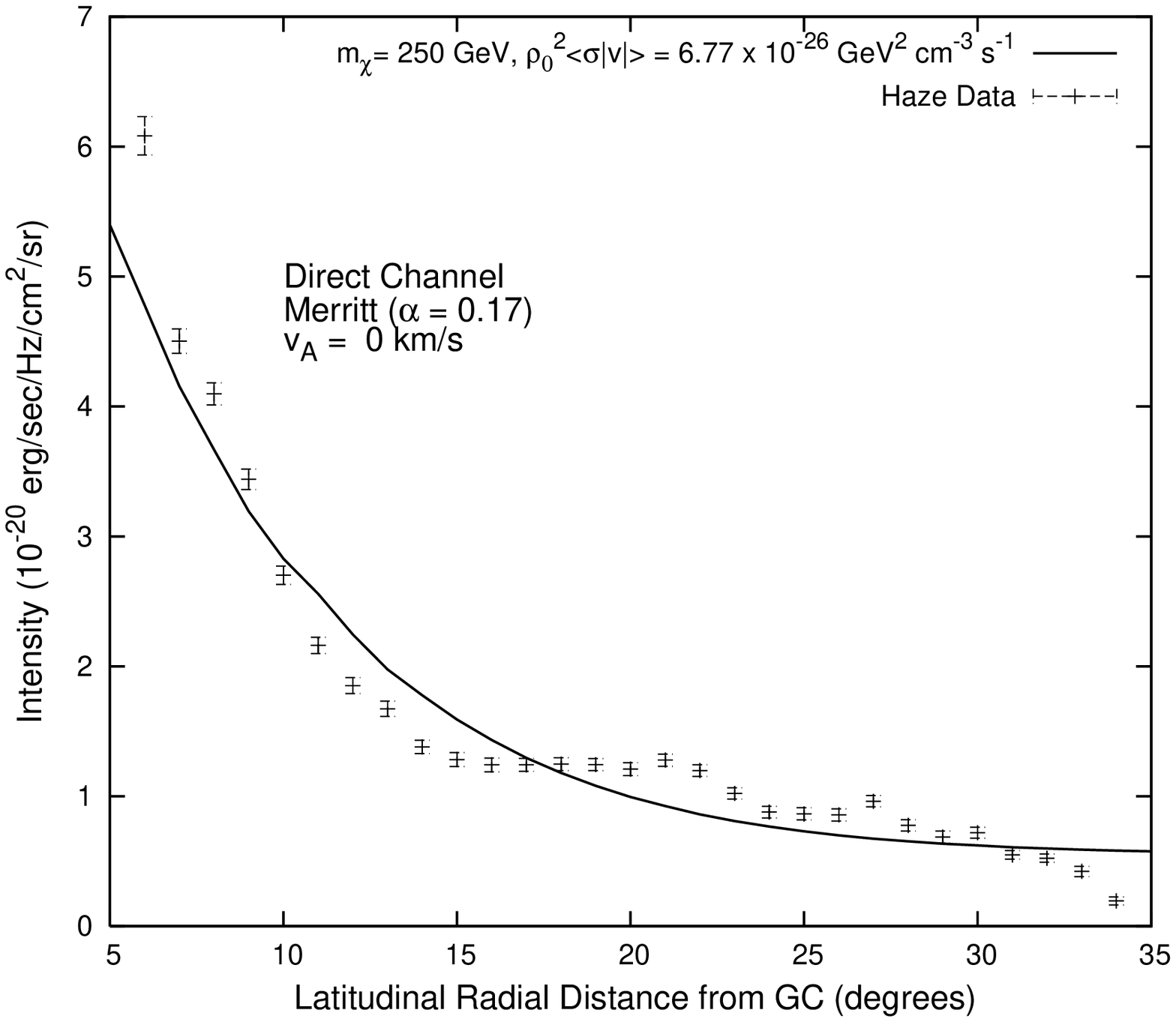,scale=.4}}\\
\subfigure[Positron fraction, muon decay channel]{\epsfig{figure=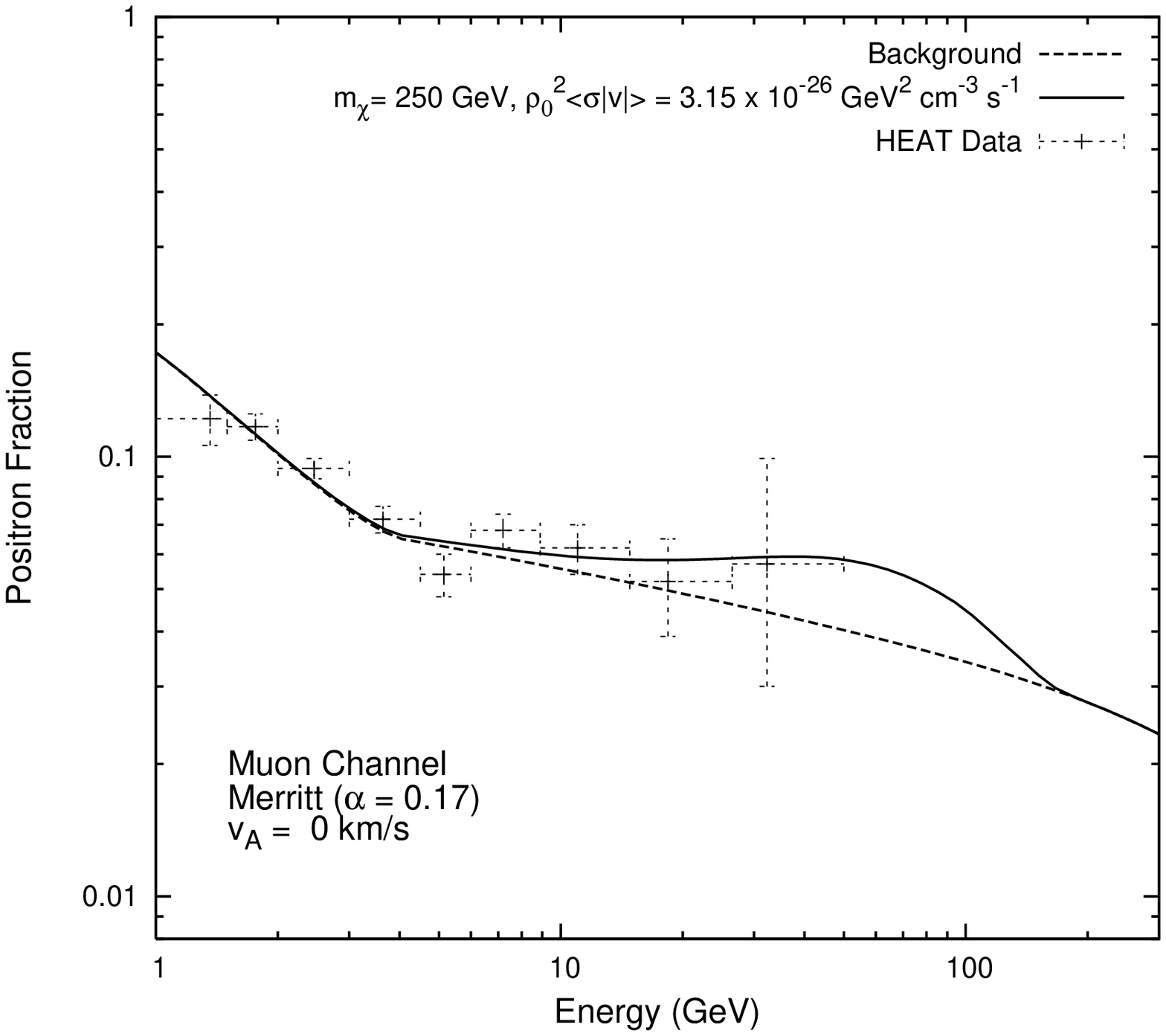,scale=.4}}
\hskip 0.15in
\subfigure[Synchrotron radiation, muon decay channel]{\epsfig{figure=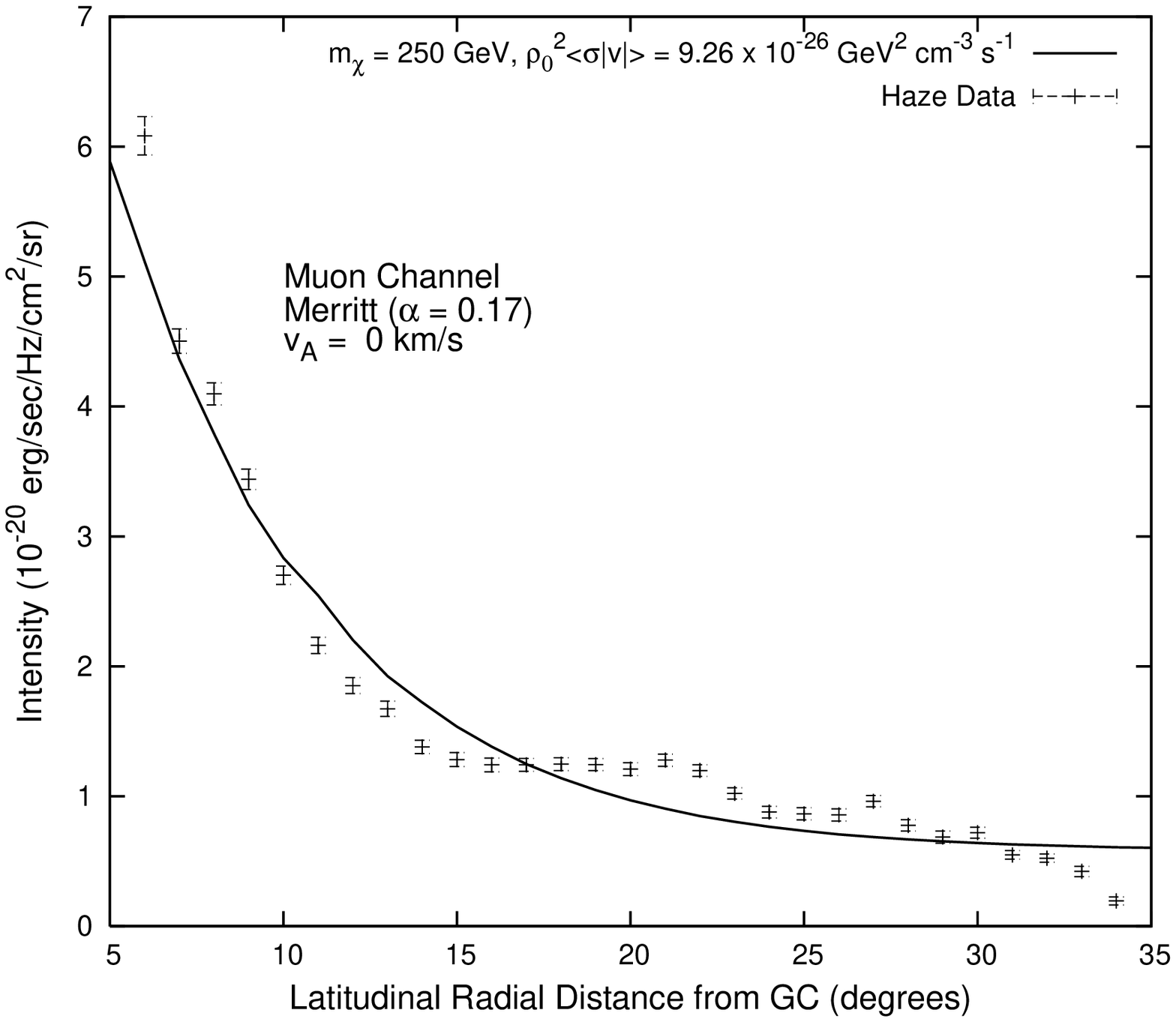,scale=.4}}
\end{center}
\caption{Canonical models for $v_A = 0$ km/s.}
\label{fig:canonical0}
\end{figure}

In order to determine the sensitivity of the signal to various unknown variables, such as particle mass and halo model, we consider a particular baseline or 
``canonical'' model about which to vary. We have two canonical models, one without and one with re-acceleration.  The model without re-acceleration is described by the conventional parameter set discussed previously with $m_{\chi} = 250$ GeV and with a Merritt halo profile with $\alpha = 0.17$ (the middle value in the acceptable range of values). These parameters are chosen because they generate a sufficiently large positron signal to explain the INTEGRAL excess \cite{Finkbeiner:2007kk}.  Fig.~\ref{fig:canonical0} shows our results for our canonical model with no re-acceleration.

\begin{figure}[htpb]
\begin{center}
\subfigure[Positron fraction, direct decay channel]{\epsfig{figure=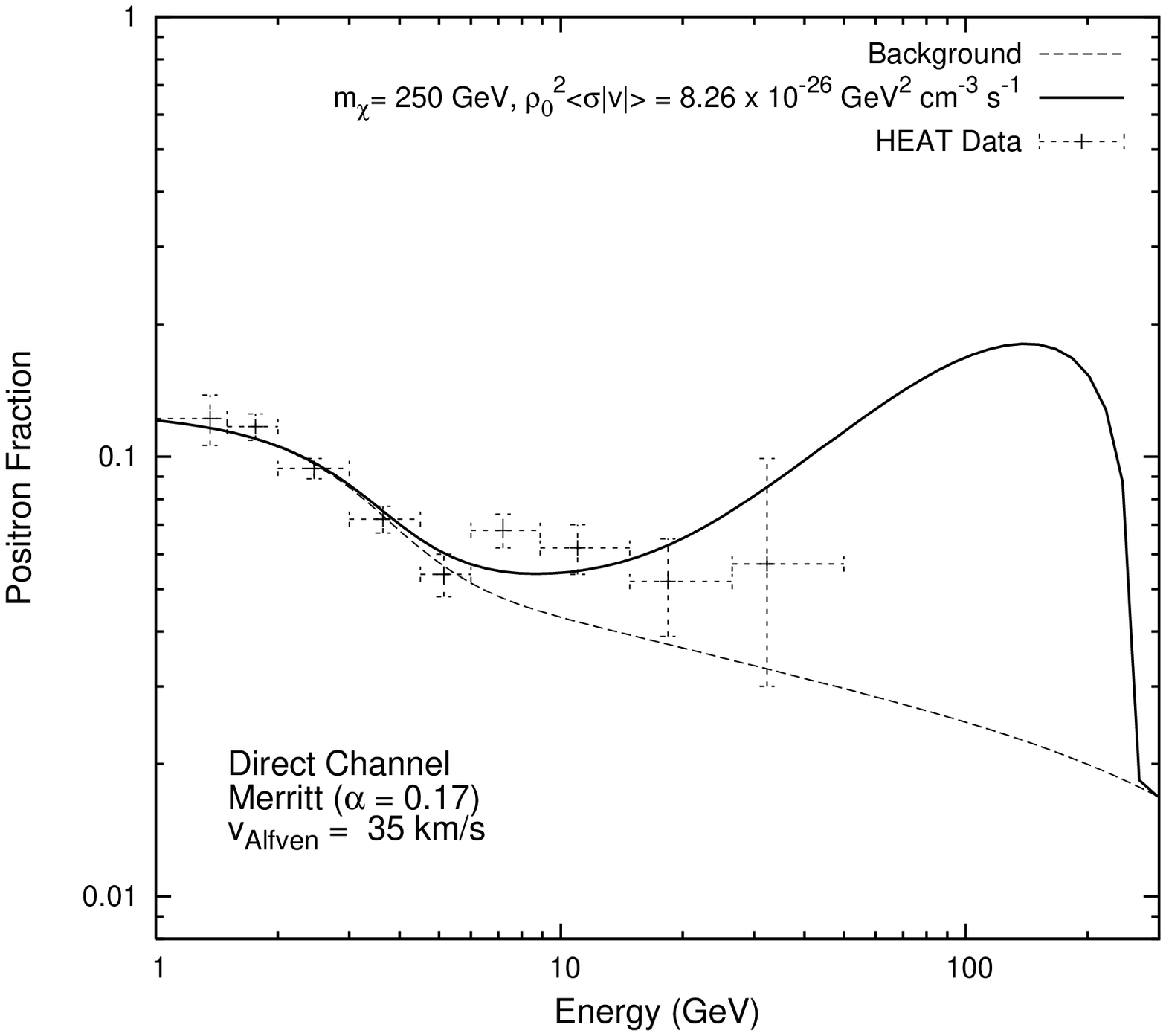,scale=.4}}
\hskip 0.15in
\subfigure[Synchrotron radiation, direct decay channel]{\epsfig{figure=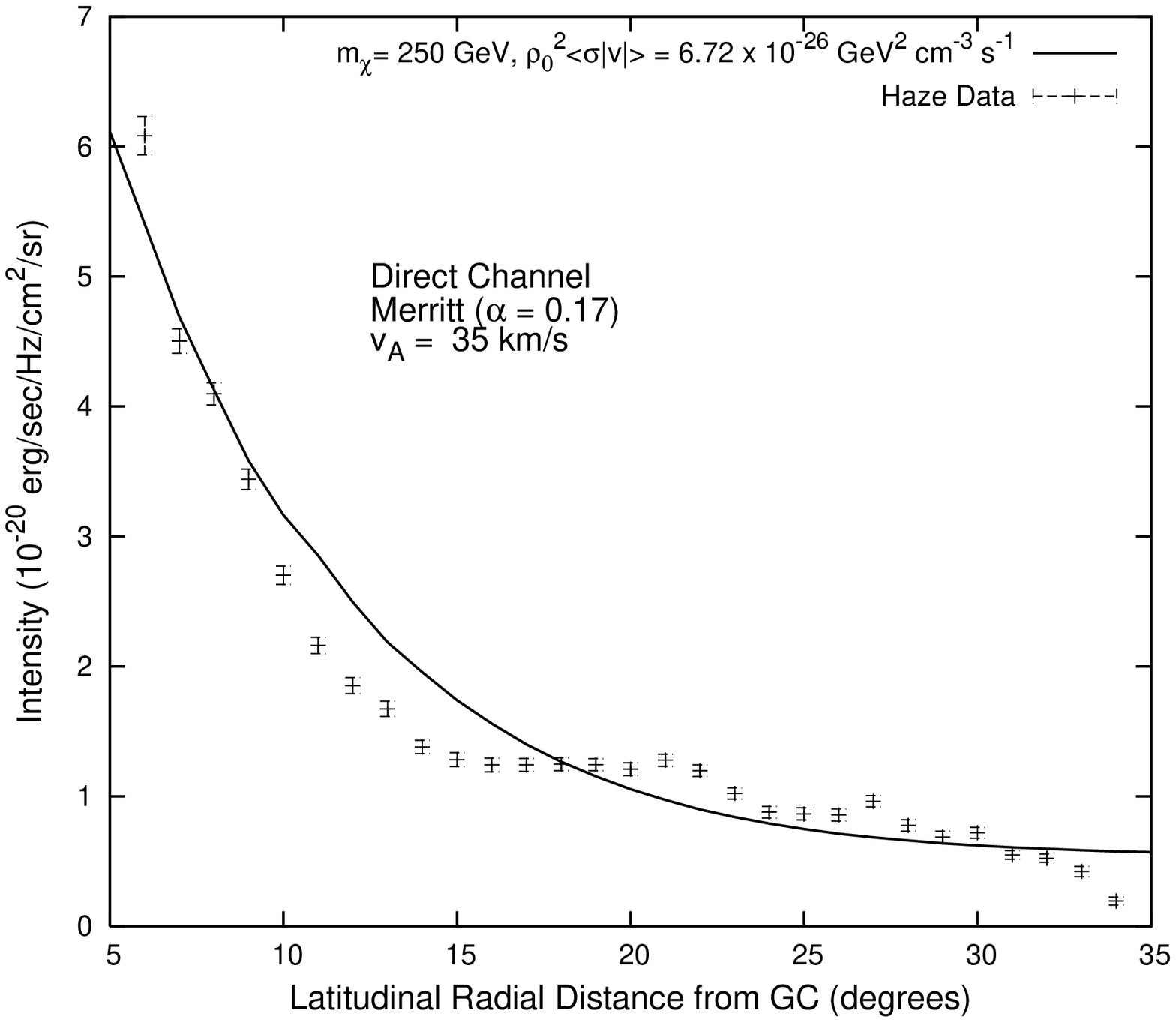,scale=.4}}\\
\subfigure[Positron fraction, muon decay channel]{\epsfig{figure=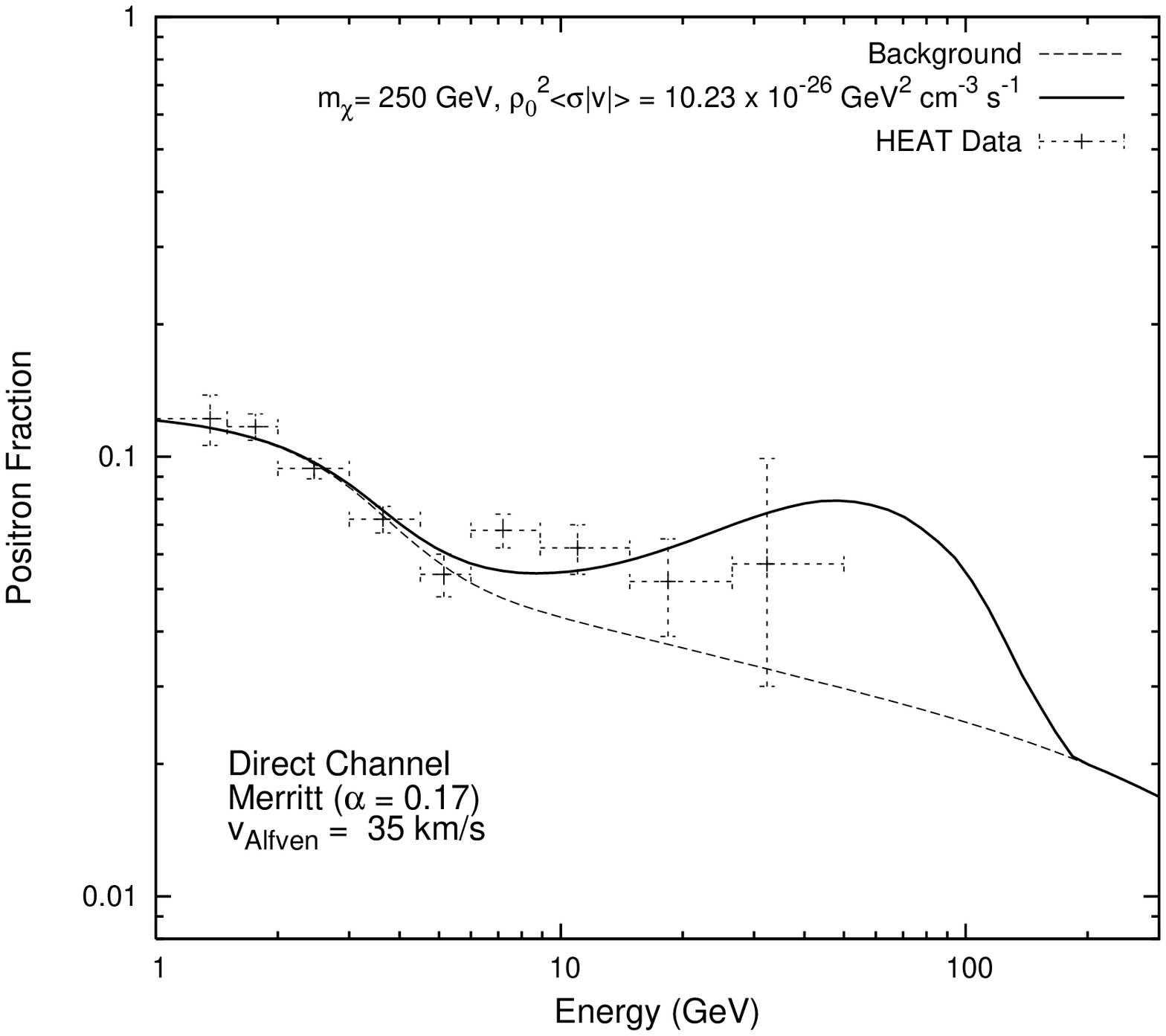,scale=.4}}
\hskip 0.15in
\subfigure[Synchrotron radiation, muon decay channel]{\epsfig{figure=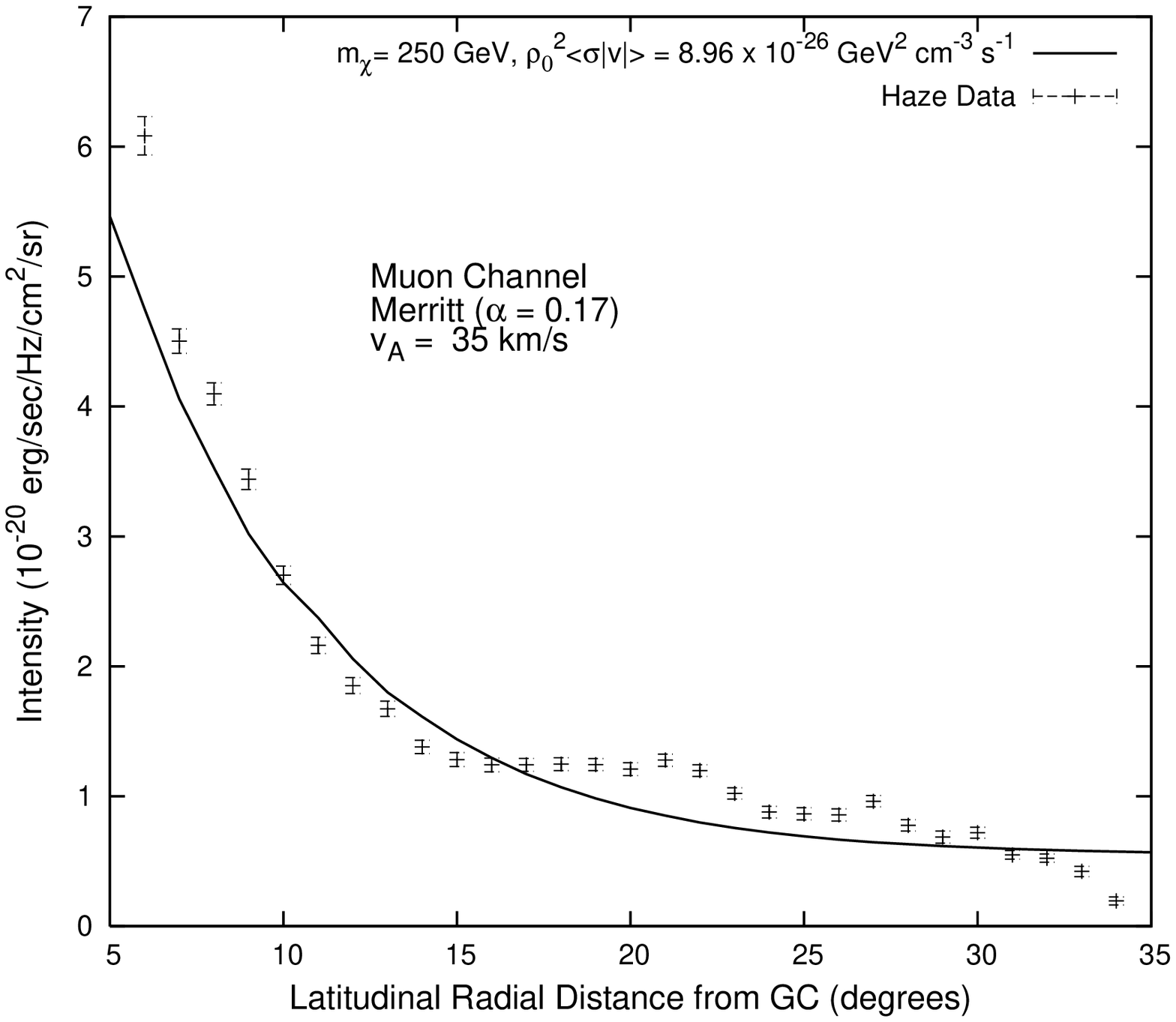,scale=.4}}
\end{center}
\caption{Canonical models for $v_A = 35$ km/s.}
\label{fig:canonical35}
\end{figure}

For a scenario with re-acceleration, we consider the same particle mass and profile and take  $v_A = 35$ km/s.  See Fig.~\ref{fig:canonical35} for the results of our canonical model with re-acceleration.  An Alfv\'{e}n velocity of 35 km/s is reasonable, and we find that this value yields the smallest differences between the needed cross sections for the haze and HEAT, for a given $m_\chi$ (see Appendix \ref{ap:bestfit}). The cross-sections differ at most by a factor of about 2 for $v_A = 35$ km/s (see Tables \ref{tab:Table direct ch vA=35} and \ref{tab:Table muon ch vA=35}), while for $v_A = 20$ km/s they more often than not differ by factors of 3 and higher (see Tables \ref{tab:Table direct ch vA=20} and \ref{tab:Table muon ch vA=35}). Throughout the following discussion, the behavior of other models is very similar to that of the canonical models, unless otherwise noted.  We refer the reader to the tables in Appendix \ref{ap:bestfit} for a listing of the best-fit values for $\rho_{0}^{2}\left\langle\sigma\left|v\right|\right\rangle$ and the corresponding $\chi^{2}$ values for all models studied.

The plots of the positron fraction include the HEAT data with error bars as calculated by the HEAT collaboration \cite{Barwick:1997ig}.  The synchrotron radiation plots include the haze data as calculated by Dobler and Finkbeiner in \cite{Dobler:2007wv}, based on WMAP measurements taken directly south of the Galactic Center.  The data are in bins separated by 1 degree in latitude with a width of 20 degrees in longitude $l\in[-10,10]$. 

\subsection{Positron Fraction}
With the results from PAMELA on the horizon, the positron production from dark matter is clearly an important avenue to pursue. However, the positron fraction contains numerous uncertainties, both in the background and in the shape of the DM signal, should there be one. In this section, we discuss a number of the different parameters and their effects on the positron fraction signal.

\subsubsection{Decay Channel Dependence}

\begin{figure}[htpb]
\begin{center}
\subfigure[$v_A = 0$ km/s, $m_{\chi} = 50$ GeV]{\epsfig{figure=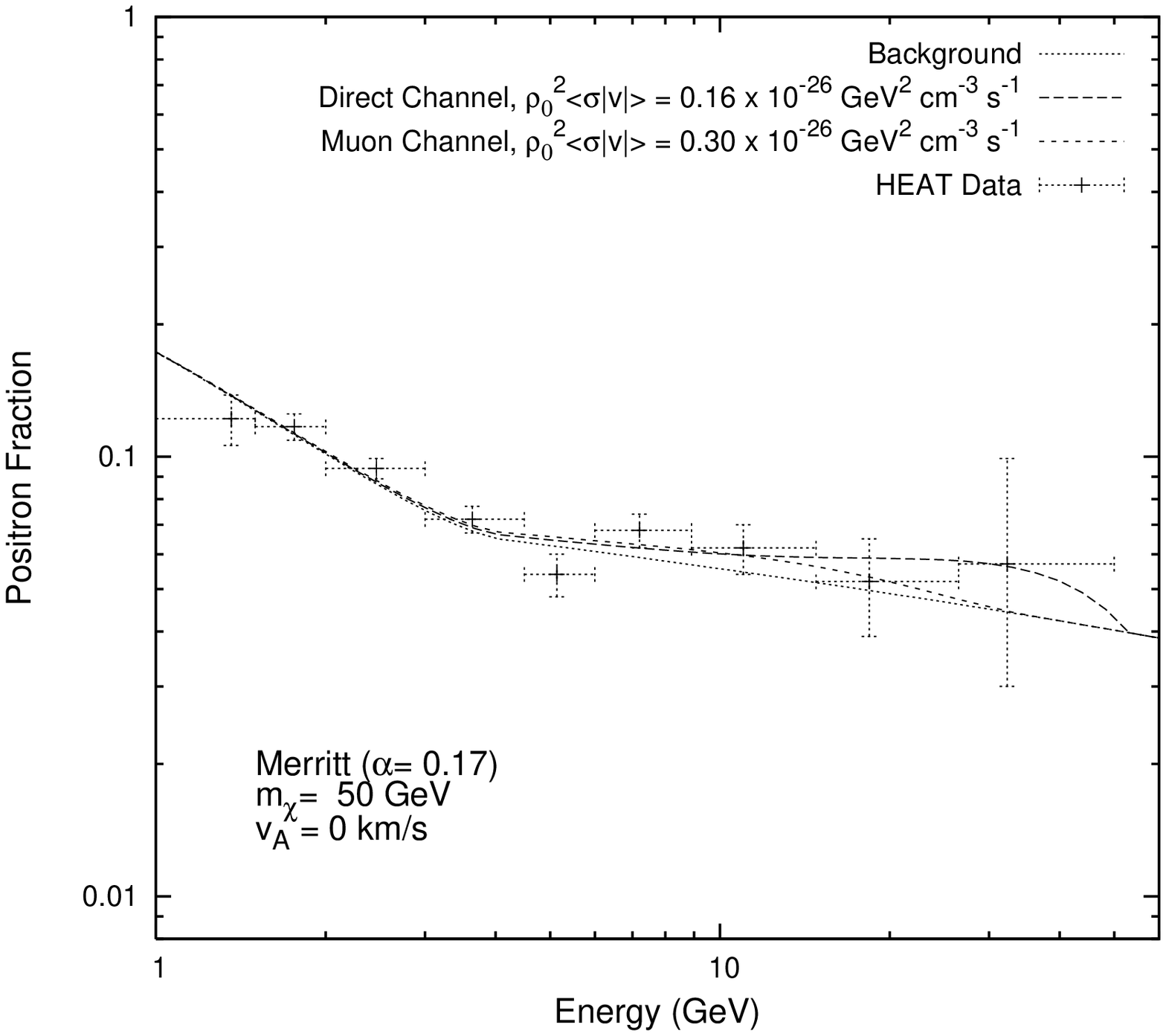,scale=.4}}
\hskip 0.15in
\subfigure[$v_A = 0$ km/s,  $m_{\chi} = 250$ GeV]{\epsfig{figure=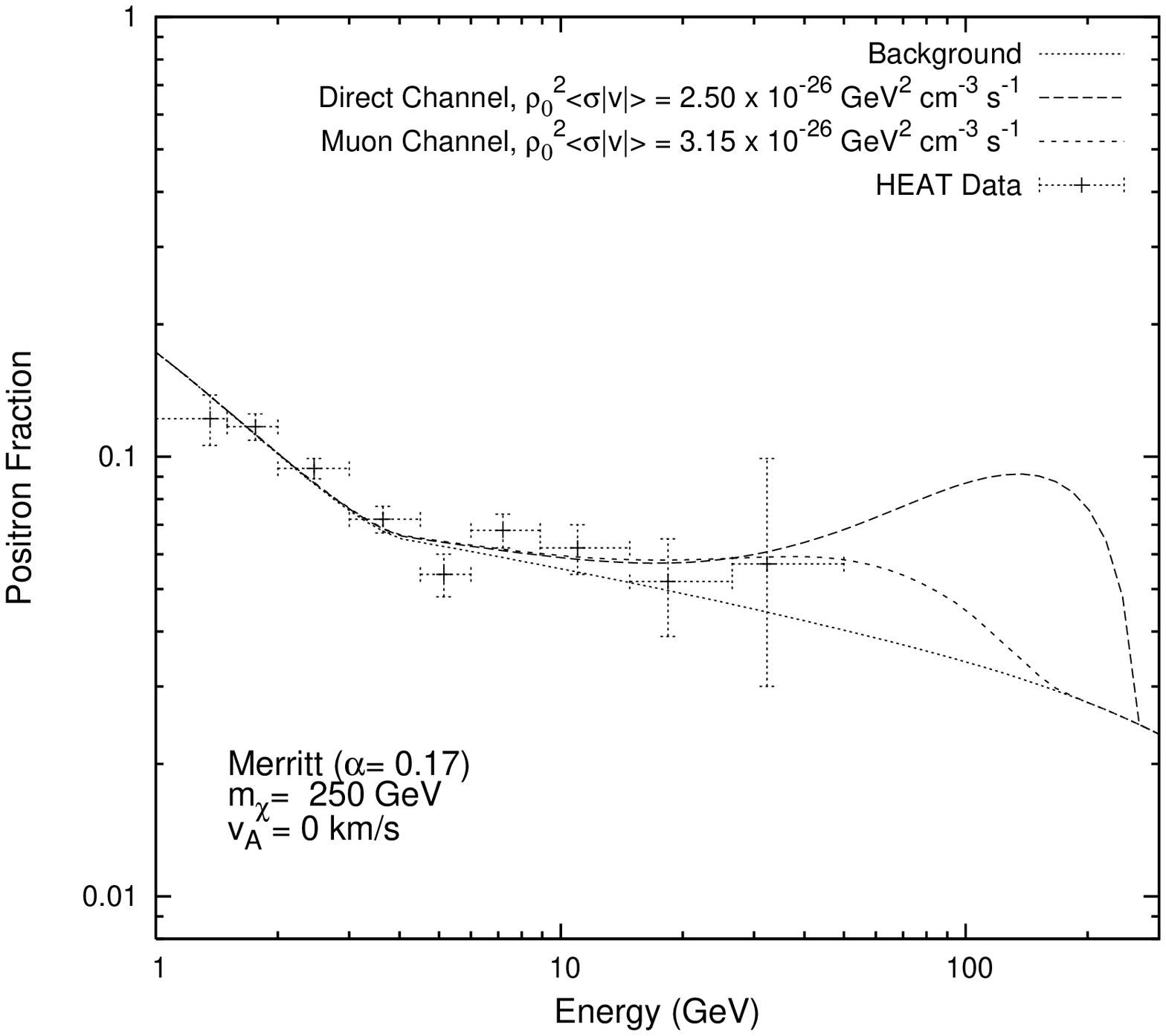,scale=.4}}\\
\subfigure[$v_A = 35$ km/s, $m_{\chi} = 50$ GeV]{\epsfig{figure=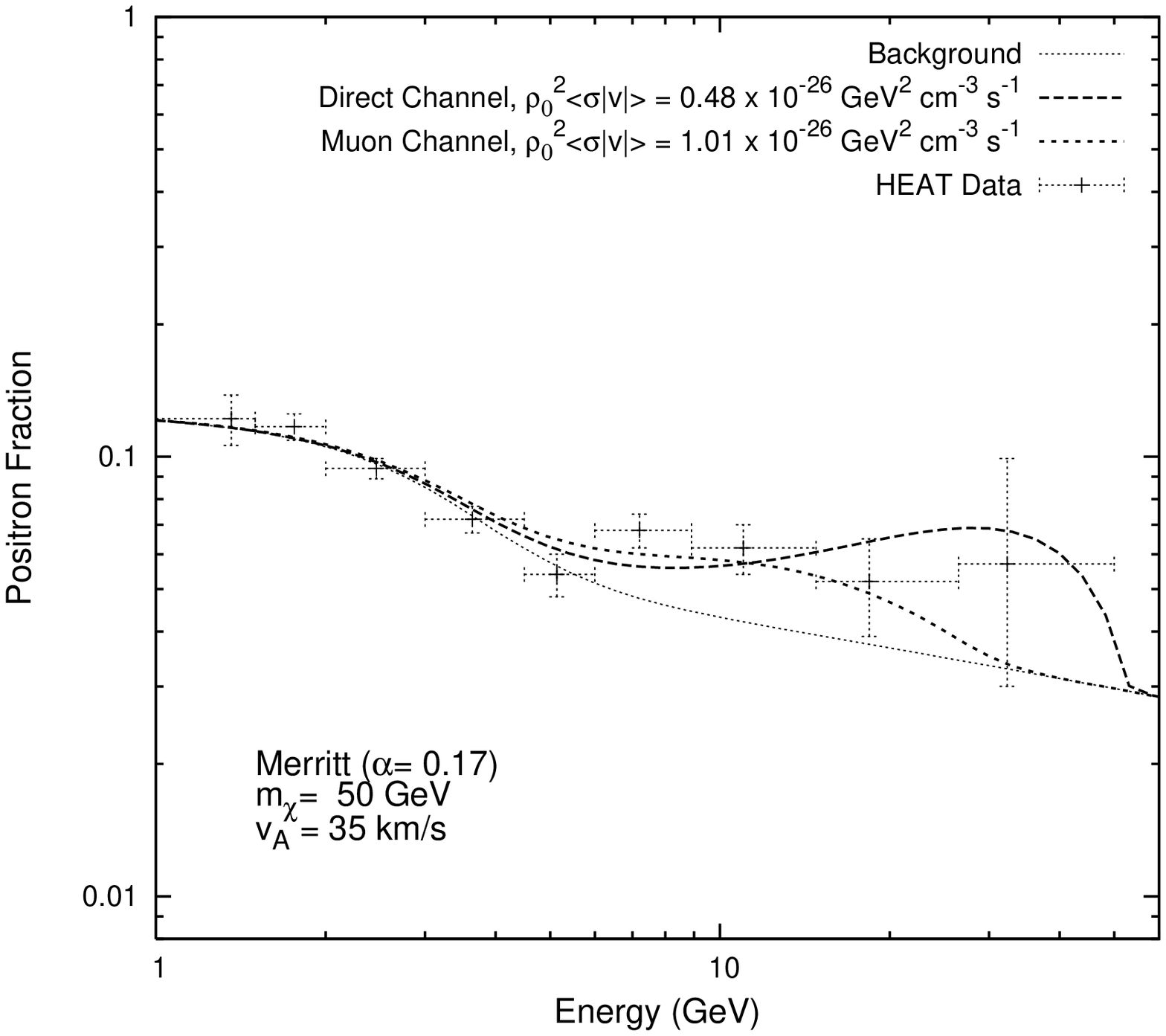,scale=.4}}
\hskip 0.15in
\subfigure[$v_A = 35$ km/s,  $m_{\chi} = 250$ GeV]{\epsfig{figure=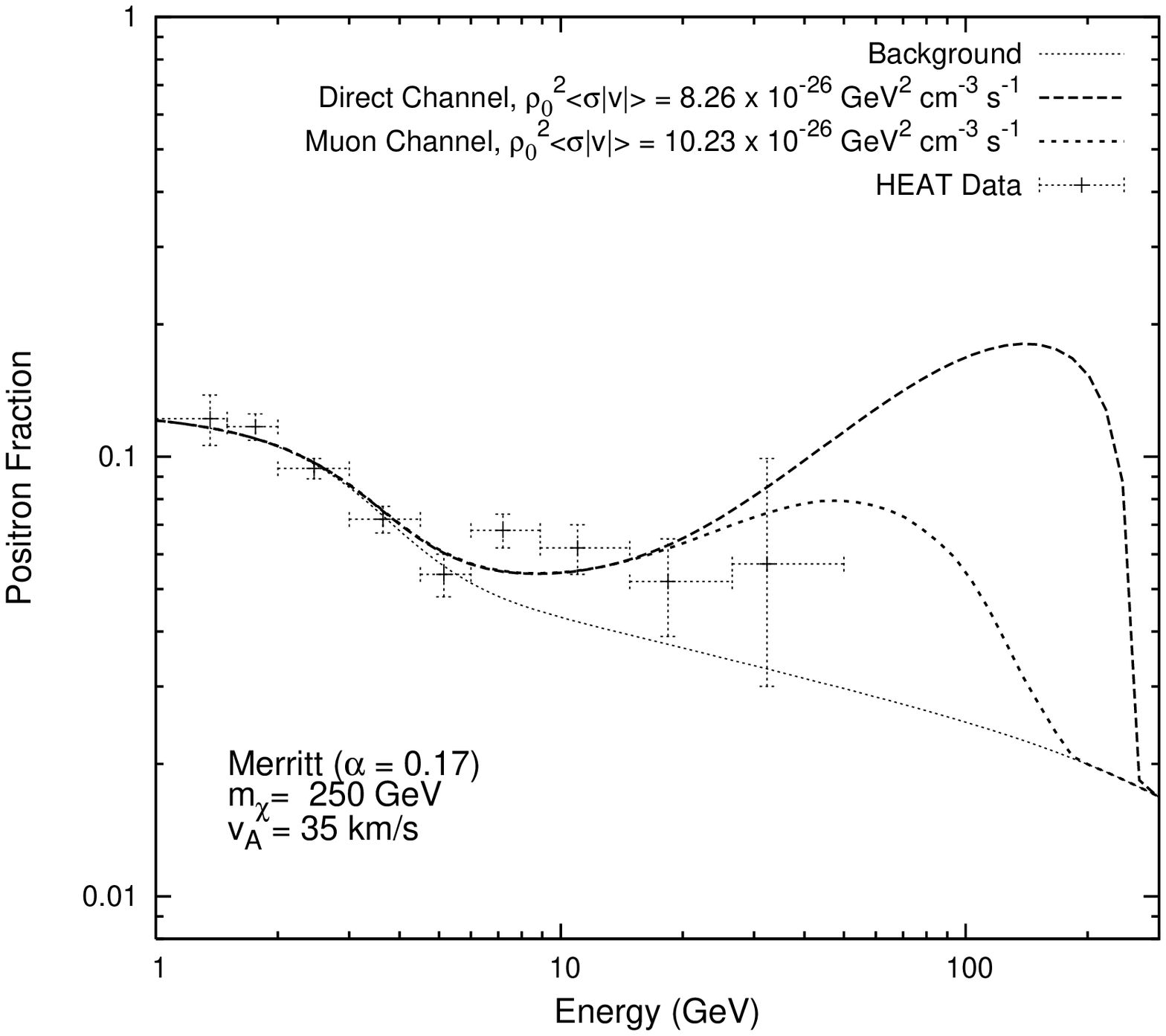,scale=.4}}
\end{center}
\caption{Positron fraction for direct decay ($2m_{e}< m_{\phi}<2m_{\mu}$) and muon decay ($2m_{\mu}\leq m_{\phi}< 2m_{\pi^{0}}$) channels.}
\label{fig:Rchandepend}
\end{figure}

The signals are noticeably different for energies close to $m_{\chi}$ for the two different decay channels, that is for the different mass ranges of the mediator $\phi$.  See Fig.~\ref{fig:Rchandepend}. The bump in the positron fraction, characteristic of the hard XDM $e^{+} e^{-}$ spectra, has a larger maximum value for the direct decay channel.  Moreover, in the muon channel the ratio starts its drop to the background value at a lower energy, so that the high energy fall-off is more gradual.  These features can be understood by comparing the injection spectra for the two decay channels as shown in Fig. \ref{fig:decaychannels}.  The flat distribution of the direct channel takes on a larger value at high energies than does the distribution of the muon channel.  More $e^{+} e^{-}$ pairs produced at high energies give rise to a larger positron fraction at these energies.  This also results in a smaller best-fit cross-section for the direct channel.  

\subsubsection{$m_\chi$ Dependence}

\begin{figure}[htpb]%
\begin{center}
\subfigure[Direct decay channel, $v_A = 0$ km/s]{\label{fig:Rdirnoreac}\epsfig{figure=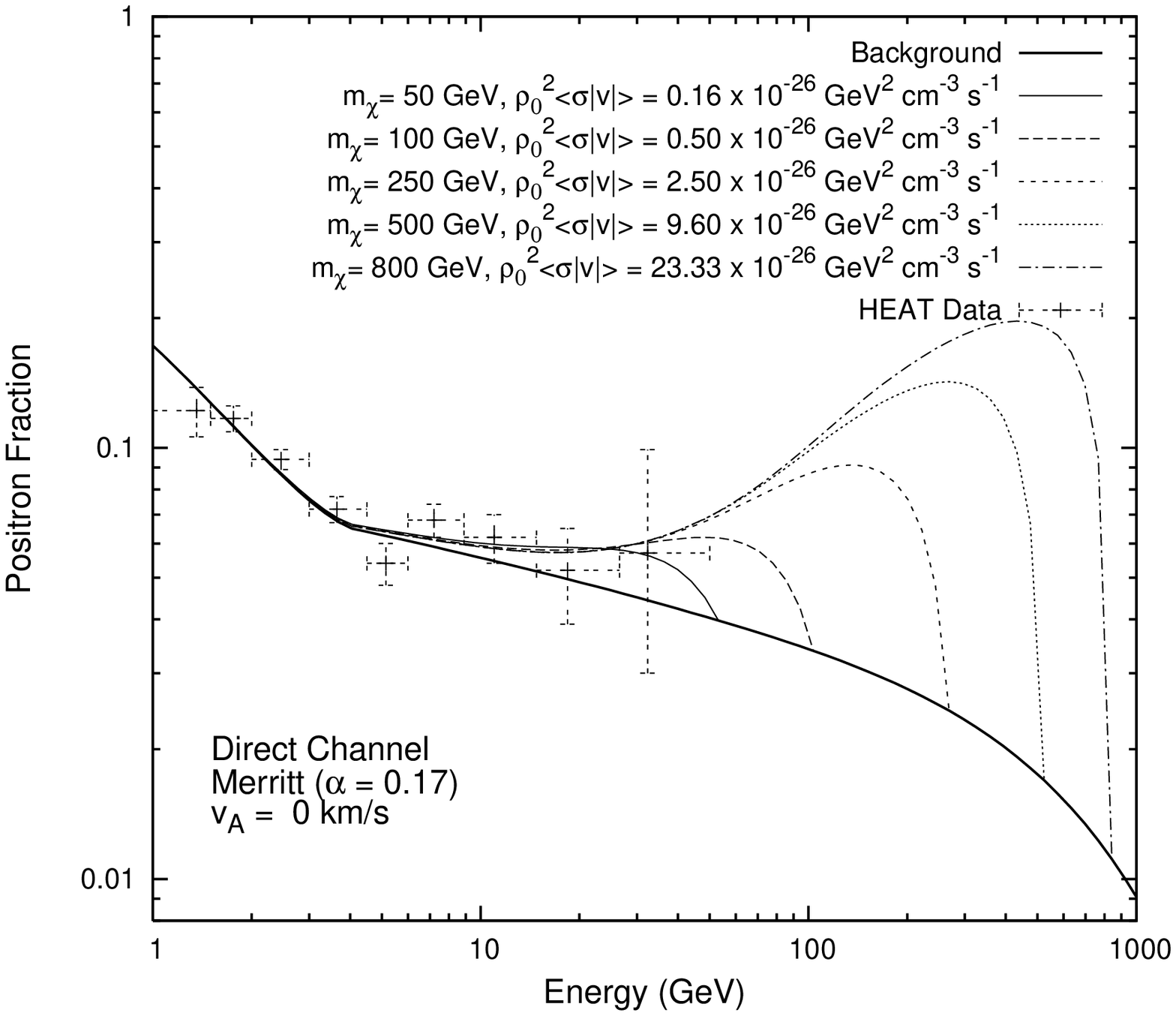,scale=.4}}
\hskip 0.15in
\subfigure[Muon decay channel, $v_A = 0$ km/s]{\epsfig{figure=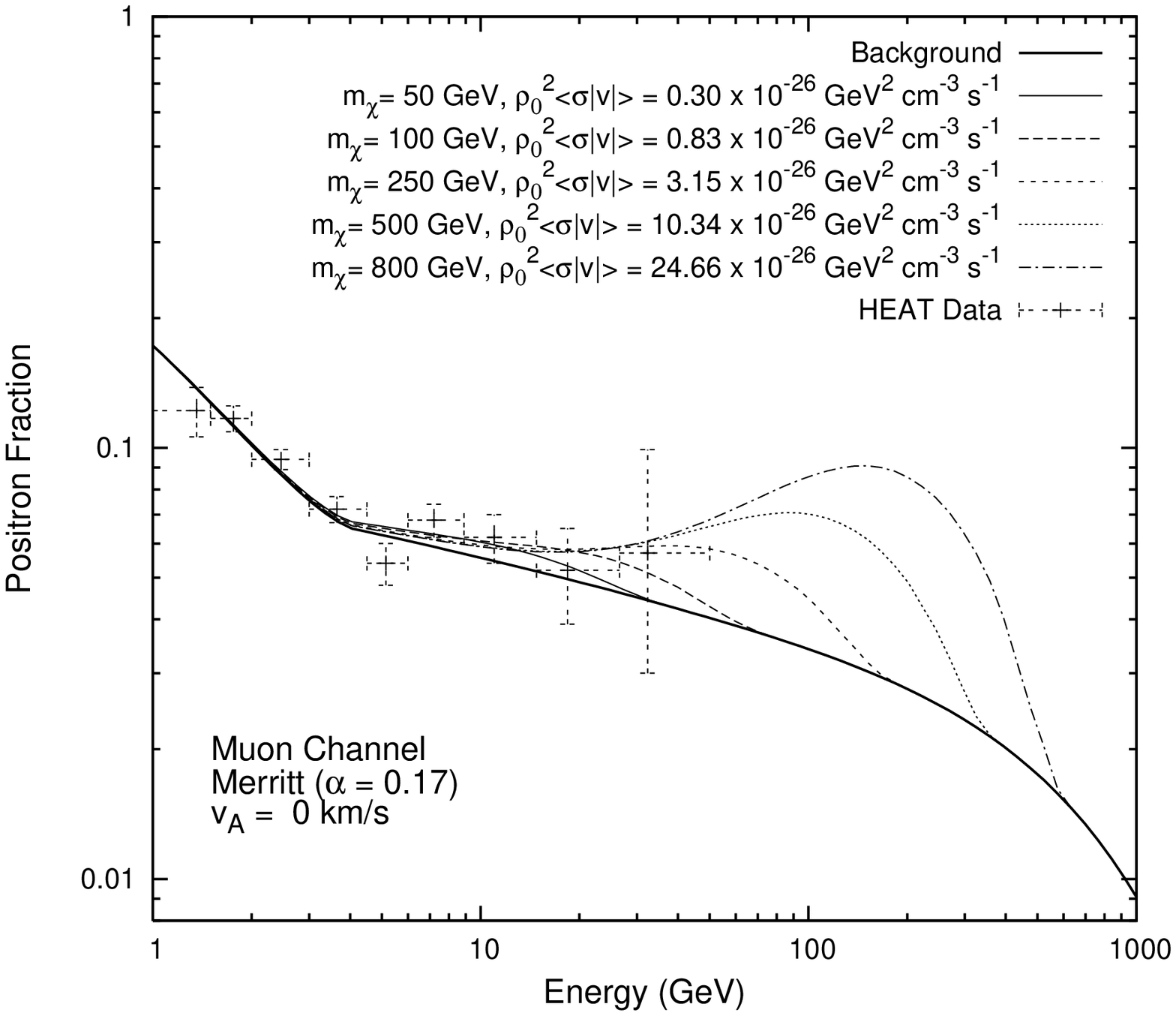,scale=.4}}\\
\subfigure[Direct decay channel, $v_A =35$ km/s]{\label{fig:Rdirreac}\epsfig{figure=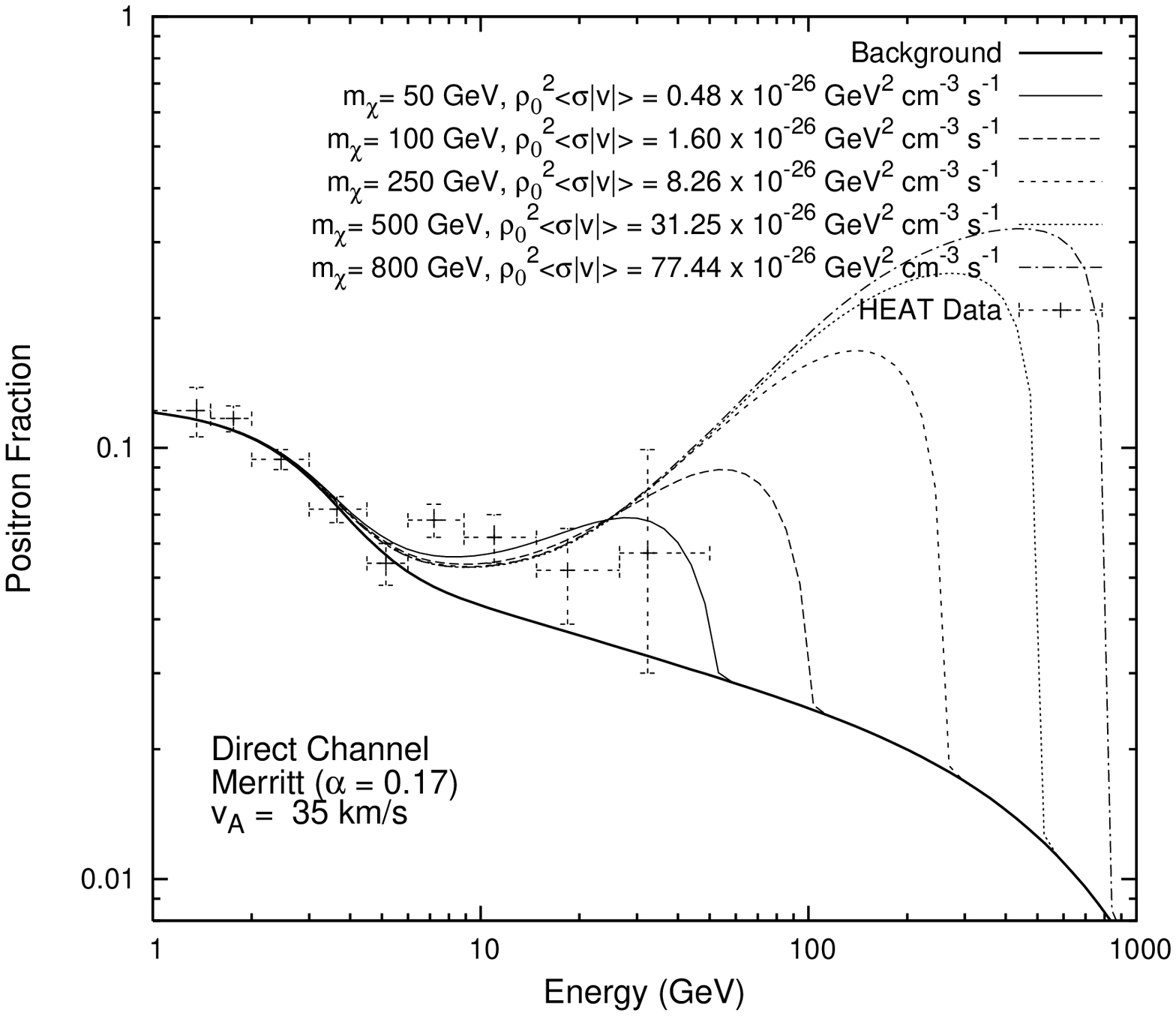,scale=.4}}
\hskip 0.15in
\subfigure[Muon decay channel, $v_A = 35$ km/s]{\epsfig{figure=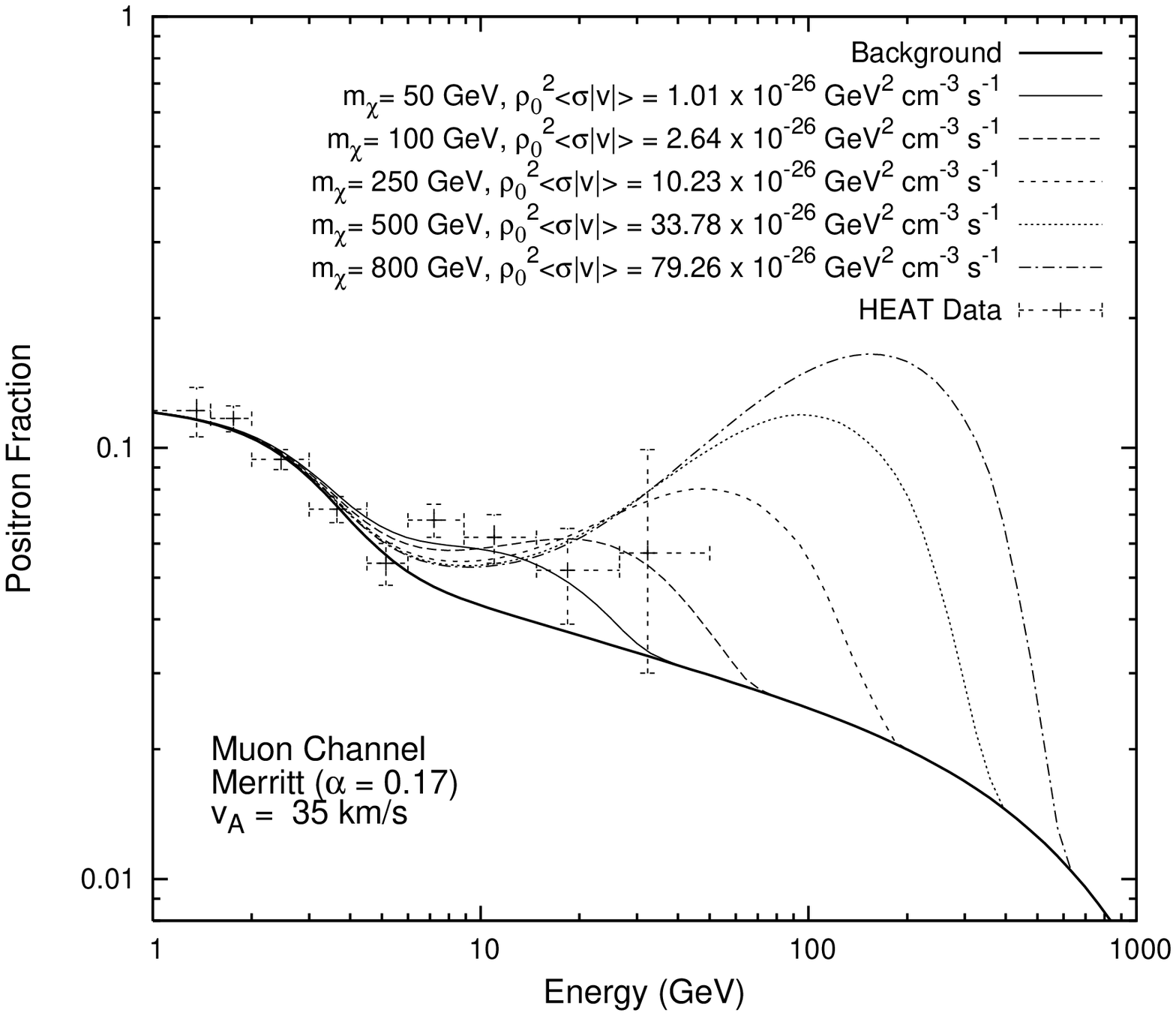,scale=.4}}
\end{center}
\caption{$m_{\chi}$ dependence of positron fraction.}%
\label{fig:ratiomassdepend}
\end{figure}

As mentioned previously, the decay of the boosted $\phi$ particle directly to light fermions results in a hard positron spectrum.  This gives rise to a characteristic enhancement in the positron fraction at energies above 5 GeV. The maximum in the positron fraction occurs at an energy that depends on the mass of $\chi$ and the decay channel; for example, for no re-acceleration and $m_{\chi} = 250$, GeV the maximum occurs at $\sim$ 140 GeV for the direct channel, and at $\sim$ 50 GeV for the muon channel.  The rapid, high energy fall-off of the positron fraction for the direct channel provides a way to determine $m_{\chi}$.   For the case with no re-acceleration, the high energy fall-off of the positron fraction occurs at $\sim m_{\chi}$ (see Fig.~\ref{fig:Rdirnoreac}), while for the case with re-acceleration, the fall-off occurs at an energy larger than $m_{\chi}$ and that depends on the strength of the re-acceleration (see Fig.~\ref{fig:Rdirreac}).  Fig.~\ref{fig:ratiomassdepend} shows the dependence of the ratio on $m_{\chi}$ for both the direct and muon decay channels for both the no re-acceleration and the re-acceleration cases.

As the mass of $\chi$ increases, the best-fit cross-section correspondingly increases. This arises from the lower number density, $n_{\chi} = \rho_{\chi}/m_{\chi}$, for fixed dark matter density. However, because higher energy particles can propagate farther, a given energy bin for HEAT is sensitive to a larger volume as one increases the energy range of the injection spectrum. This partially compensates for the decreased number density, but significantly larger cross sections are still required for higher DM masses.

\subsubsection{Dependence on Halo Profile}

\begin{figure}[htpb]
\begin{center}
\subfigure[Direct decay channel, $v_A = 0$ km/s]{\epsfig{figure=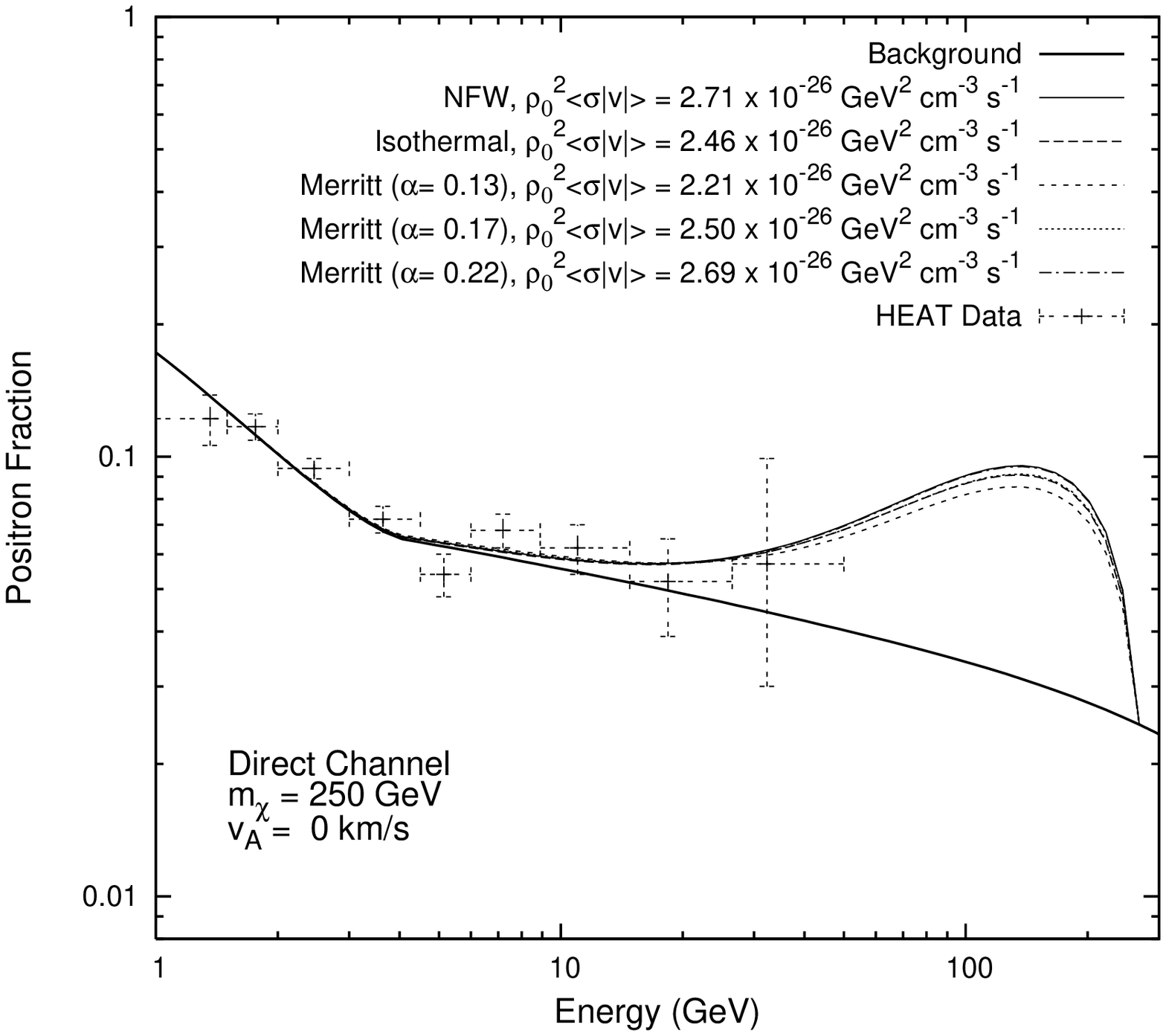,scale=.4}}
\hskip 0.15in
\subfigure[Muon decay channel, $v_A = 0$ km/s]{\epsfig{figure=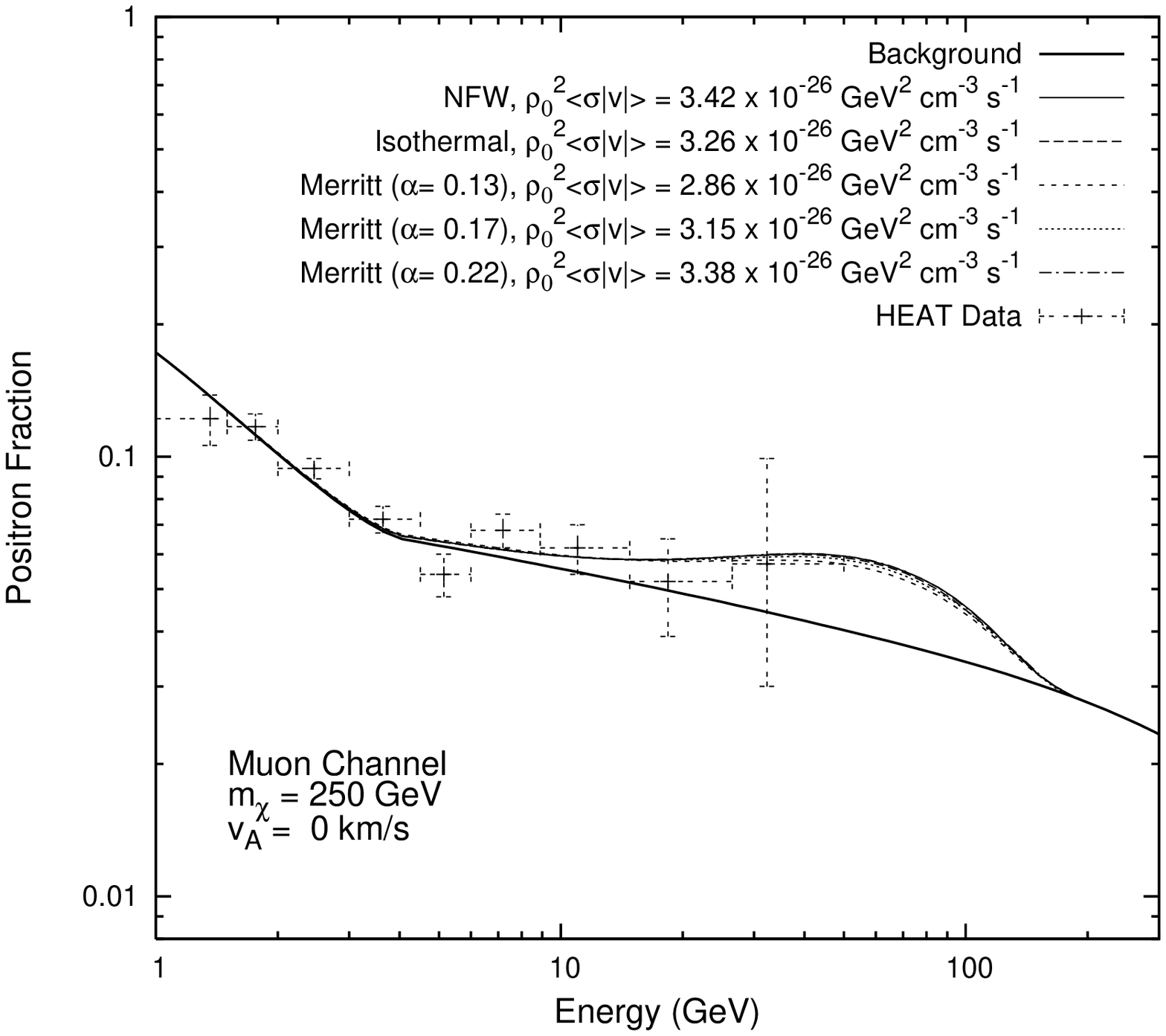,scale=.4}}\\
\subfigure[Direct decay channel, $v_A = 35$ km/s]{\epsfig{figure=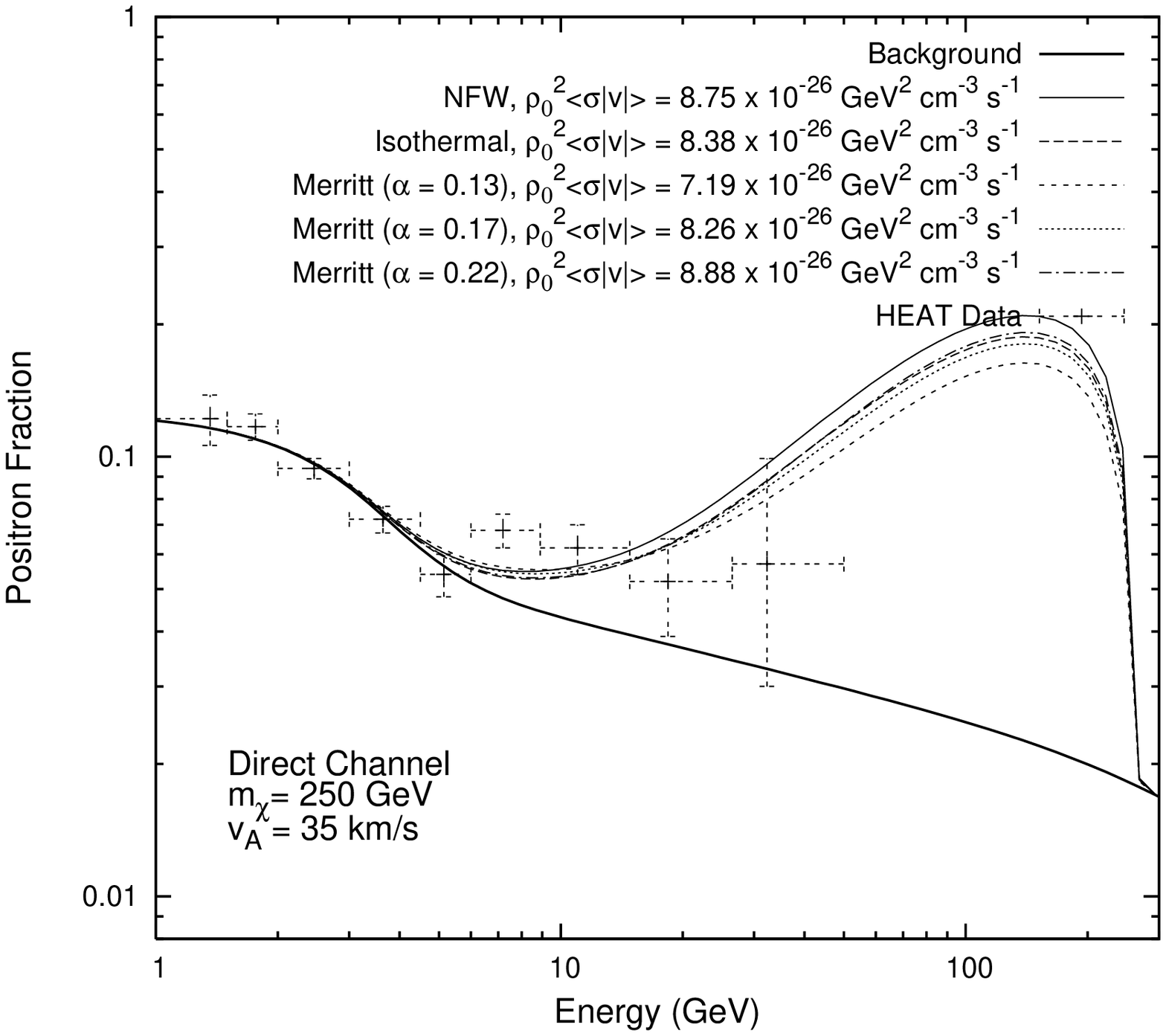,scale=.4}}
\hskip 0.15in 
\subfigure[Muon decay channel, $v_A = 35$ km/s]{\epsfig{figure=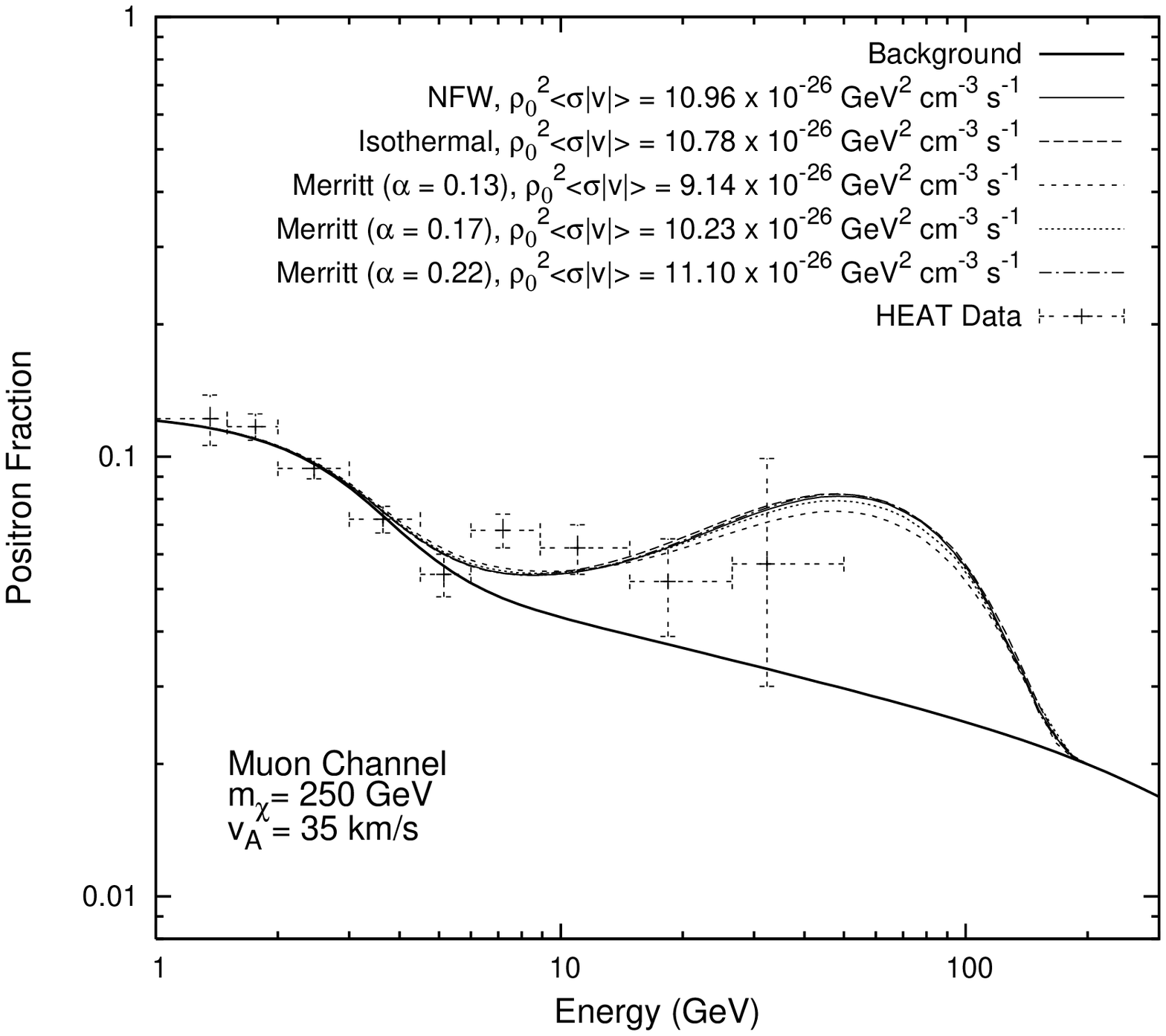,scale=.4}}
\end{center}
\caption{Positron fraction for all halo models for $m_{\chi}=250$ GeV.}
\label{fig:ratiohalodepend}
\end{figure}

The positron fraction is fairly robust with respect to the halo profile.  This is what we would expect, as the local positron fraction at high energies depends mainly on the $e^{+} e^{-}$ pairs produced nearby.  Electrons and positrons lose energy very quickly through a variety of mechanisms as they travel through the Galaxy, so the fluxes of electrons and positrons that we measure locally at energies larger than 1 GeV are produced fairly nearby.  Since the DM profiles are very similar at our location in the Galaxy, we expect the fluxes of DM-produced particles to be similar.  Fig.~\ref{fig:ratiohalodepend} shows the positron fraction for all five halo models.  The $e^{+} e^{-}$ ratios are virtually indistinguishable, and this is true for all values of $m_{\chi}$, though we emphasize that the cross-sections are different.  As the Merritt parameter $\alpha$ increases, the cross-section increases.  This is easily understood, as $\alpha$ is inversely related to the cuspiness of the profile.  A smaller DM density results in fewer $e^{+} e^{-}$ pairs produced, requiring a larger cross-section to make the fit to the HEAT data.

\subsubsection{Dependence on Alfv\'{e}n Velocity}

\begin{figure}[htpb]
\begin{center}
\subfigure[Direct decay channel]{\epsfig{figure=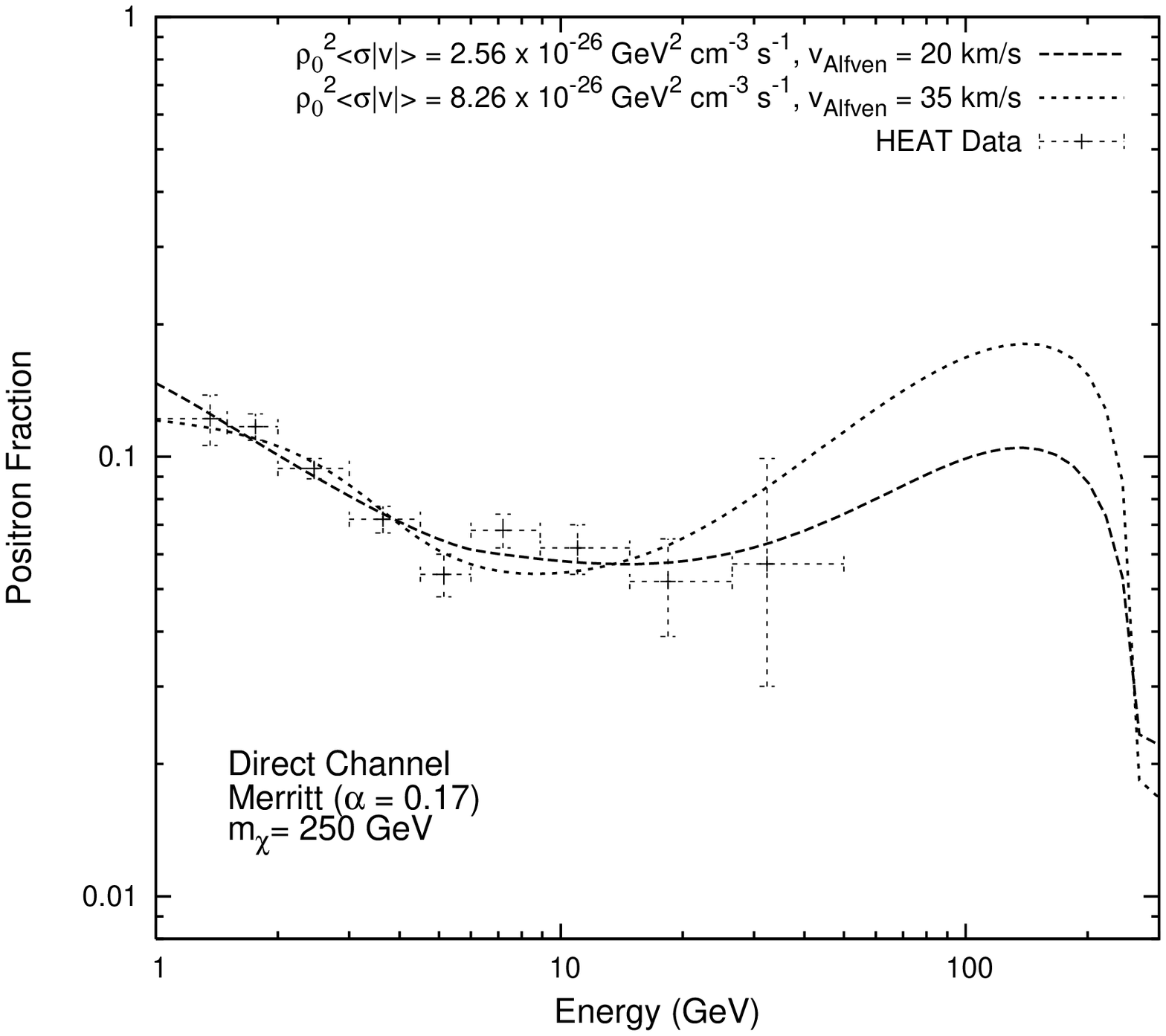,scale=.4}}
\hskip 0.15in 
\subfigure[Muon decay channel]{\epsfig{figure=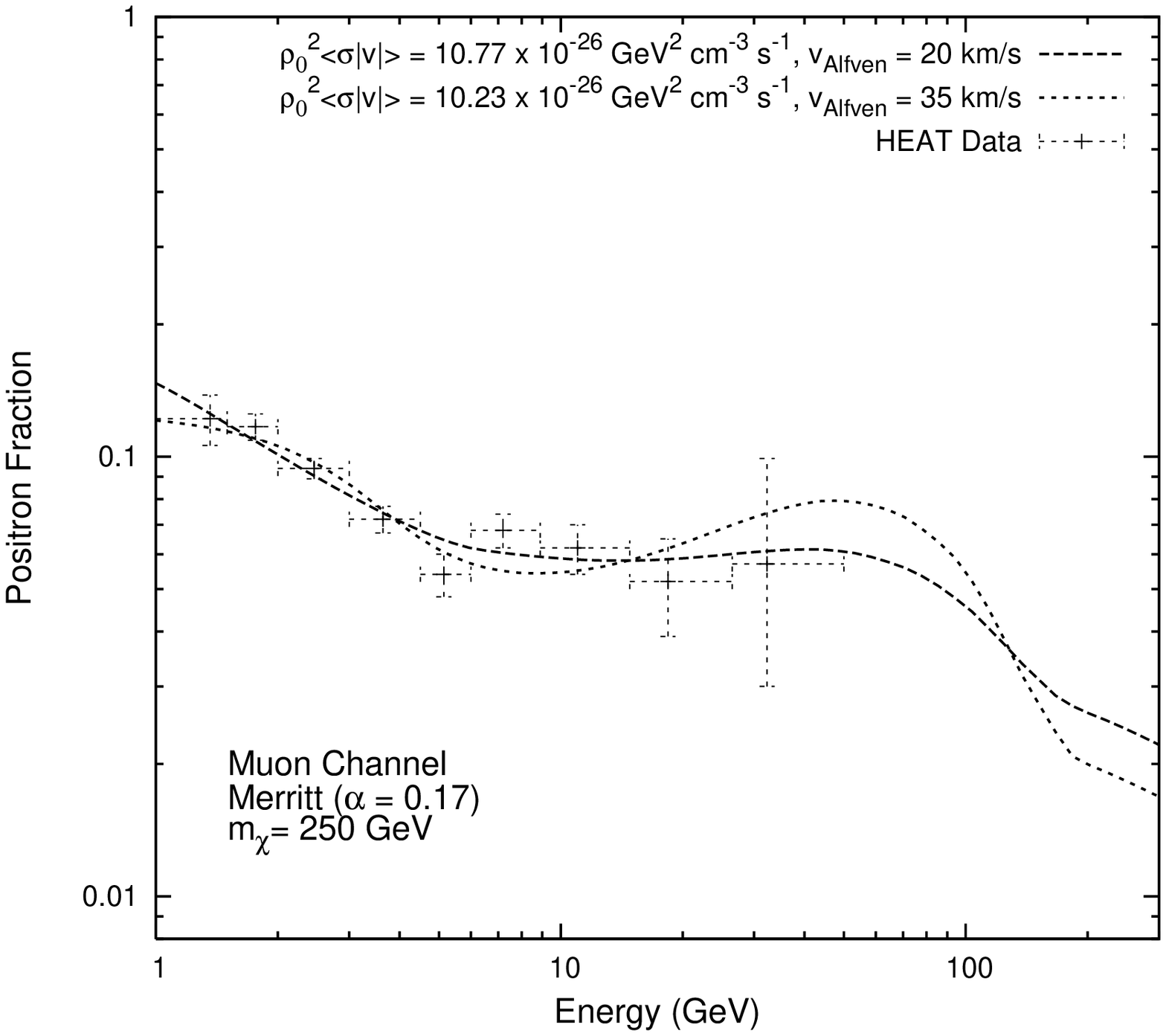,scale=.4}}\\
\end{center}
\caption{$v_A$ dependence of the positron fraction.}
\label{fig:RvAdepend}
\end{figure}

Fig.~\ref{fig:RvAdepend} shows the positron fraction for the two different values of non-zero $v_A$ studied, 20 km/s and 35 km/s.  It appears that the re-acceleration parameter $v_A$ has a strong affect on the positron fraction, and one might conclude that this is due to an underlying effect on the dark matter contribution to the ratio.  However, the plots are misleading.  As mentioned earlier, the effect of re-acceleration on the particle fluxes is small at energies above about 10 GeV, so the dark matter contribution to the positron fraction above 10 GeV is the same regardless of the value of $v_A$.  The differences in the ratios at high energies as shown in Fig.~\ref{fig:RvAdepend} are due to the differences in the backgrounds at those energies (see Fig.~\ref{fig:backgrounds}).  The background positron ratio, which depends on the primary electron flux and secondary electron and positron fluxes, is determined by fitting the data at energies below about 7 GeV to the first five HEAT data points.  Re-acceleration has a significant effect on particle fluxes at these energies, so to fit the data for different values of $v_A$, we must vary the primary electron flux through its power law index, the break rigidity of this index, and the normalization of the flux.  (The secondary fluxes are also affected by re-acceleration, but we don't directly vary these through the inputs to GALPROP.)  As the strength of re-acceleration increases, the power law index and flux normalization must increase to maintain the fit to data (see Table \ref{tab:elecspecparams}).  The changes in the background at low energies necessary to fit to the HEAT data come with corresponding changes in the background at high energies.  It is the resulting differences in the background positron fraction at high energies that give rise to the differences in the total positron fraction at those energies.

\subsubsection{Implications for PAMELA}

Soon, results from PAMELA should provide data for high energy positrons up to $\sim$ 270 GeV. 
It is somewhat natural that if the HEAT excess is, in fact, arising from dark matter, that PAMELA should see additional signals. Except for extremely light particles, the increased sensitivity and statistics of the PAMELA experiment should improve the understanding of this situation. Whether we can claim a signal as a discovery of dark matter we shall return to in a moment. However, even absent the HEAT signal, the haze motivates us to consider what might be seen at PAMELA.

Let us assume that the haze is, in fact, arising from dark matter annihilations as we describe here. A simple examination of the tables in Appendix \ref{ap:bestfit} shows that the cross sections that explain the haze are in general {\em comparable to or larger than} the cross sections needed to generate a signal consistent with HEAT. Consequently, for a wide range of parameters, a strong signal would be expected at PAMELA, if the scenario as described explains the haze.

\begin{figure}[htpb]
\begin{center}
\subfigure[$m_{\chi}=50$ GeV, $v_A = 0$ km/s]{\epsfig{figure=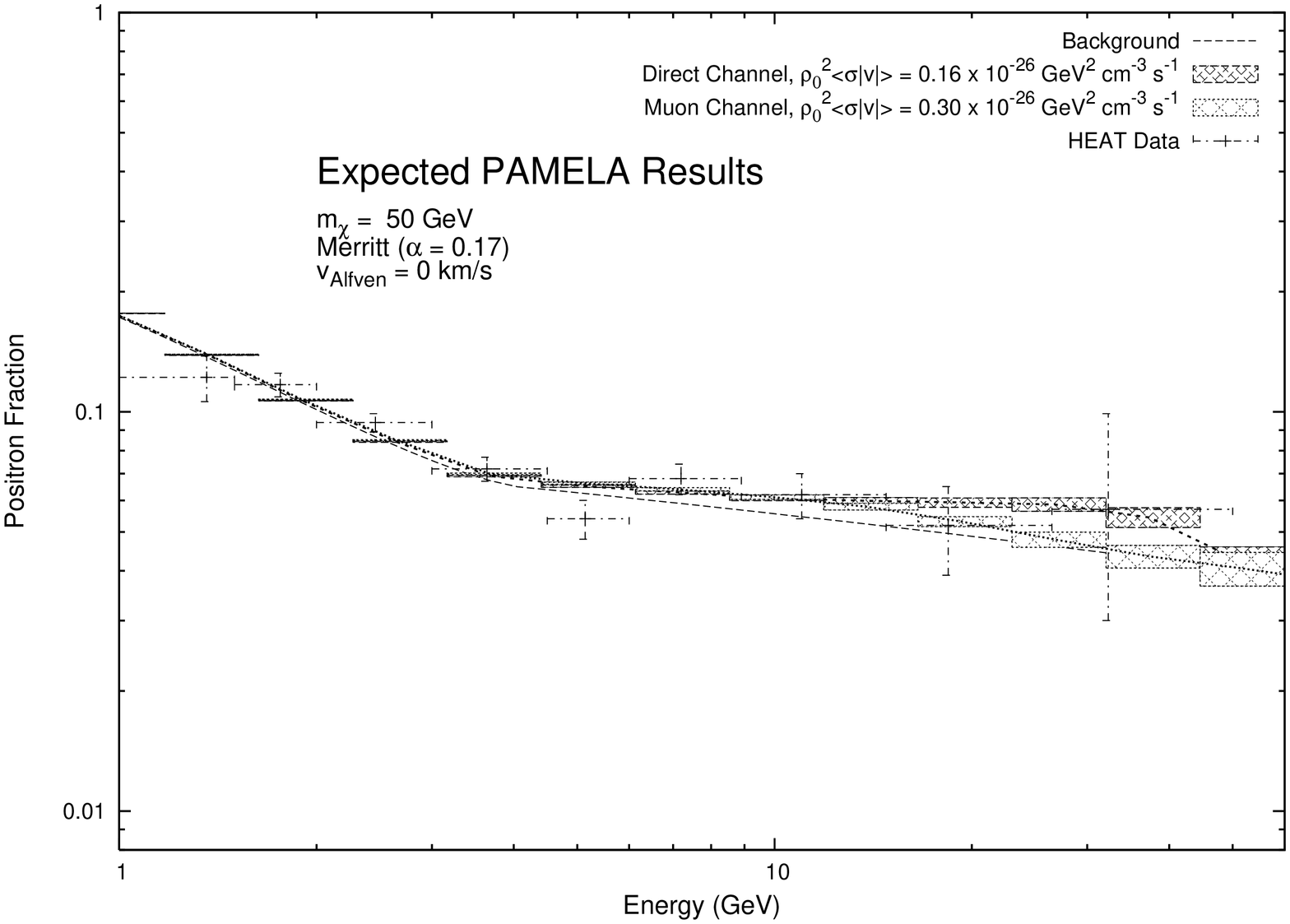,width=0.32\textwidth,angle=-90}}
\hskip 0.15in 
\subfigure[$m_{\chi}=50$ GeV, $v_A = 35$ km/s]{\epsfig{figure=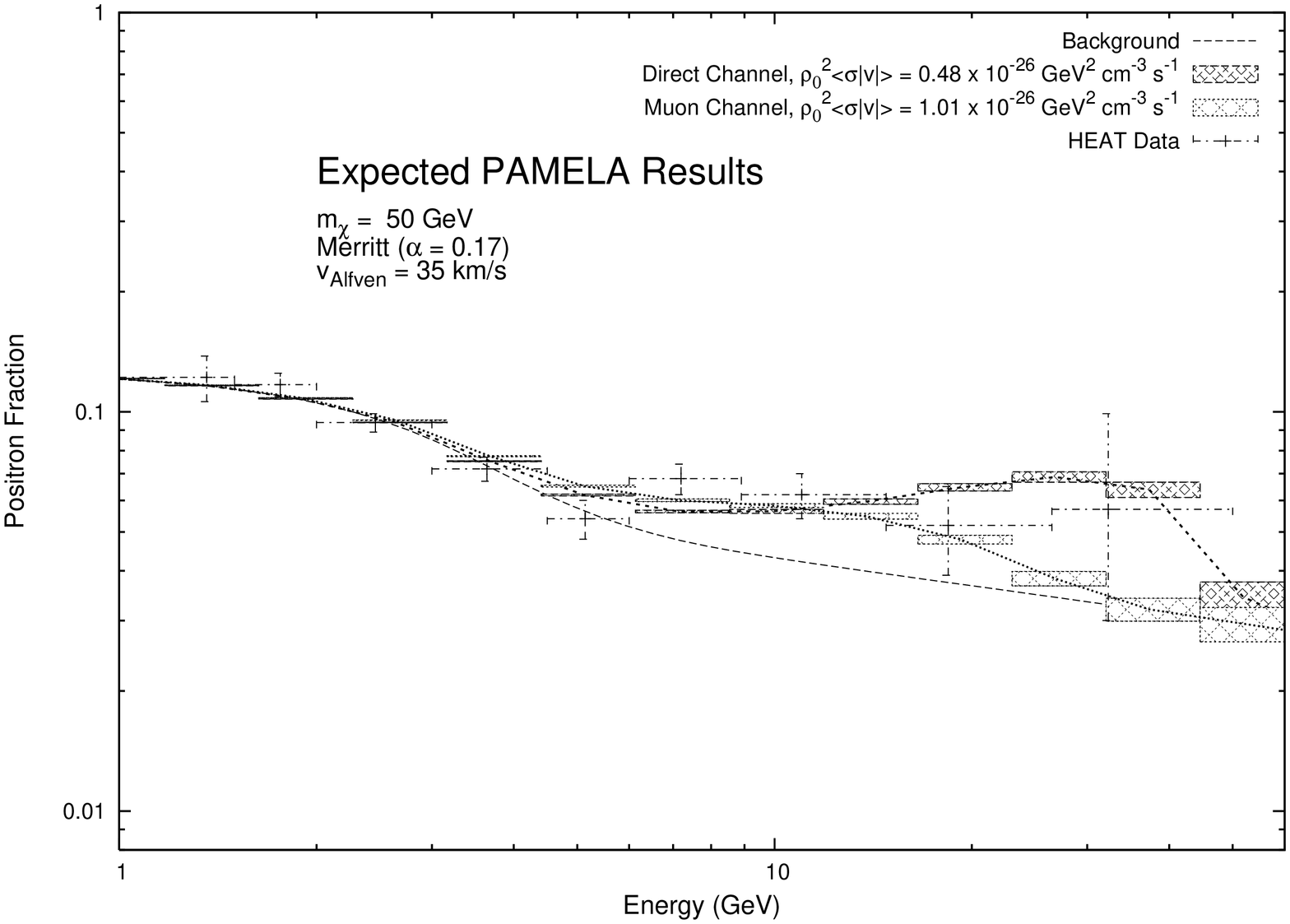,width=0.32\textwidth,angle=-90}}\\
\subfigure[$m_{\chi}=250$ GeV, $v_A = 0$ km/s]{\label{PAM250noreacc}\epsfig{figure=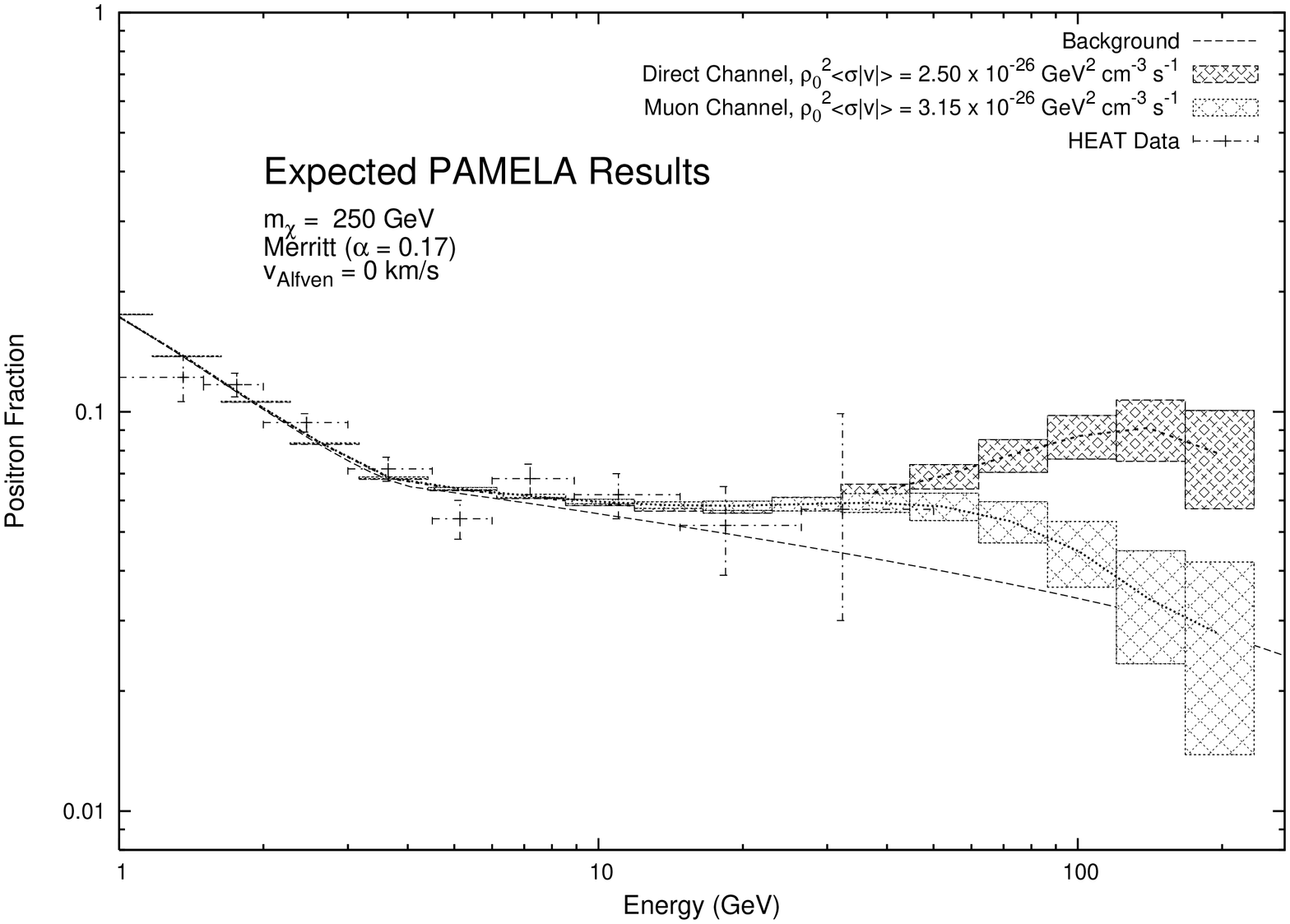,width=0.32\textwidth,angle=-90}}
\hskip 0.15in 
\subfigure[$m_{\chi}=250$ GeV, $v_A = 35$ km/s]{\label{PAM250reacc}\epsfig{figure=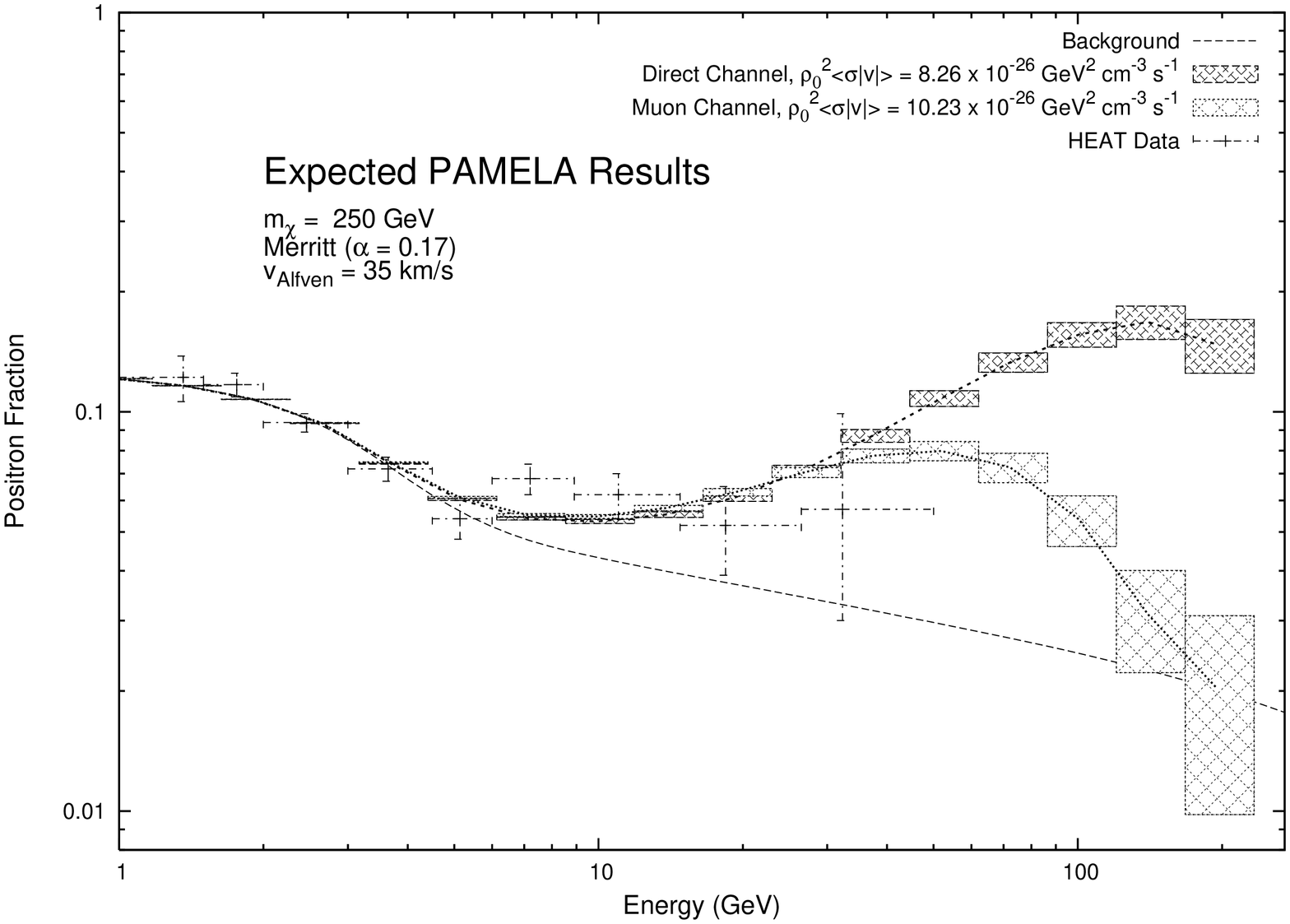,width=0.32\textwidth,angle=-90}}\\
\end{center}
\caption{Predictions for the PAMELA experiment results in the XDM scenario.}
\label{fig:PAMELA}
\end{figure}

\begin{figure}[htpb]
\begin{center}
\subfigure[$m_{\chi}=250$ GeV, $v_A = 0$ km/s]{\label{PAMprelim250noreacc}\epsfig{figure=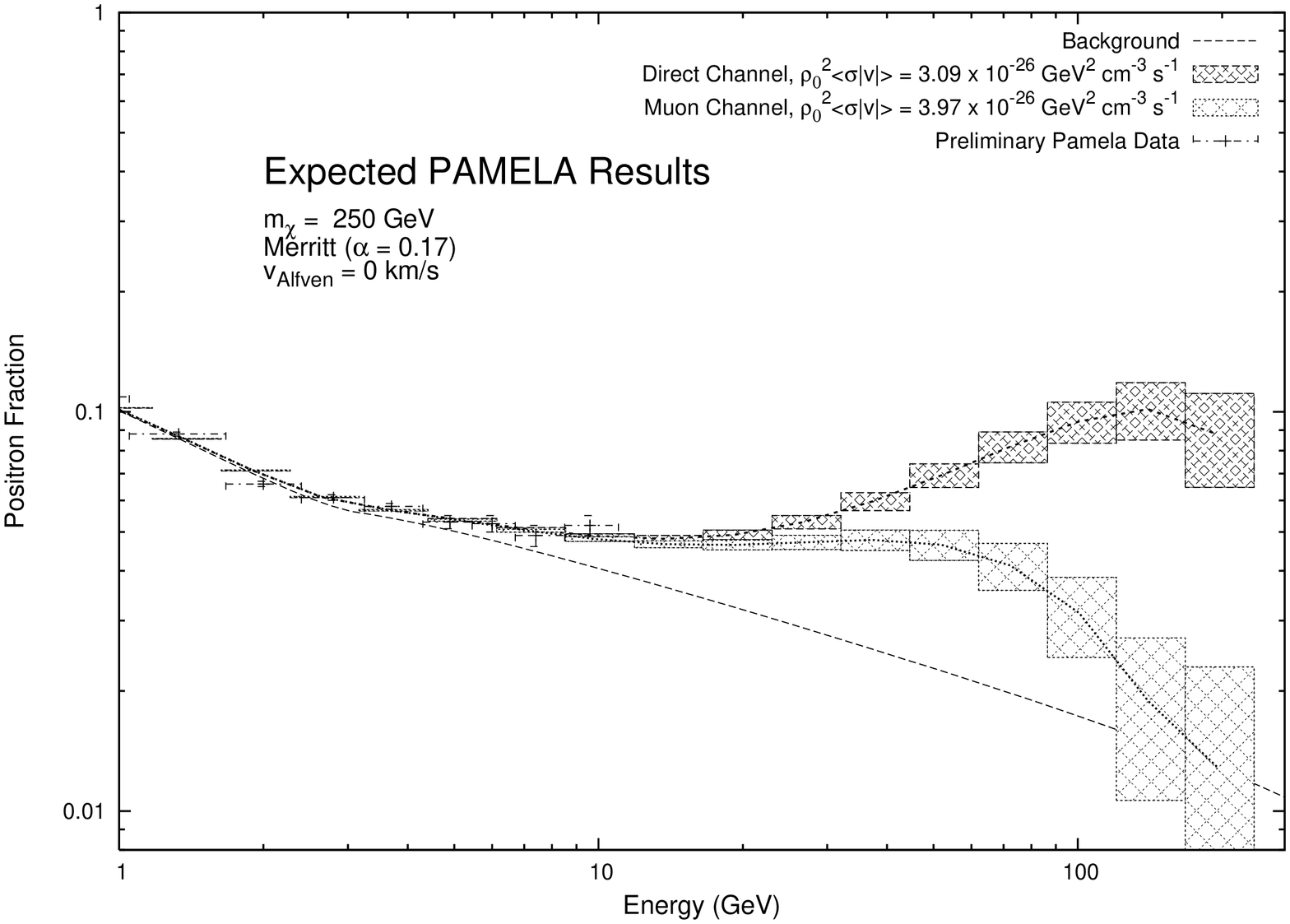,width=0.32\textwidth,angle=-90}}
\hskip 0.15in 
\subfigure[$m_{\chi}=250$ GeV, $v_A = 20$ km/s]{\label{PAMprelim250reacc}\epsfig{figure=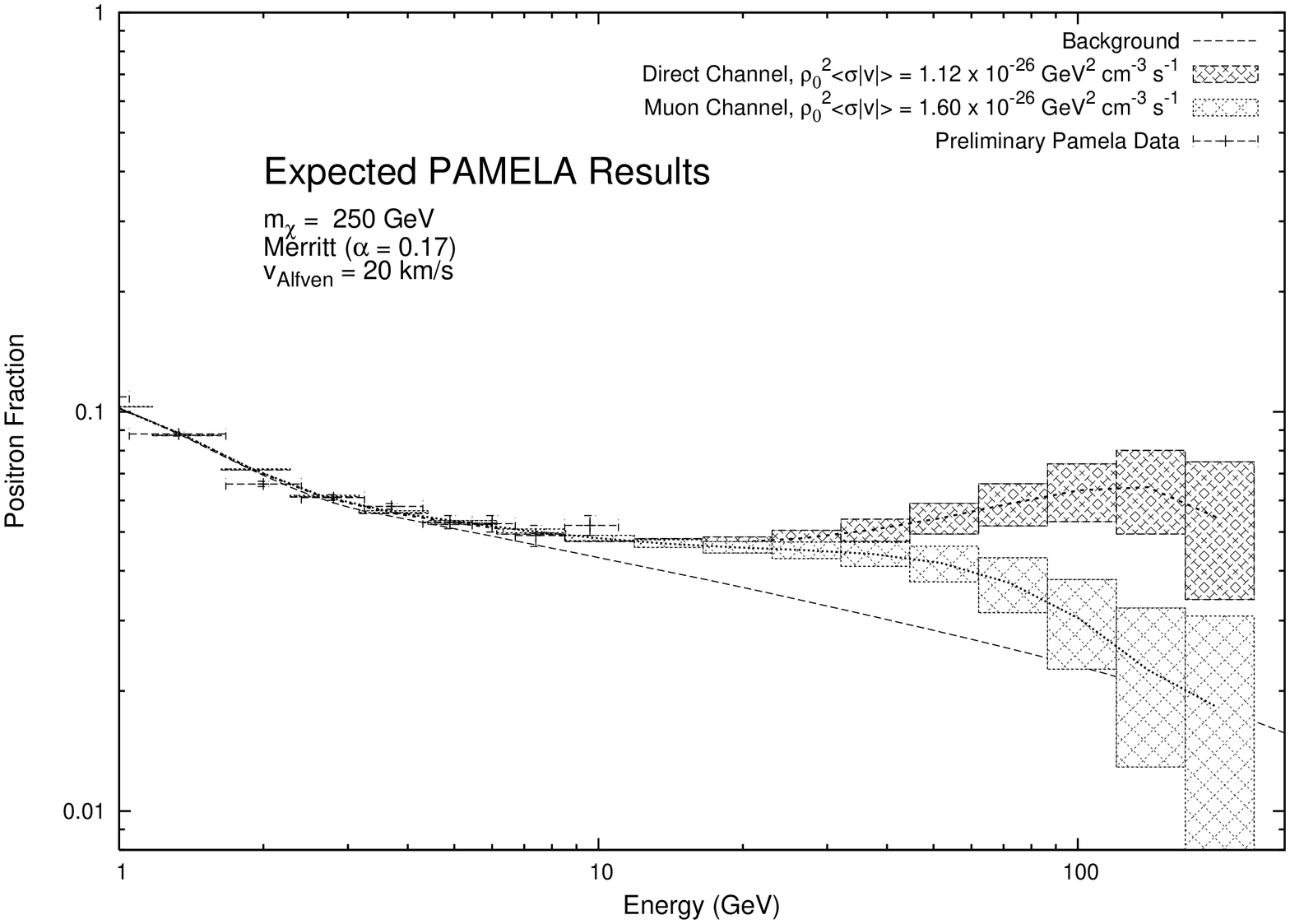,width=0.32\textwidth,angle=-90}}\\
\end{center}
\caption{Predictions for the PAMELA experiment results in the XDM scenario using the preliminary PAMELA data.}
\label{fig:PAMELApreliminary}
\end{figure}

We show representative data for PAMELA in Figs.~\ref{fig:PAMELA} and \ref{fig:PAMELApreliminary}.  In Fig.~\ref{fig:PAMELA} we consider two cases, $m_\chi = $ 50 and 250 GeV, and present predictions for the PAMELA results after three years of data collection using the previously discussed backgrounds fit to the HEAT data.  In Fig.~\ref{fig:PAMELApreliminary} we present similar predictions for $m_\chi = $ 250 GeV using \textit{different} backgrounds fit to the preliminary PAMELA data \cite{pamprelim}.  For these backgrounds, the parameters defining the primary electron and proton injection spectra and the propagation parameters, e.g. the diffusion coefficient and the strength of re-acceleration, are listed in Table~\ref{tab:pamelaparams}.  The backgrounds are in agreement with the local measurements of the electron and proton flux data as well as the B/C ratio data.

It is clear that for the higher mass situation (see Figs.~\ref{PAM250noreacc}, \ref{PAM250reacc}, and \ref{fig:PAMELApreliminary}), regardless of whether one includes re-acceleration or not, a strong signal should be seen, which would be difficult to ascribe to uncertainties in the background.  There appears to be some degeneracy in the data between lighter mass candidates and the decay channel considered. In particular, because the muon channel does not have a hard cutoff, it may be difficult to establish the mass of the WIMP with any accuracy.  Additionally, it may be difficult to distinguish electron from muon decay channels if the energy cutoff is above the PAMELA threshold. 

For lighter mass particles, the situation is not as clear. For instance, if we consider the situation without significant re-acceleration, it is impossible to claim discovery of a $50 \gev$ WIMP at PAMELA in this scenario, while with re-acceleration it is. This arises simply from the shape of the assumed primary electron spectrum, which gives rise to the background spectrum.  For re-acceleration, the background is lower, so the dark matter component needed to fit the HEAT data is larger than that for no re-acceleration.  Therefore, the signal in the re-acceleration case is larger.  Discovery may still be possible for lighter mass particles in a scenario with no re-acceleration as measurements of other cosmic rays pin down additional astrophysical parameters, and as the low energy positron and electron spectra are established.

\begin{figure}[htpb]
\begin{center}
\subfigure{\epsfig{figure=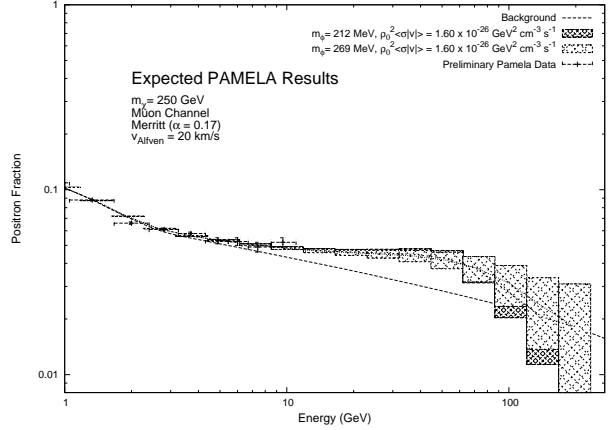,width=0.32\textwidth,angle=-90}}
\end{center}
\caption{Dependence of the predictions for the PAMELA experiment results on $m_{\phi}$.}
\label{fig:PAMELAphimass}
\end{figure}

Fig.~\ref{fig:PAMELAphimass} shows our predicted PAMELA results for $m_{\chi}=$ 250 GeV for two different values of $m_{\phi}$, 212 MeV and 269 MeV, the limits of the range of $\phi$ masses for which the decay into muons is the dominant decay channel.  The decay though a lighter scalar mediator with the same value of $\rho_{0}^{2}\left\langle\sigma\left|v\right|\right\rangle$ results in a positron fraction with a larger peak value and a faster decline to the background value.  These differences in signal are small for large values of $m_{\chi}$ and our backgrounds and cannot be distinguished by PAMELA.  However, for smaller values of $m_{\chi}$ and a different background ratio these differences are more pronounced and possibly distinguishable.

We want to emphasize that the background is dependent on the choice of propagation parameters, even within the constraints of local CR data.  However, regardless of the differences in the background, we show that a significant signal can be seen at PAMELA for $m_{\chi}\leq$ 250 GeV with values of $\rho_{0}^{2}\left\langle\sigma\left|v\right|\right\rangle$ of the the order $10^{-26} \rm \ GeV^{2} cm^{-3} s^{-1}$.

\begin{table}
\begin{center}
\caption{GALDEF parameters for the background fits to preliminary PAMELA data}
\label{tab:pamelaparams}
\begin{tabular}{l|c|c}
\hline\hline
 & $ v_A=0 \ \rm km/s $ &  $v_A=20 \ \rm km/s $\\
 \hline
Electron Injection Index Below Lower Break Rigidity & 1.80 & 2.18 \\
Electron Injection Index Above Lower Break Rigidity & 2.60 & 2.54 \\
Electron Injection Index Above Upper Break Rigidity & 5.00 & 5.00 \\
Electron Lower Break Rigidity (MV) & $3.0\times 10^3$ & $3.0\times 10^3$ \\
Electron Upper Break Rigidity (MV) & $1.0\times10^9$ & $1.0\times 10^9$ \\
Electron Flux Normalization ($\rm cm^{-2} sr^{-1} s^{-1} MeV^{-1}$) &  $0.2760\times 10^{-9}$ & $0.2148\times 10^{-9}$ \\
Nucleus Injection Index Below Break Rigidity & 1.90 & 1.98 \\
Nucleus Injection Index Above Break Rigidity & 2.60 & 2.42 \\
Nucleus Break Rigidity (MV) & $20.0 \times 10^3$ & $9.0 \times 10^3$ \\
Proton Flux Normalization ($\rm cm^{-2} sr^{-1} s^{-1} MeV^{-1}$) & $4.41\times 10^{-9}$ & $4.41\times 10^{-9}$ \\
$z_{max}$ (kpc) (half of Diffusion Zone Width) & 10.0 & 10.0 \\
Diffusion Coefficient Normalization ($\times10^{28} \rm cm^{2}s^{-1}$) & $5.80\times 10^{28}$ & $5.80\times 10^{28}$ \\
Diffusion Coefficient Index Below Break Rigidity & 0.34 & 0.34 \\
Diffusion Coefficient Index Above Break Rigidity & 0.40 & 0.40 \\
Diffusion Coefficient Break Rigidity (MV) & $6.0\times 10^3$ & $1.0\times 10^3$ \\
\hline
\hline
\end{tabular}
\end{center}
\end{table}

\subsection{DM Synchrotron Component}

\subsubsection{Decay Channel Dependence}
\begin{figure}[htpb]
\begin{center}
\subfigure[$v_A = 0$ km/s]{\epsfig{figure=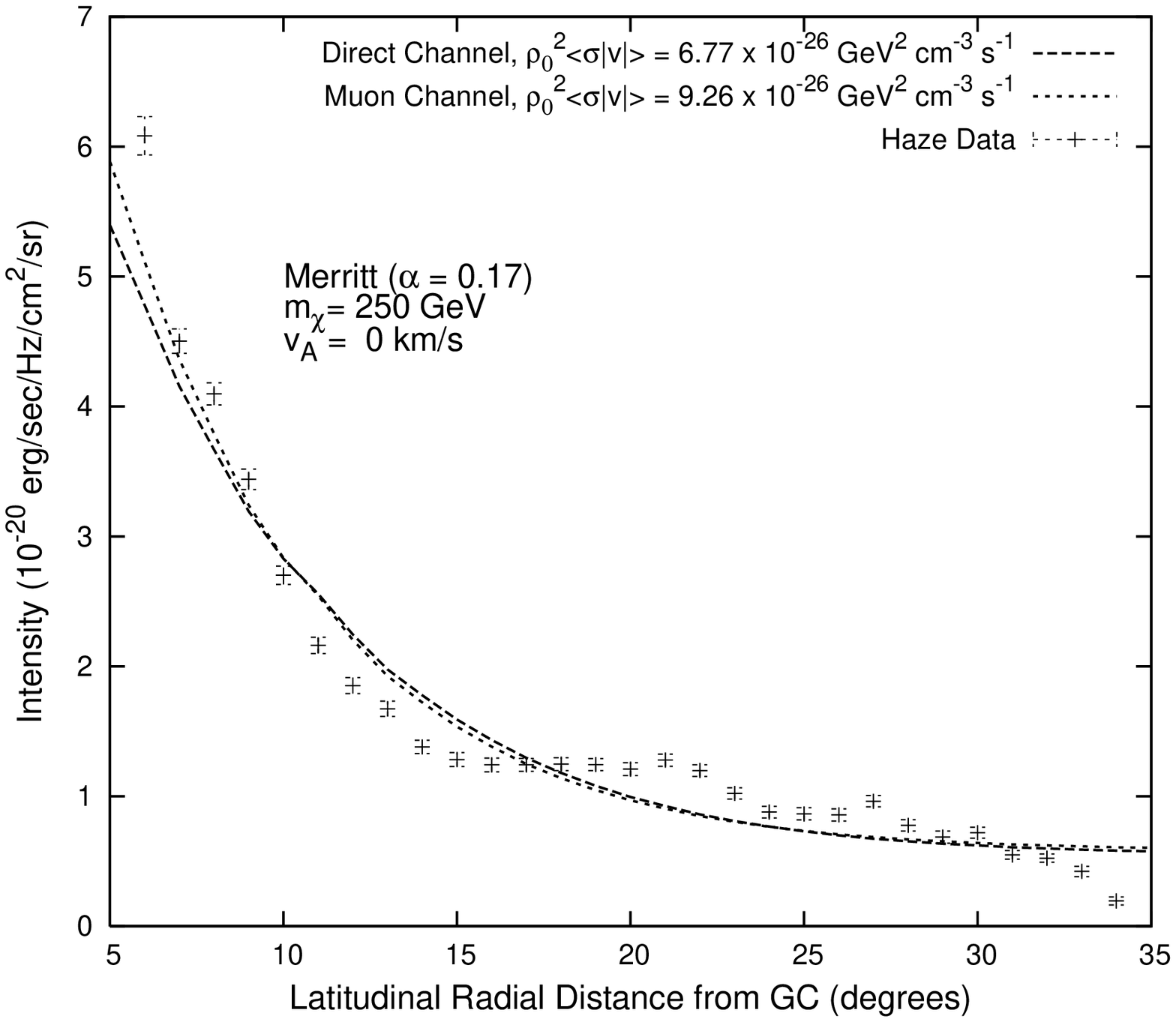,scale=.4}}
\hskip 0.15in 
\subfigure[$v_A = 35$ km/s]{\epsfig{figure=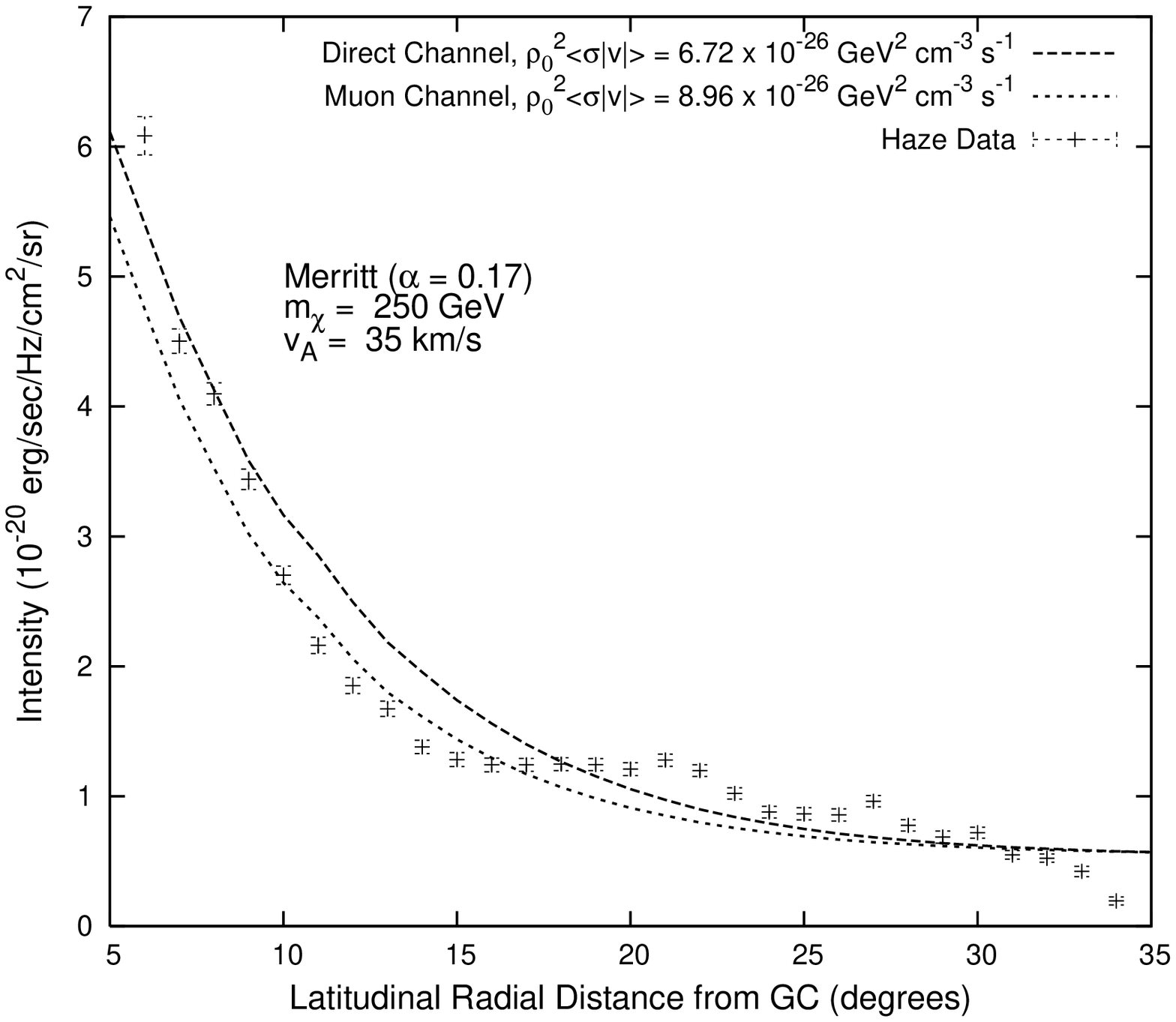,scale=.4}}\\
\end{center}
\caption{\label{fig:channeldepend}250 GeV XDM contribution to synchrotron radiation at 22.5 GHz for the direct and muon decay channels.}
\end{figure}

While the different decay channels lead to quite distinct results for the positron fraction, the differences in the synchrotron radiation for the direct and muon channels are more subtle.  See Fig.~\ref{fig:channeldepend}.  The muon channel produces an intensity distribution that is more peaked in the center of the Galaxy than that produced by the direct channel.  This leads to slightly better agreement with the haze data in almost all cases, particularly for larger values of $m_{\chi}$.  The best-fit cross-sections for the muon channel are larger than those for the direct channel for a specific $m_{\chi}$ and DM halo profile (refer eg. to Tables~\ref{tab:Table direct ch vA=0} and \ref{tab:Table muon ch vA=0}).  Since the contribution to the number of high energy $e^{+} e^{-}$ pairs is larger for the direct channel annihilation process, one would expect more synchrotron radiation.  The power spectrum of synchrotron radiation scales as the square of the boost of the particle \cite{Schlickeiser}, so it is consistent that the harder electron and positron spectra produce more radiation.\\ 

\subsubsection{ $m_\chi$ Dependence}

\begin{figure}[htpb]%
\begin{center}
\subfigure[Direct decay channel, $v_A = 0$ km/s]{\epsfig{figure=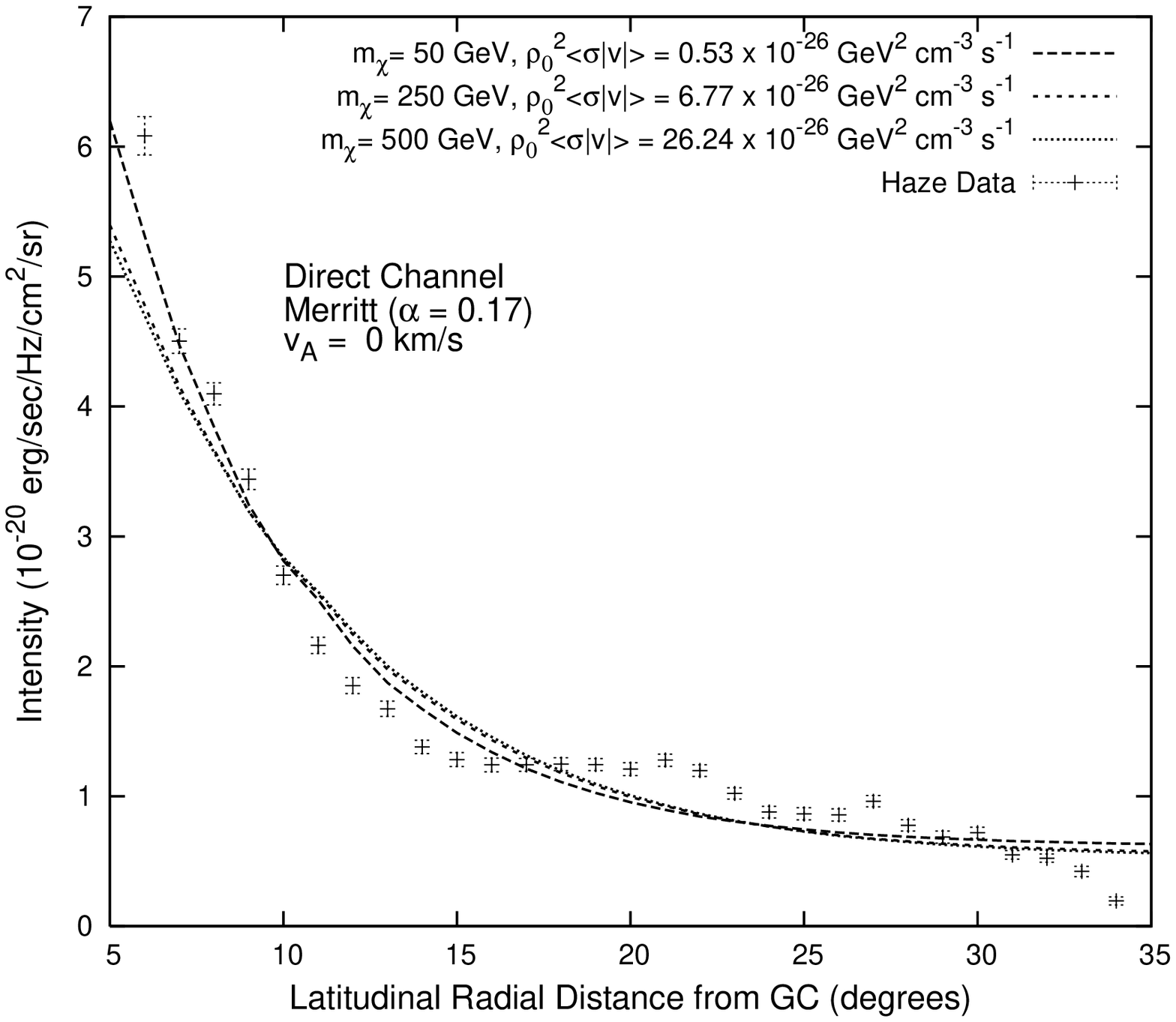,scale=.4}}
\hskip 0.15in
\subfigure[Muon decay channel, $v_A = 0$ km/s]{\epsfig{figure=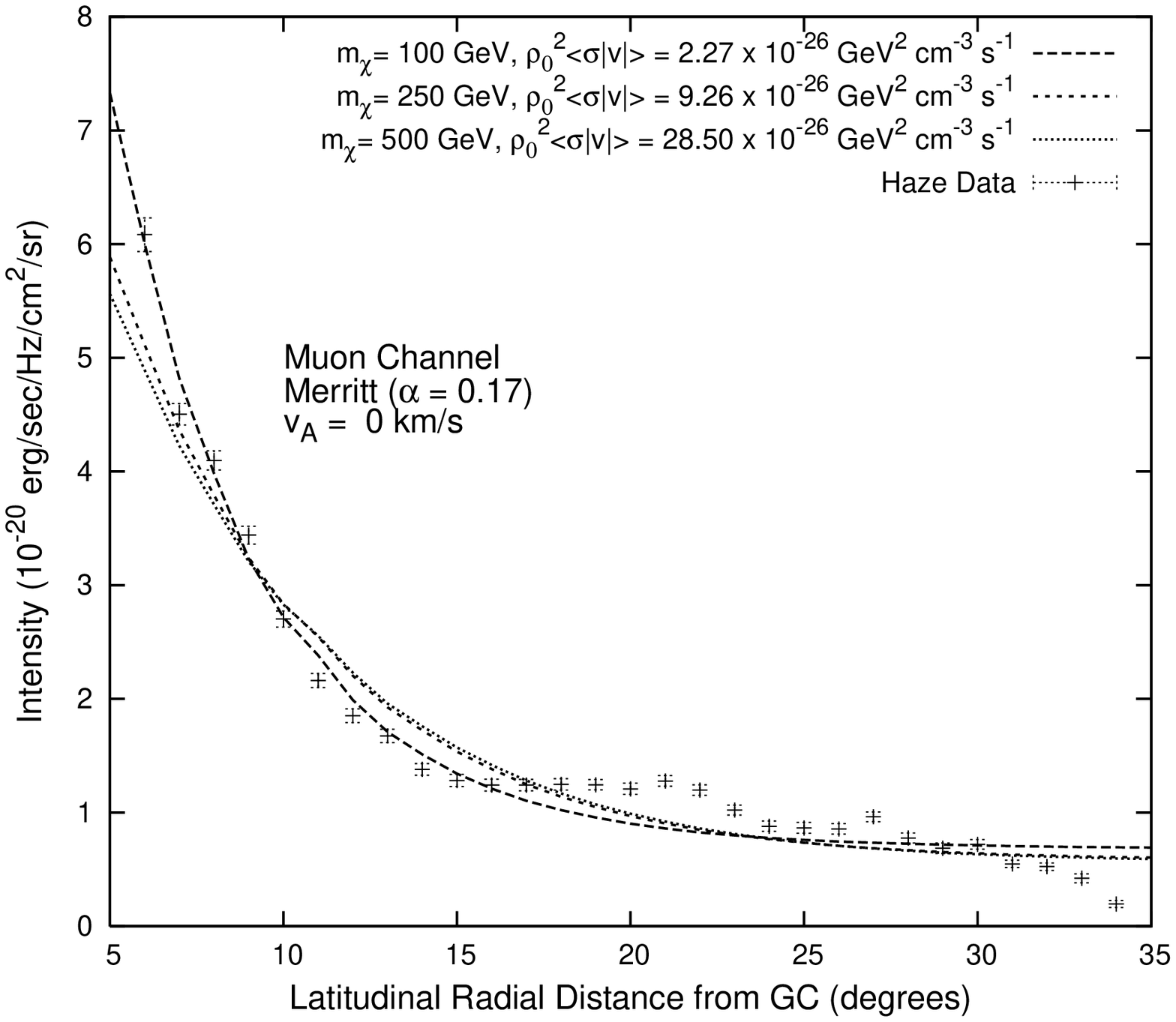,scale=.4}}\\
\subfigure[Direct decay channel, $v_A = 35$ km/s]{\epsfig{figure=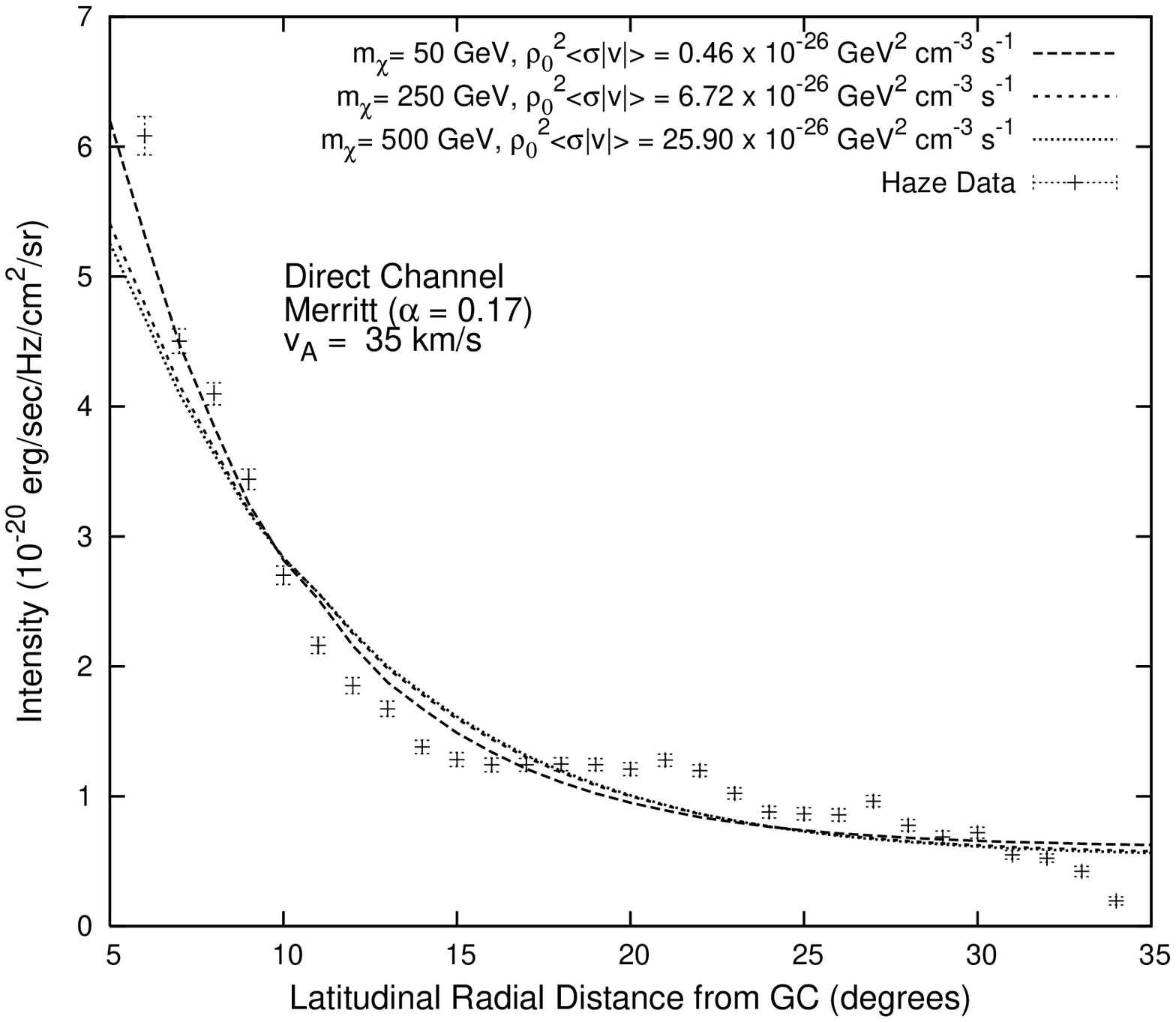,scale=.4}}
\hskip 0.15in
\subfigure[Muon decay channel, $v_A = 35$ km/s]{\epsfig{figure=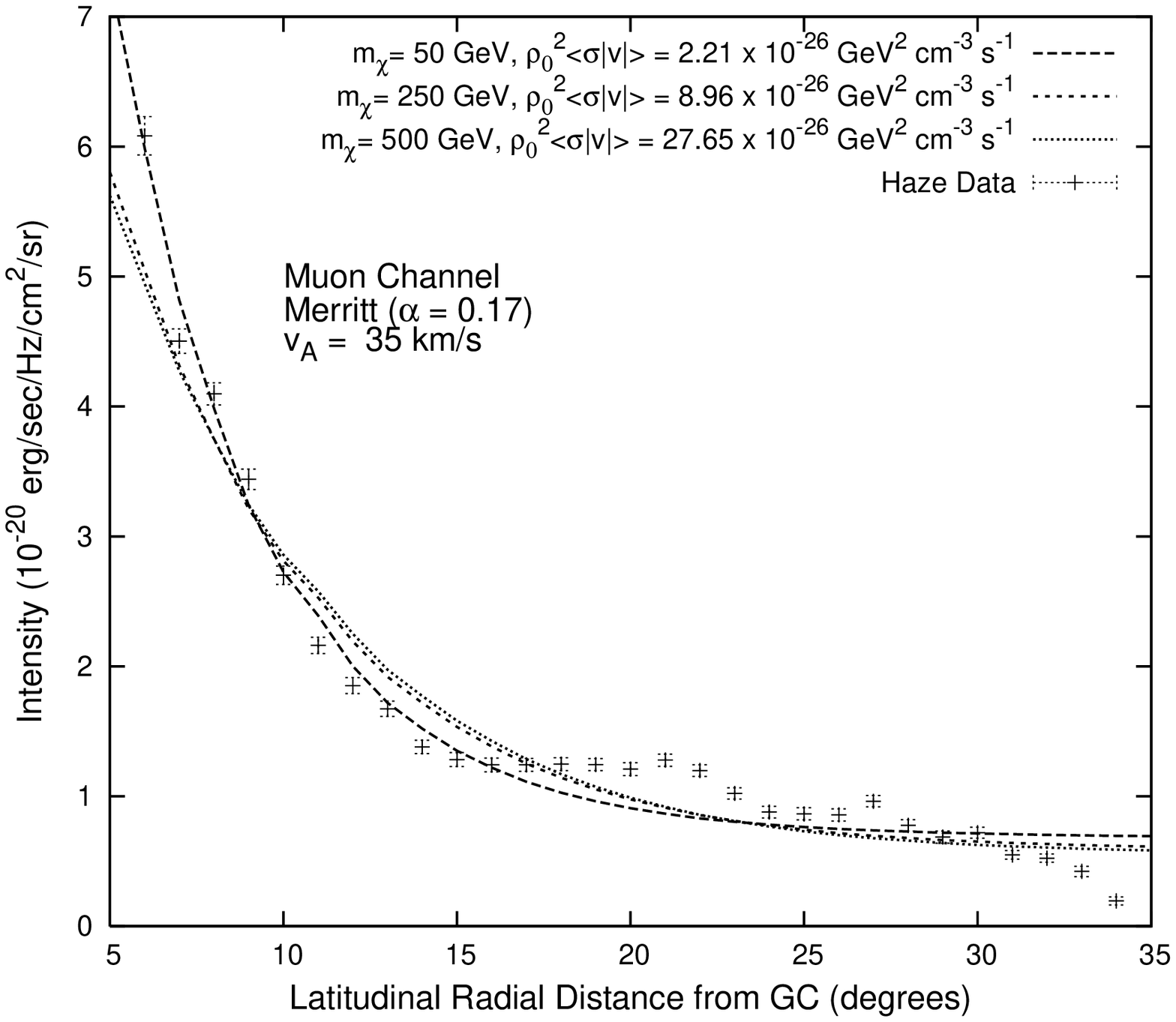,scale=.4}}
\end{center}
\caption{$m_{\chi}$ dependence of XDM contribution to synchrotron radiation at 22.5 GHz.}
\label{fig:massdepend}
\end{figure}

Fig.~\ref{fig:massdepend} shows the XDM contribution to the synchrotron radiation for various dark matter masses.  The intensity distribution is more peaked in the center of the Galaxy for smaller values of $m_{\chi}$ for both production channels, leading to slightly better agreement with the haze data in almost all cases. This cuspiness arises simply because the lower energy particles do not propagate away from the Galactic Center as readily.  As is the case for the positron fraction, the best-fit cross-section increases with increasing $m_{\chi}$.  Again, this can be understood as arising from the lower number density of dark matter particles, though this is somewhat compensated by the higher energy particles providing more energy for synchrotron radiation.

\subsubsection{Dependence on Halo Profile}

\begin{figure}[htpb]
\begin{center}
\subfigure[Direct decay channel, $v_A = 0$ km/s]{\epsfig{figure=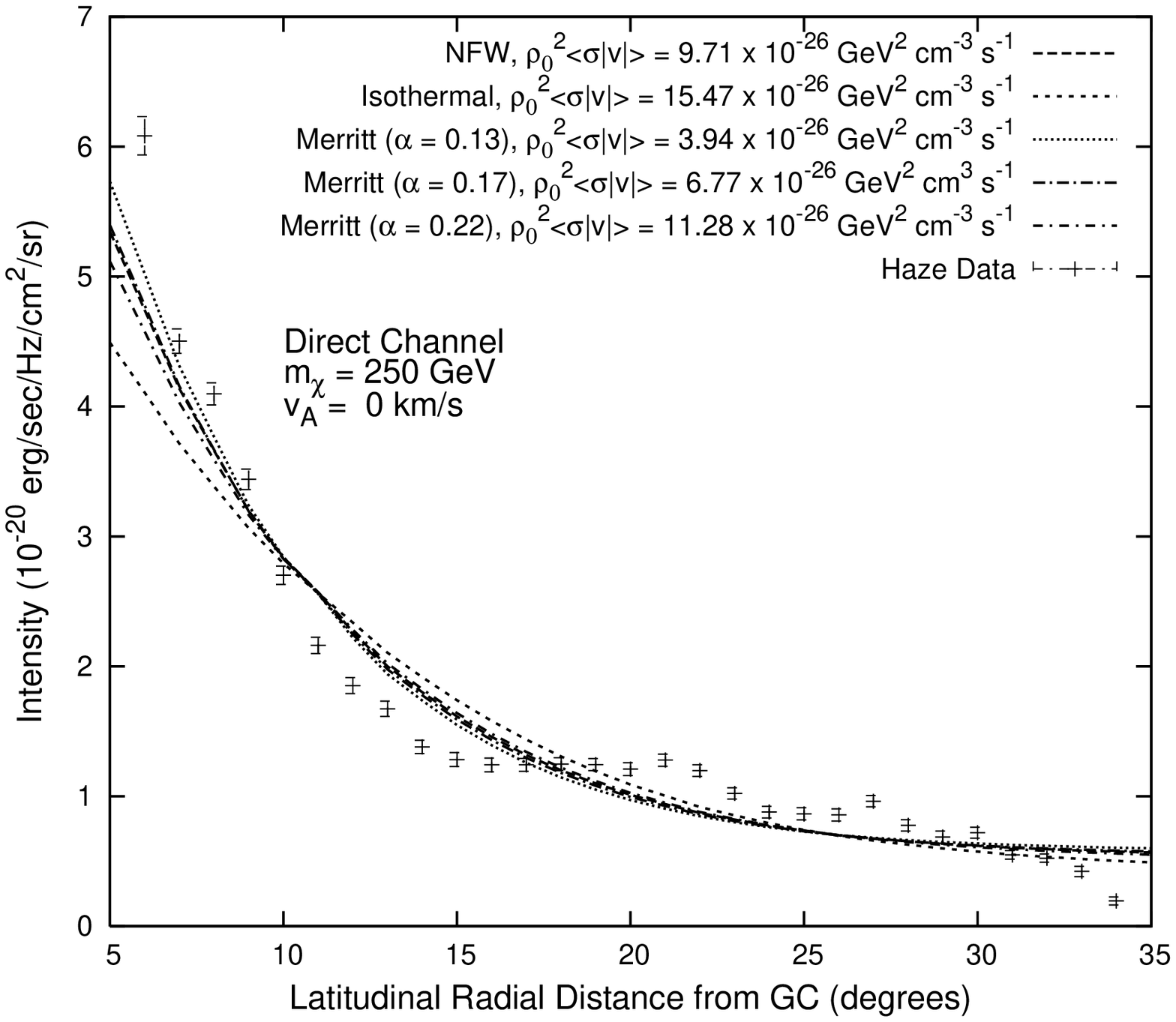,scale=.4}}
\hskip 0.15in
\subfigure[Muon decay channel, $v_A = 0$ km/s]{\epsfig{figure=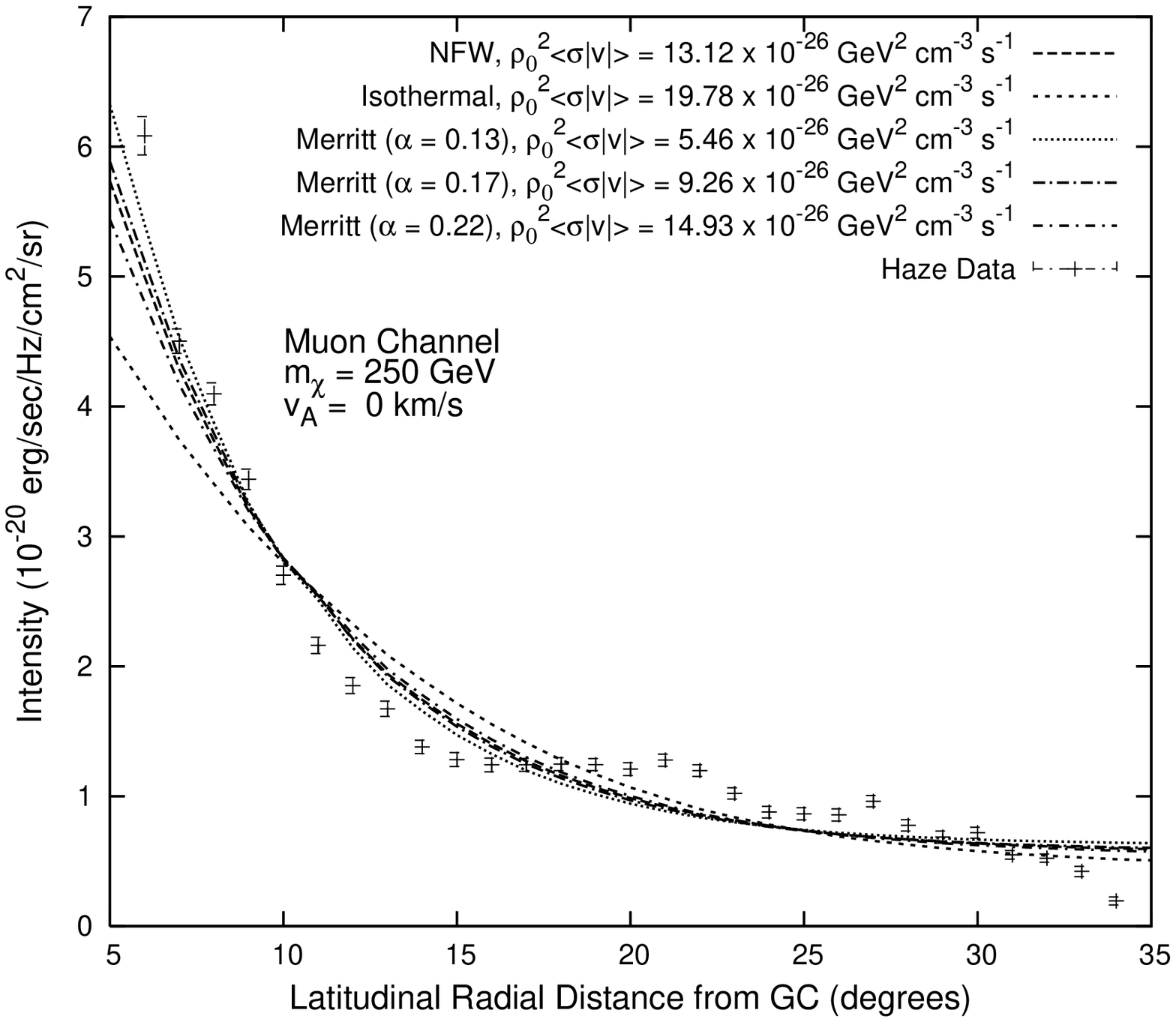,scale=.4}}\\
\subfigure[Direct decay channel, $v_A = 35$ km/s]{\epsfig{figure=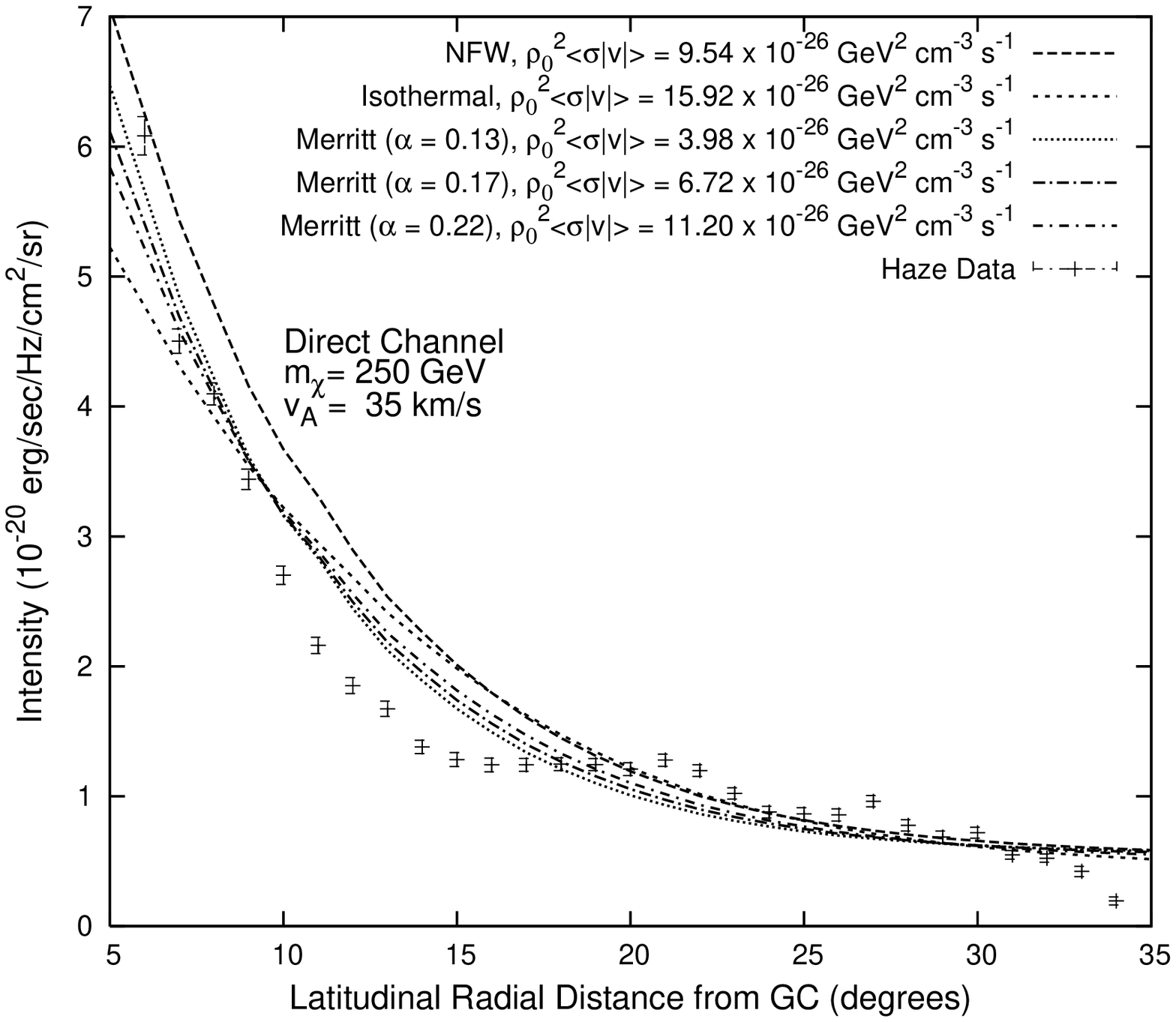,scale=.4}}
\hskip 0.15in
\subfigure[Muon decay channel, $v_A = 35$ km/s]{\epsfig{figure=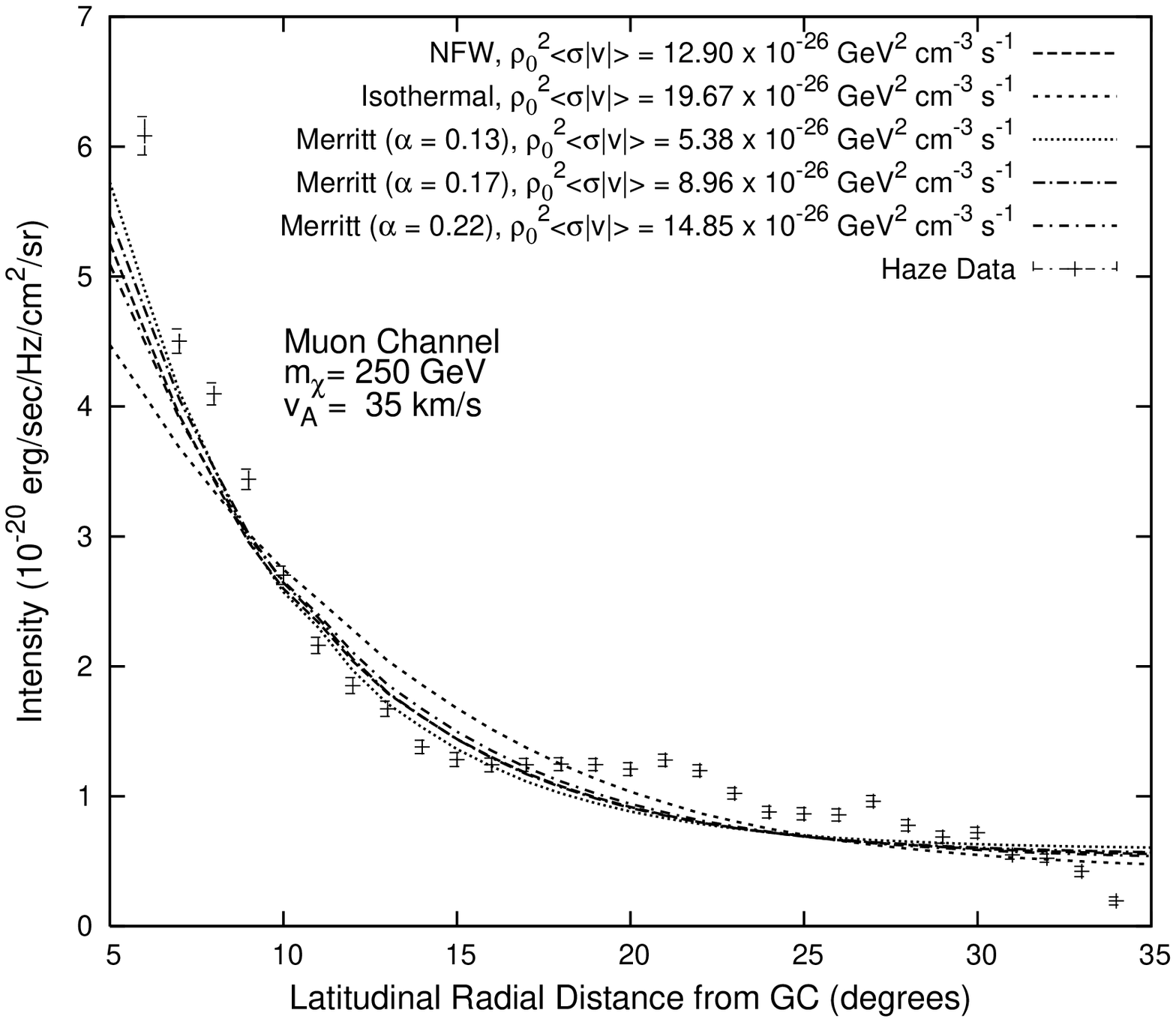,scale=.4}}
\end{center}
\caption{XDM contribution to synchrotron radiation at 22.5 GHz for various halo profiles.}
\label{fig:halodepend}
\end{figure}

The shape of the intensity distribution of 22.5 GeV synchrotron radiation is relatively model-independent, with the notable exception of the isothermal model.  See Fig.~\ref{fig:halodepend}.  The isothermal model produces a significantly flatter distribution.  We investigated two additional halo profiles which we do not present here, the Evans profile and the alternative profile \cite{Moskalenko:1999sb}, both of which are included with the GALPROP code.  The Evans model produced a DM synchrotron distribution that was even flatter than that produced by the isothermal model, while the alternative model produced a more peaked distribution that gave good agreement with the haze data for all $m_{\chi}$.  Ignoring the Isothermal model, we note that the N.F.W. profile requires a larger cross-section to fit the haze data.  For example, the best-fit value of $\rho_{0}^{2}\left\langle\sigma\left|v\right|\right\rangle$ is $9.54\times 10^{-26} \rm GeV^{2}cm^{-3}s^{-1}$ for N.F.W. versus $6.72\times 10^{-26} \rm GeV^{2}cm^{-3}s^{-1}$ for Merritt ($\alpha = .17$) for the direct channel decay with $m_{\chi}$ = 250 GeV and $v_A = 35$ km/s.  At small radial distances ($r \lsim$ 8 kpc), the N.F.W. profile gives a smaller value for the dark matter density than does the Merritt profile (see Fig.~\ref{fig:profiles}), so fewer $e^{+} e^{-}$ pairs are produced and thus less synchrotron radiation.

Fig.~\ref{fig:halodepend} shows the dependence of the XDM contribution to the synchrotron radiation on the Merritt parameter $\alpha$. The intensity is more peaked in the center of the Galaxy for smaller values of $\alpha$,  and this results in better fits to the haze data.  Additionally, the best-fit cross-section is smaller for smaller $\alpha$.  We understand this in terms of the effect of the parameter on the shape of the Merritt profile; a profile with smaller $\alpha$ is more cuspy.  The additional contribution to synchrotron radiation in the center of the galaxy for the cuspier profile reduces the cross-section necessary for a good fit to the haze.  Note that the constant offset in our fitting (see Section \ref{sec:DMsynchrad}) compensates for the reduction at larger galactic angles.

\subsubsection{Dependence on Alfv\'{e}n Velocity}
Re-acceleration has a significant effect on the particle fluxes only below about 10 GeV, so the intensity distributions of synchrotron radiation for different values of $v_A$ are identical. The effects of re-acceleration on the positron fraction at high energies came about from changes to the background ratio.  Since the DM synchrotron component is independent of the background $e^+e^-$ fluxes, the effects of re-acceleration are minimal.  A comparison of the cross sections for different values of $v_A$ in Appendix \ref{ap:bestfit} shows that the best-fit values are essentially identical.

\section{Conclusions}
\label{sec:conc}
Dark matter remains an elusive component of the matter in the universe. Nonetheless, there are reasons for optimism that we may soon, if not already, have indications from astrophysics as to its nature. Hints from the WMAP haze and from excess positrons in the HEAT experiment both suggest new primary sources for high energy positrons and electrons.  One explanation would be annihilating dark matter.

We have considered the implications of ``exciting'' dark matter for these experiments and for the future results from PAMELA, and have found that the HEAT excess, as well as the haze, both arise naturally in this framework. Because the dark matter annihilates into a light state which then decays directly or indirectly to electrons and positrons, the resulting $e^+e^-$ particles are highly boosted, and thus more relevant for experiment. In this way XDM falls into a broader framework of ``secluded'' dark matter \cite{Pospelov:2007mp} where the annihilation products of the WIMP are not SM states themselves, but do couple to SM states.  Some interesting avenues to consider include a broader range of states into which $\phi$ can decay and an expanded model framework.  We will pursue these in future work.

If local dark matter densities are high ($\rho_0 \sim 1 \GeV \cm^{-3}$), the natural range of s-wave cross sections ($\left\langle\sigma_{ann}\left|v\right|\right\rangle \sim 2.5 \times 10^{-26}\; \rm cm^3 s^{-1}$)  needed to explain the appropriate relic abundance also yields signals in agreement with the haze and HEAT, and with significant implications for PAMELA. It is important to note that in this scenario one needn't resort to large astrophysical boosts to explain HEAT, and that in most cases, the cross section which explains the haze has a significant signal at PAMELA. In particular, if one restricts oneself to the range of parameters which also explain the INTEGRAL signal (namely $m_\chi \gsim 400\gev$), if XDM explains the haze, a PAMELA signal is expected for the halo models we consider here.

Many of the uncertainties associated with the signal arise from the uncertainty in the primary electron spectrum, i.e. its shape and normalization.  Measuring the $e^+e^-$ fluxes at PAMELA, as well as other cosmic ray observables, should reduce some of the astrophysical uncertainties, which should help to establish the significance of the signal beyond what we can discuss here.  However, it is clear that PAMELA, with its measurements that are dominantly sensitive to the local environment, will not be able to significantly distinguish between the halo models we considered here.  In the event of a clear signal, however, PAMELA should be able to make some statement of the mass of the dark matter particle, even if just a lower bound.

In conclusion, we have found that XDM provides a simple framework for explaining all existing astrophysical electronic anomalies: the low energy INTEGRAL 511 keV line from positron annihilation, the multi-GeV HEAT positron excess, and the possible multi-GeV electron and positron source generating the WMAP haze. Such signals all lie naturally within the same parameter range, and, in fact, the light mediator necessary for the success of XDM implies, in general, highly boosted annihilation products, which contribute to the significance of the haze and HEAT signals. Upcoming results from PAMELA should shed light on whether this scenario is realized in nature.

\vskip 0.2 in
\noindent {\bf Acknowledgments}
\vskip 0.05in
\noindent It is a pleasure to acknowledge very useful conversations with Doug Finkbeiner, Igor Moskalenko, Nikhil Padmanhaban, Andrew Strong, Aaron Pierce and Dan Phalen.  NW is supported by NSF CAREER grant PHY-0449818, and IC, LG and NW are supported by DOE OJI grant \# DE-FG02-06ER41417.

\newpage

\appendix

\section{``Muon'' decay channel spectra}
\label{ap:muonspec}
The probability density for electrons and positrons created via the direct channel is a uniform distribution given by
\begin{equation}
\label{eq:one}
f(E)=\frac{1}{E_{max}-E_{min}} = \frac{1}{\frac{m_{\chi}}{2}(1+\Lambda)-\frac{m_{\chi}}{2}(1-\Lambda)} = \frac{1}{m_{\chi} \Lambda}
\end{equation}
where \[\Lambda = \frac{\sqrt{(m_{\chi}^{2}-m_{\phi}^{2})(m_{\phi}^{2}-(2 m_{e})^2)}}{m_\chi m_\phi}.\]
For $m_{\chi} \gg m_{\phi} \gg m_{e}$ this reduces to \[f(E)\approx \frac{1}{m_{\chi}}.\]

\begin{figure}[t]
\begin{center}
\includegraphics[scale=.4]{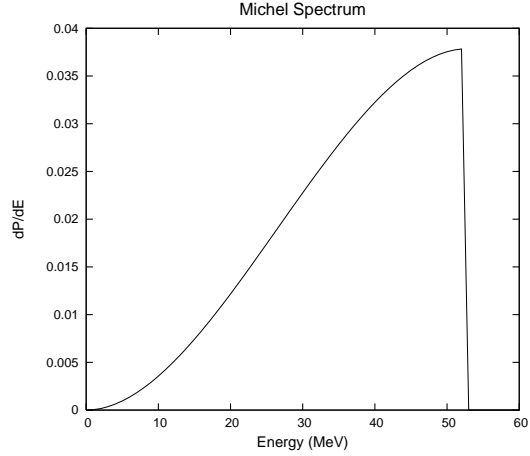}
\end{center}
\caption{\label{fig:michelspec}Energy spectrum of electrons and positrons from muon decay.}
\end{figure}

 The spectrum for the muon channel is a bit more involved. The distribution of injection energies for the final state electrons and positrons is the convolution of the spectrum of the positrons in the muon frame with the spectrum of the muons in the Galaxy, or DM, frame.  The positron spectrum resulting from muon decay has been well-studied and is known as the Michel spectrum \cite{Michel:1949qe}.  It is given by (ignoring radiative corrections)
\begin{equation}
f(E,E_{max})=N E^{2}\,(\frac{3}{2} E_{max}-E)
\end{equation}
where N is a normalization factor and $E_{max}=\frac{m_{\mu}}{2}$ is the endpoint of the spectrum.  It is shown in Fig.~\ref{fig:michelspec}.
  
The spectrum for the muons in the DM frame is identical to that for the positrons in the direct channel (see Eq.~(\ref{eq:one})) with the replacement $m_e \rightarrow m_{\mu}$.  The convolution of the two spectra was done numerically.  The injection spectra for $e^{+}$ $e^{-}$ pairs produced via both channels for $m_{\phi}=0.1$ GeV and $m_{\chi}=0.25$ GeV are shown in Fig. \ref{fig:decaychannels}.

With the following approximations, $m_{e} = 0$ and $\gamma_{1} > 10$ (see below), valid for the $\phi$ and $\chi$ masses of interest to us, the injection spectrum for the muon channel can be calculated analytically.  In this case the probability density is given by

\begin{eqnarray*}
\label{eq: muon ch. inj. sp.}
f(E) &=& \Bigl[\frac{5}{6W} \Bigl( \frac{1}{(\gamma_{2}-\gamma_{1})}ln( \frac{\gamma_{2}}{\gamma_{1}}) \Bigr) \\
&+& \frac{E^{2}}{W^{3}} \Bigr( -\frac{3}{8\gamma_{2}^{3}} - \frac{3}{16\gamma_{1}^{2}(\gamma_{2}-\gamma_{1})} - \frac{3\gamma_{1}}{8\gamma_{2}^{3}(\gamma_{2}-\gamma_{1})} + \frac{9}{16\gamma_{2}^{2}(\gamma_{2}-\gamma_{1})}  \Bigr)\\
 &+& \frac{E^{3}}{W^{4}} \Bigr( \frac{1}{12\gamma_{2}^{4}} + \frac{\gamma_{1}}{12\gamma_{2}^{4}(\gamma_{2}-\gamma_{1})} - \frac{1}{9\gamma_{2}^{3}(\gamma_{2}-\gamma_{1})} + \frac{1}{36\gamma_{1}^{3}(\gamma_{2}-\gamma_{1})}\Bigr)\Bigr] \;\Theta(2W\gamma_{1}-E)\\
 &+& \Big[ \frac{1}{W} \Big( \frac{5}{6\gamma_{2}} - \frac{49}{36(\gamma_{2}-\gamma_{1})} +\frac{5}{6(\gamma_{2}-\gamma_{1})}ln(\frac{2\gamma_{2}W}{E}) + \frac{5\gamma_{1}}{6\gamma_{2}(\gamma_{2}-\gamma_{1})} \Bigr)\\ 
&+& \frac{E^{2}}{W^{3}}\Bigl( \frac{-3}{8\gamma_{2}^{3}} - \frac{3\gamma_{1}}{8\gamma_{2}^{3}(\gamma_{2}-\gamma_{1})} + \frac{9}{16\gamma_{2}^{2}(\gamma_{2}-\gamma_{1})}\Big)\\
&+& \frac{E^{3}}{W^{4}}\Big(\frac{1}{12\gamma_{2}^{4}} + \frac{\gamma_{1}}{12\gamma_{2}^{4}(\gamma_{2}-\gamma_{1})} -\frac{1}{9\gamma_{2}^{3}(\gamma_{2}-\gamma_{1})}\Big) \Bigr] \;\Theta(E-2W\gamma_{1})\;\Theta(2W\gamma_{2}-E),
\end{eqnarray*}

where
\[W = \frac{m_{\mu}}{2},\; \gamma_{1}=\frac{m_{\chi}}{2m_{\mu}}\Bigl(1 - \sqrt{1-\Bigl(\frac{2m_{\mu}}{m_{\phi}}\Bigr)^{2}}\sqrt{1-\Bigl(\frac{m_{\phi}}{m_{\chi}}\Bigr)^{2}}\Bigr) \mbox{, and} \;\gamma_{2}=\frac{m_{\chi}}{m_{\mu}} -\gamma_{1} \;.\]

\newpage
\section{Best Fit Parameters}
\label{ap:bestfit}

\begin{table}[htpb]
\begin{center}
\begin{tabular} {|c|c|c|c|c|c|}
\hline
Halo Model & $M_{\chi}$(GeV) & $\rho_{0}^{2}\left\langle\sigma_{HZ}\left|v\right|\right\rangle$&$\chi^{2}_{HZ}$&$\rho_{0}^{2}\left\langle\sigma_{HT}\left|v\right|\right\rangle$ & $\chi^{2}_{HT}$ \\
 & & $(\times10^{-26} \rm GeV^{2}cm^{-3}s^{-1})$ & & $(\times10^{-26} \rm GeV^{2}cm^{-3}s^{-1})$ & \\
\hline\hline
N.F.W. & 50 & 0.75 & 683 & 0.17 & 8.48 \\
\hline
 & 100 & 2.04 & 728 & 0.53 & 8.64 \\
\hline
 & 250 & 9.71 & 784 & 2.71 & 8.75 \\
\hline
 & 500 & 37.26 & 811 & 10.38 & 8.77 \\
\hline
 & 800 & 92.79 & 827 & 2.013 & 25.97 \\
\hline\hline
Isothermal & 50 & 1.11 & 980 & 0.17 & 8.48 \\
\hline
 & 100 & 3.11 & 1036 & 0.52 & 8.65 \\
\hline
 & 250 & 15.47 & 1038 & 2.46 & 8.80 \\
\hline
 & 500 & 60.81 & 1017 & 9.84 & 8.84 \\
\hline
 & 800 & 153.96 & 1001 & 24.61 & 8.87 \\
\hline\hline
Mrt $\alpha=0.13$ & 50 & 0.30 & 689 & 0.14 & 8.44 \\
\hline
 & 100 & 0.85 & 696 & 0.45 & 8.52 \\
\hline
 & 250 & 3.94 & 739 & 2.21 & 8.57 \\
\hline
 & 500 & 14.88 & 769 & 8.16 & 8.57 \\
\hline
 & 800 & 36.72 & 787 & 19.44 & 8.57 \\
\hline\hline
Mrt $\alpha=0.17$ & 50 & 0.53 & 687 & 0.16 & 8.45 \\
\hline
 & 100 & 1.44 & 723 & 0.50 & 8.58 \\
\hline
 & 250 & 6.77 & 776 & 2.50 & 8.67 \\
\hline
 & 500 & 26.24 & 804 & 9.60 & 8.69 \\
\hline
 & 800 & 64.80 & 820 & 23.33 & 8.70 \\
\hline\hline
Mrt $\alpha=0.22$ & 50 & 0.85 & 723 & 0.18 & 8.49 \\
\hline
 & 100 & 2.35 & 778 & 0.54 & 8.65 \\
\hline
 & 250 & 11.28 & 830 & 2.69 & 8.77 \\
\hline
 & 500 & 43.84 & 850 & 10.56 & 8.80 \\
\hline
 & 800 & 109.73 & 861 & 26.35 & 8.81 \\
\hline
\end{tabular}
\end{center}
\caption{\label{tab:Table direct ch vA=0}Table of best-fit values of $\rho_{0}^{2}\left\langle\sigma\left|v\right|\right\rangle$ and corresponding $\chi^{2}$ for the direct decay channel, $m_{\phi} = .1$ GeV, for $v_A=0$ km/sec.  We give separately the best-fit values for $\rho_{0}^{2}\left\langle\sigma_{HZ}\left|v\right|\right\rangle$ in units of $10^{-26} \rm GeV^{2} cm^{-3} s^{-1}$ to the haze data at 22.5 GHz and $\rho_{0}^{2}\left\langle\sigma_{HT}\left|v\right|\right\rangle$ to the HEAT data. We also give the corresponding $\chi^2_{HZ}=\sum\chi^{2}$ for 29 data points of the haze and  $\chi^2_{HT} = \sum\chi^{2}$ for 9 data points of the HEAT positron fraction.}
\label{tab:Table direct ch. vA=0}
\end{table}

\newpage

\begin{table}[htpb]
\begin{center}
\begin{tabular} {|c|c|c|c|c|c|}
\hline
Halo Model & $M_{\chi}$(GeV) & $\rho_{0}^{2}\left\langle\sigma_{HZ}\left|v\right|\right\rangle$&$\chi^{2}_{HZ}$&$\rho_{0}^{2}\left\langle\sigma_{HT}\left|v\right|\right\rangle$ & $\chi^{2}_{HT}$ \\
 & & $(\times10^{-26} \rm GeV^{2}cm^{-3}s^{-1})$ & & $(\times10^{-26} \rm GeV^{2}cm^{-3}s^{-1})$ & \\
\hline\hline
N.F.W. & 50 & 3.28 & 687 & 0.34 & 8.52 \\
\hline
 & 100 & 4.35 & 666 & 0.88 & 8.31 \\
\hline
 & 250 & 13.12 & 715 & 3.42 & 8.56 \\
\hline
 & 500 & 38.91 & 755 & 10.98 & 8.69 \\
\hline
 & 800 & 94.66 & 779 & 26.50 & 8.74 \\
\hline\hline
Isothermal & 50 & 4.96 & 797 & 0.31 & 8.55 \\
\hline
 & 100 & 6.41 & 921 & 0.85 & 8.32 \\
\hline
 & 250 & 19.78 & 1017 & 3.26 & 8.57 \\
\hline
 & 500 & 63.49 & 1036 & 10.83 & 8.73 \\
\hline
 & 800 & 150.78 & 1034 & 25.21 & 8.79 \\
\hline\hline
Mrt $\alpha=0.13$& 50 & 1.41 & 812 & 0.27 & 8.62 \\
\hline
 & 100 & 1.82 & 700 & 0.74 & 8.35 \\
\hline
 & 250 & 5.46 & 689 & 2.86 & 8.47 \\
\hline
 & 500 & 16.78 & 713 & 9.33 & 8.55 \\
\hline
 & 800 & 38.59 & 734 & 21.44 & 8.57 \\
\hline\hline
Mrt $\alpha=0.17$& 50 & 2.27 & 721 & 0.30 & 8.55 \\
\hline
 & 100 & 3.02 & 677 & 0.83 & 8.32 \\
\hline
 & 250 & 9.26 & 711 & 3.15 & 8.51 \\
\hline
 & 500 & 28.50 & 747 & 10.34 & 8.63 \\
\hline
 & 800 & 67.01 & 771 & 24.66 & 8.67 \\
\hline\hline
Mrt $\alpha=0.22$& 50 & 3.79 & 682 & 0.34 & 8.52 \\
\hline
 & 100 & 4.88 & 696 & 0.90 & 8.32 \\
\hline
 & 250 & 14.93 & 762 & 3.38 & 8.57 \\
\hline
 & 500 & 47.15 & 804 & 10.72 & 8.71 \\
\hline
 & 800 & 110.42 & 825 & 26.80 & 8.77  \\
\hline
\end{tabular}
\end{center}
\caption{As table \ref{tab:Table direct ch vA=0}, but for the muon decay channel, $m_{\phi} = .25$ GeV, for $v_A=0$ km/sec.}
\label{tab:Table muon ch vA=0}
\end{table}

\newpage
\begin{table}[htpb]
\begin{center}
\begin{tabular} {|c|c|c|c|c|c|}
\hline
Halo Model & $M_{\chi}$(GeV) & $\rho_{0}^{2}\left\langle\sigma_{HZ}\left|v\right|\right\rangle$&$\chi^{2}_{HZ}$&$\rho_{0}^{2}\left\langle\sigma_{HT}\left|v\right|\right\rangle$ & $\chi^{2}_{HT}$ \\
 & & $(\times10^{-26} \rm GeV^{2}cm^{-3}s^{-1})$ & & $(\times10^{-26} \rm GeV^{2}cm^{-3}s^{-1})$ & \\
\hline\hline
N.F.W. & 50 & 0.74 & 684 & 0.18 & 7.07 \\
\hline
 & 100 & 2.03 & 729 & 0.56 & 7.23 \\
\hline
 & 250 & 9.64 & 784 & 2.83 & 7.35 \\
\hline
 & 500 & 37.16 & 812 & 11.18 & 7.37 \\
\hline
 & 800 & 92.10 & 827 & 27.57 & 7.37 \\
\hline\hline
Isothermal & 50 & 1.09 & 980 & 0.17 & 7.02 \\
\hline
 & 100 & 3.09 & 1035 & 0.53 & 7.19 \\
\hline
 & 250 & 15.25 & 1037 & 2.80 & 7.37 \\
\hline
 & 500 & 60.50 & 1017 & 10.62 & 7.42 \\
\hline
 & 800 & 153.24 & 1001 & 26.40 & 7.45 \\
\hline\hline
Mrt $\alpha=0.13$ & 50 & 0.30 & 689 & 0.14 & 7.24 \\
\hline
 & 100 & 0.83 & 697 & 0.45 & 7.25 \\
\hline
 & 250 & 3.94 & 739 & 2.16 & 7.26 \\
\hline
 & 500 & 14.82 & 769 & 8.16 & 7.23 \\
\hline
 & 800 & 36.58 & 788 & 20.11 & 7.20 \\
\hline\hline
Mrt $\alpha=0.17$ & 50 & 0.51 & 687 & 0.16 & 7.11 \\
\hline
 & 100 & 1.42 & 723 & 0.52 & 7.21 \\
\hline
 & 250 & 6.70 & 777 & 2.56 & 7.30 \\
\hline
 & 500 & 26.02 & 804 & 9.85 & 7.30 \\
\hline
 & 800 & 64.53 & 821 & 24.67 & 7.29 \\
\hline\hline
Mrt $\alpha=0.22$ & 50 & 0.85 & 724 & 0.18 & 7.06 \\
\hline
 & 100 & 2.34 & 778 & 0.56 & 7.23 \\
\hline
 & 250 & 11.20 & 830 & 2.86 & 7.37 \\
\hline
 & 500 & 43.60 & 851 & 11.23 & 7.39 \\
\hline
 & 800 & 109.71 & 861 & 28.33 & 7.40 \\
\hline
\end{tabular}
\end{center}
\caption{As table \ref{tab:Table direct ch vA=0}, but for the direct decay channel, $m_{\phi} = .1$ GeV, for $v_A=20$ km/sec.}
\label{tab:Table direct ch vA=20}
\end{table}

\newpage

\begin{table}[htpb]
\begin{center}
\begin{tabular} {|c|c|c|c|c|c|}
\hline
Halo Model & $M_{\chi}$(GeV) & $\rho_{0}^{2}\left\langle\sigma_{HZ}\left|v\right|\right\rangle$&$\chi^{2}_{HZ}$&$\rho_{0}^{2}\left\langle\sigma_{HT}\left|v\right|\right\rangle$ & $\chi^{2}_{HT}$ \\
 & & $(\times10^{-26} \rm GeV^{2}cm^{-3}s^{-1})$ & & $(\times10^{-26} \rm GeV^{2}cm^{-3}s^{-1})$ & \\
\hline\hline
N.F.W. & 50 & 3.29 & 687 & 0.31 & 7.60 \\
\hline
 & 100 & 4.23 & 667 & 0.87 & 6.99 \\
\hline
 & 250 & 13.16 & 715 & 3.55 & 7.14 \\
\hline
 & 500 & 40.88 & 756 & 11.84 & 7.29 \\
\hline
 & 800 & 94.06 & 780 & 27.53 & 7.35 \\
\hline\hline
Isothermal & 50 & 4.86 & 798 & 0.28 & 7.63 \\
\hline
 & 100 & 6.32 & 921 & 0.83 & 6.97 \\
\hline
 & 250 & 19.71 & 1018 & 3.47 & 7.10 \\
\hline
 & 500 & 62.87 & 1036 & 11.32 & 7.28 \\
\hline
 & 800 & 149.70 & 1033 & 27.66 & 7.36 \\
\hline\hline
Mrt $\alpha=0.13$ & 50 & 1.36 & 803 & 0.24 & 7.79 \\
\hline
 & 100 & 1.79 & 700 & 0.68 & 7.23 \\
\hline
 & 250 & 5.28 & 691 & 2.84 & 7.21 \\
\hline
 & 500 & 16.90 & 715 & 9.42 & 7.26 \\
\hline
 & 800 & 38.19 & 735 & 21.99 & 7.26 \\
\hline\hline
Mrt $\alpha=0.17$ & 50 & 2.29 & 722 & 0.27 & 7.70 \\
\hline
 & 100 & 3.01 & 678 & 0.80 & 7.08 \\
\hline
 & 250 & 9.15 & 711 & 3.29 & 7.14 \\
\hline
 & 500 & 28.51 & 748 & 10.77 & 7.26 \\
\hline
 & 800 & 66.53 & 771 & 25.13 & 7.29 \\
\hline\hline
Mrt $\alpha=0.22$ & 50 & 3.65 & 681 & 0.30 & 7.59 \\
\hline
 & 100 & 4.86 & 696 & 0.86 & 6.98 \\
\hline
 & 250 & 14.78 & 763 & 3.59 & 7.14 \\
\hline
 & 500 & 47.52 & 804 & 12.12 & 7.30 \\
\hline
 & 800 & 109.65 & 825 & 28.27 & 7.36 \\
\hline
\end{tabular}
\end{center}
\caption{As table \ref{tab:Table direct ch vA=0}, but for the muon decay channel, $m_{\phi} = .25$ GeV, for $v_A=20$ km/sec.}
\label{tab:Table muon ch vA=20}
\end{table}

\newpage

\begin{table}[htpb]
\begin{center}
\begin{tabular} {|c|c|c|c|c|c|}
\hline
Halo Model & $M_{\chi}$(GeV) & $\rho_{0}^{2}\left\langle\sigma_{HZ}\left|v\right|\right\rangle$&$\chi^{2}_{HZ}$&$\rho_{0}^{2}\left\langle\sigma_{HT}\left|v\right|\right\rangle$ & $\chi^{2}_{HT}$ \\
 & & $(\times10^{-26} \rm GeV^{2}cm^{-3}s^{-1})$ & & $(\times10^{-26} \rm GeV^{2}cm^{-3}s^{-1})$ & \\
\hline\hline
N.F.W. & 50 & 0.74 & 684 & 0.56 & 8.16 \\
\hline
 & 100 & 2.04 & 730 & 1.73 & 9.73 \\
\hline
 & 250 & 9.54 & 785 & 8.75 & 10.61 \\
\hline
 & 500 & 36.90 & 812 & 34.39 & 10.77 \\
\hline
 & 800 & 91.75 & 827 & 85.77 & 10.82 \\
\hline\hline
Isothermal & 50 & 1.09 & 980 & 0.56 & 7.94 \\
\hline
 & 100 & 3.03 & 1035 & 1.66 & 9.72 \\
\hline
 & 250 & 15.92 & 1036 & 8.38 & 10.97 \\
\hline
 & 500 & 59.98 & 1017 & 32.83 & 11.35 \\
\hline
 & 800 & 151.42 & 1001 & 82.61 & 11.52 \\
\hline\hline
Mrt $\alpha=0.13$ & 50 & 0.27 & 688 & 0.42 & 8.36 \\
\hline
 & 100 & 0.83 & 698 & 1.44 & 9.01 \\
\hline
 & 250 & 3.98 & 745 & 7.19 & 9.24 \\
\hline
 & 500 & 15.04 & 772 & 26.28 & 9.18 \\
\hline
 & 800 & 35.55 & 792 & 64.22 & 9.16 \\
\hline\hline
Mrt $\alpha=0.17$ & 50 & 0.46 & 687 & 0.48 & 8.11 \\
\hline
 & 100 & 1.41 & 724 & 1.60 & 9.35 \\
\hline
 & 250 & 6.72 & 777 & 8.26 & 10.03 \\
\hline
 & 500 & 25.90 & 805 & 31.25 & 10.13 \\
\hline
 & 800 & 64.45 & 821 & 77.44 & 10.18 \\
\hline\hline
Mrt $\alpha=0.22$ & 50 & 0.75 & 725 & 0.52 & 8.13 \\
\hline
 & 100 & 2.34 & 780 & 1.74 & 9.80 \\
\hline
 & 250 & 11.20 & 830 & 8.88 & 10.79 \\
\hline
 & 500 & 43.46 & 851 & 34.80 & 11.01 \\
\hline
 & 800 & 108.90 & 861 & 86.89 & 11.10 \\
\hline
\end{tabular}
\end{center}
\caption{As table \ref{tab:Table direct ch vA=0}, but for the direct decay channel, $m_{\phi} = .1$ GeV, for $v_A=35$ km/sec.}
\label{tab:Table direct ch vA=35}
\end{table}

\newpage

\begin{table}[htpb]
\begin{center}
\begin{tabular} {|c|c|c|c|c|c|}
\hline
Halo Model & $M_{\chi}$(GeV) & $\rho_{0}^{2}\left\langle\sigma_{HZ}\left|v\right|\right\rangle$&$\chi^{2}_{HZ}$&$\rho_{0}^{2}\left\langle\sigma_{HT}\left|v\right|\right\rangle$ & $\chi^{2}_{HT}$ \\
 & & $(\times10^{-26} \rm GeV^{2}cm^{-3}s^{-1})$ & & $(\times10^{-26} \rm GeV^{2}cm^{-3}s^{-1})$ & \\
\hline\hline
N.F.W. & 50 & 3.23 & 683 & 1.10 & 8.81 \\
\hline
 & 100 & 4.32 & 667 & 2.90 & 7.17 \\
\hline
 & 250 & 12.90 & 716 & 10.96 & 9.11 \\
\hline
 & 500 & 40.36 & 757 & 36.53 & 10.22 \\
\hline
 & 800 & 93.78 & 780 & 86.62 & 10.58 \\
\hline\hline
Isothermal & 50 & 4.77 & 799 & 1.10 & 8.50 \\
\hline
 & 100 & 6.27 & 921 & 2.80 & 6.91 \\
\hline
 & 250 & 19.67 & 1017 & 10.78 & 9.03 \\
\hline
 & 500 & 62.40 & 1035 & 34.65 & 10.40 \\
\hline
 & 800 & 149.14 & 1033 & 83.94 & 10.91 \\
\hline\hline
Mrt $\alpha=0.13$ & 50 & 1.31 & 798 & 0.87 & 10.47 \\
\hline
 & 100 & 2.30 & 699 & 2.32 & 8.11 \\
\hline
 & 250 & 5.38 & 690 & 9.14 & 8.73 \\
\hline
 & 500 & 16.42 & 714 & 30.11 & 9.20 \\
\hline
 & 800 & 37.70 & 736 & 70.83 & 9.29 \\
\hline\hline
Mrt $\alpha=0.17$ & 50 & 2.21 & 717 & 1.01 & 9.29 \\
\hline
 & 100 & 2.98 & 677 & 2.64 & 7.41 \\
\hline
 & 250 & 8.96 & 713 & 10.23 & 8.88 \\
\hline
 & 500 & 27.65 & 750 & 33.78 & 9.78 \\
\hline
 & 800 & 65.47 & 772 & 79.26 & 10.04 \\
\hline\hline
Mrt $\alpha=0.22$ & 50 & 3.68 & 680 & 1.14 & 8.63 \\
\hline
 & 100 & 4.86 & 696 & 2.90 & 7.08 \\
\hline
 & 250 & 14.85 & 763 & 11.10 & 9.16 \\
\hline
 & 500 & 46.66 & 805 & 36.72 & 10.35 \\
\hline
 & 800 & 109.12 & 826 & 87.69 & 10.76 \\
\hline
\end{tabular}
\end{center}
\caption{As table \ref{tab:Table direct ch vA=0}, but for the muon decay channel, $m_{\phi} = .25$ GeV, for $v_A=35$ km/sec.}
\label{tab:Table muon ch vA=35}
\end{table}

\bibliography{xdmpositron}

\begin{thebibliography}{71}
\expandafter\ifx\csname natexlab\endcsname\relax\def\natexlab#1{#1}\fi
\expandafter\ifx\csname bibnamefont\endcsname\relax
  \def\bibnamefont#1{#1}\fi
\expandafter\ifx\csname bibfnamefont\endcsname\relax
  \def\bibfnamefont#1{#1}\fi
\expandafter\ifx\csname citenamefont\endcsname\relax
  \def\citenamefont#1{#1}\fi
\expandafter\ifx\csname url\endcsname\relax
  \def\url#1{\texttt{#1}}\fi
\expandafter\ifx\csname urlprefix\endcsname\relax\def\urlprefix{URL }\fi
\providecommand{\bibinfo}[2]{#2}
\providecommand{\eprint}[2][]{\url{#2}}

\bibitem[{\citenamefont{Zwicky}(1933)}]{Zwicky:1933gu}
\bibinfo{author}{\bibfnamefont{F.}~\bibnamefont{Zwicky}},
  \bibinfo{journal}{Helv. Phys. Acta} \textbf{\bibinfo{volume}{6}},
  \bibinfo{pages}{110} (\bibinfo{year}{1933}).

\bibitem[{\citenamefont{Rubin et~al.}(1980)\citenamefont{Rubin, Thonnard, and
  Ford}}]{Rubin:1980zd}
\bibinfo{author}{\bibfnamefont{V.~C.} \bibnamefont{Rubin}},
  \bibinfo{author}{\bibfnamefont{N.}~\bibnamefont{Thonnard}}, \bibnamefont{and}
  \bibinfo{author}{\bibfnamefont{J.}~\bibnamefont{Ford}, \bibfnamefont{W.~K.}},
  \bibinfo{journal}{Astrophys. J.} \textbf{\bibinfo{volume}{238}},
  \bibinfo{pages}{471} (\bibinfo{year}{1980}).

\bibitem[{\citenamefont{Rubin et~al.}(1985)\citenamefont{Rubin, Burstein, Ford,
  and Thonnard}}]{Rubin:1985ze}
\bibinfo{author}{\bibfnamefont{V.~C.} \bibnamefont{Rubin}},
  \bibinfo{author}{\bibfnamefont{D.}~\bibnamefont{Burstein}},
  \bibinfo{author}{\bibfnamefont{J.}~\bibnamefont{Ford}, \bibfnamefont{W.~K.}},
  \bibnamefont{and} \bibinfo{author}{\bibfnamefont{N.}~\bibnamefont{Thonnard}},
  \bibinfo{journal}{Astrophys. J.} \textbf{\bibinfo{volume}{289}},
  \bibinfo{pages}{81} (\bibinfo{year}{1985}).

\bibitem[{\citenamefont{Tyson et~al.}(1998)\citenamefont{Tyson, Kochanski, and
  Dell'Antonio}}]{Tyson:1998vp}
\bibinfo{author}{\bibfnamefont{J.~A.} \bibnamefont{Tyson}},
  \bibinfo{author}{\bibfnamefont{G.~P.} \bibnamefont{Kochanski}},
  \bibnamefont{and} \bibinfo{author}{\bibfnamefont{I.~P.}
  \bibnamefont{Dell'Antonio}}, \bibinfo{journal}{Astrophys. J.}
  \textbf{\bibinfo{volume}{498}}, \bibinfo{pages}{L107} (\bibinfo{year}{1998}),
  \eprint{astro-ph/9801193}.

\bibitem[{\citenamefont{Cen and Ostriker}(1994)}]{Cen:1993az}
\bibinfo{author}{\bibfnamefont{R.-Y.} \bibnamefont{Cen}} \bibnamefont{and}
  \bibinfo{author}{\bibfnamefont{J.~P.} \bibnamefont{Ostriker}},
  \bibinfo{journal}{Astrophys. J.} \textbf{\bibinfo{volume}{429}},
  \bibinfo{pages}{4} (\bibinfo{year}{1994}), \eprint{astro-ph/9404012}.

\bibitem[{\citenamefont{Spergel et~al.}(2007)}]{Spergel:2006hy}
\bibinfo{author}{\bibfnamefont{D.~N.} \bibnamefont{Spergel}}
  \bibnamefont{et~al.} (\bibinfo{collaboration}{WMAP}),
  \bibinfo{journal}{Astrophys. J. Suppl.} \textbf{\bibinfo{volume}{170}},
  \bibinfo{pages}{377} (\bibinfo{year}{2007}), \eprint{astro-ph/0603449}.

\bibitem[{\citenamefont{Cole et~al.}(2005)}]{Cole:2005sx}
\bibinfo{author}{\bibfnamefont{S.}~\bibnamefont{Cole}} \bibnamefont{et~al.}
  (\bibinfo{collaboration}{The 2dFGRS}), \bibinfo{journal}{Mon. Not. Roy.
  Astron. Soc.} \textbf{\bibinfo{volume}{362}}, \bibinfo{pages}{505}
  (\bibinfo{year}{2005}), \eprint{astro-ph/0501174}.

\bibitem[{\citenamefont{Tegmark et~al.}(2006)}]{Tegmark:2006az}
\bibinfo{author}{\bibfnamefont{M.}~\bibnamefont{Tegmark}} \bibnamefont{et~al.}
  (\bibinfo{collaboration}{SDSS}), \bibinfo{journal}{Phys. Rev.}
  \textbf{\bibinfo{volume}{D74}}, \bibinfo{pages}{123507}
  (\bibinfo{year}{2006}), \eprint{astro-ph/0608632}.

\bibitem[{\citenamefont{Clowe et~al.}(2006)}]{Clowe:2006eq}
\bibinfo{author}{\bibfnamefont{D.}~\bibnamefont{Clowe}} \bibnamefont{et~al.},
  \bibinfo{journal}{Astrophys. J.} \textbf{\bibinfo{volume}{648}},
  \bibinfo{pages}{L109} (\bibinfo{year}{2006}), \eprint{astro-ph/0608407}.

\bibitem[{\citenamefont{{Kolb} and {Turner}}(1990)}]{kolbandturner}
\bibinfo{author}{\bibfnamefont{E.~W.} \bibnamefont{{Kolb}}} \bibnamefont{and}
  \bibinfo{author}{\bibfnamefont{M.~S.} \bibnamefont{{Turner}}},
  \emph{\bibinfo{title}{{The early universe}}} (\bibinfo{publisher}{Frontiers
  in Physics, Reading, MA: Addison-Wesley}, \bibinfo{year}{1990}).

\bibitem[{\citenamefont{Ahrens et~al.}(2002)}]{Ahrens:2002eb}
\bibinfo{author}{\bibfnamefont{J.}~\bibnamefont{Ahrens}} \bibnamefont{et~al.}
  (\bibinfo{collaboration}{AMANDA}), \bibinfo{journal}{Phys. Rev.}
  \textbf{\bibinfo{volume}{D66}}, \bibinfo{pages}{032006}
  (\bibinfo{year}{2002}), \eprint{astro-ph/0202370}.

\bibitem[{\citenamefont{Ahrens et~al.}(2004)}]{Ahrens:2003ix}
\bibinfo{author}{\bibfnamefont{J.}~\bibnamefont{Ahrens}} \bibnamefont{et~al.}
  (\bibinfo{collaboration}{IceCube}), \bibinfo{journal}{Astropart. Phys.}
  \textbf{\bibinfo{volume}{20}}, \bibinfo{pages}{507} (\bibinfo{year}{2004}),
  \eprint{astro-ph/0305196}.

\bibitem[{\citenamefont{Picozza et~al.}(2007)}]{Picozza:2006nm}
\bibinfo{author}{\bibfnamefont{P.}~\bibnamefont{Picozza}} \bibnamefont{et~al.},
  \bibinfo{journal}{Astropart. Phys.} \textbf{\bibinfo{volume}{27}},
  \bibinfo{pages}{296} (\bibinfo{year}{2007}), \eprint{astro-ph/0608697}.

\bibitem[{\citenamefont{Gehrels and Michelson}(1999)}]{Gehrels:1999ri}
\bibinfo{author}{\bibfnamefont{N.}~\bibnamefont{Gehrels}} \bibnamefont{and}
  \bibinfo{author}{\bibfnamefont{P.}~\bibnamefont{Michelson}},
  \bibinfo{journal}{Astropart. Phys.} \textbf{\bibinfo{volume}{11}},
  \bibinfo{pages}{277} (\bibinfo{year}{1999}).

\bibitem[{\citenamefont{Hinton}(2004)}]{Hinton:2004eu}
\bibinfo{author}{\bibfnamefont{J.~A.} \bibnamefont{Hinton}}
  (\bibinfo{collaboration}{The HESS}), \bibinfo{journal}{New Astron. Rev.}
  \textbf{\bibinfo{volume}{48}}, \bibinfo{pages}{331} (\bibinfo{year}{2004}),
  \eprint{astro-ph/0403052}.

\bibitem[{\citenamefont{Weidenspointner
  et~al.}(2006)}]{Weidenspointner:2006nua}
\bibinfo{author}{\bibfnamefont{G.}~\bibnamefont{Weidenspointner}}
  \bibnamefont{et~al.} (\bibinfo{year}{2006}), \eprint{astro-ph/0601673}.

\bibitem[{\citenamefont{Barbiellini et~al.}(1996)}]{Barbiellini:1996ba}
\bibinfo{author}{\bibfnamefont{G.}~\bibnamefont{Barbiellini}}
  \bibnamefont{et~al.}, \bibinfo{journal}{Astron. Astrophys.}
  \textbf{\bibinfo{volume}{309}}, \bibinfo{pages}{L15} (\bibinfo{year}{1996}).

\bibitem[{\citenamefont{Protheroe}(1982)}]{Protheroe:1982pp}
\bibinfo{author}{\bibfnamefont{R.~J.} \bibnamefont{Protheroe}},
  \bibinfo{journal}{Astrophys. J.} \textbf{\bibinfo{volume}{254}},
  \bibinfo{pages}{391} (\bibinfo{year}{1982}).

\bibitem[{\citenamefont{Barwick et~al.}(1997)}]{Barwick:1997ig}
\bibinfo{author}{\bibfnamefont{S.~W.} \bibnamefont{Barwick}}
  \bibnamefont{et~al.} (\bibinfo{collaboration}{HEAT}),
  \bibinfo{journal}{Astrophys. J.} \textbf{\bibinfo{volume}{482}},
  \bibinfo{pages}{L191} (\bibinfo{year}{1997}), \eprint{astro-ph/9703192}.

\bibitem[{\citenamefont{Coutu et~al.}(1999)}]{Coutu:1999ws}
\bibinfo{author}{\bibfnamefont{S.}~\bibnamefont{Coutu}} \bibnamefont{et~al.},
  \bibinfo{journal}{Astropart. Phys.} \textbf{\bibinfo{volume}{11}},
  \bibinfo{pages}{429} (\bibinfo{year}{1999}), \eprint{astro-ph/9902162}.

\bibitem[{\citenamefont{Finkbeiner}(2004{\natexlab{a}})}]{Finkbeiner:2003im}
\bibinfo{author}{\bibfnamefont{D.~P.} \bibnamefont{Finkbeiner}},
  \bibinfo{journal}{Astrophys. J.} \textbf{\bibinfo{volume}{614}},
  \bibinfo{pages}{186} (\bibinfo{year}{2004}{\natexlab{a}}),
  \eprint{astro-ph/0311547}.

\bibitem[{\citenamefont{Finkbeiner}(2004{\natexlab{b}})}]{Finkbeiner:2004us}
\bibinfo{author}{\bibfnamefont{D.~P.} \bibnamefont{Finkbeiner}}
  (\bibinfo{year}{2004}{\natexlab{b}}), \eprint{astro-ph/0409027}.

\bibitem[{\citenamefont{Hooper et~al.}(2007)\citenamefont{Hooper, Finkbeiner,
  and Dobler}}]{Hooper:2007kb}
\bibinfo{author}{\bibfnamefont{D.}~\bibnamefont{Hooper}},
  \bibinfo{author}{\bibfnamefont{D.~P.} \bibnamefont{Finkbeiner}},
  \bibnamefont{and} \bibinfo{author}{\bibfnamefont{G.}~\bibnamefont{Dobler}},
  \bibinfo{journal}{Phys. Rev.} \textbf{\bibinfo{volume}{D76}},
  \bibinfo{pages}{083012} (\bibinfo{year}{2007}), \eprint{arXiv:0705.3655
  [astro-ph]}.

\bibitem[{\citenamefont{Weidenspointner et~al.}(2007)}]{Weidenspointner:2007ru}
\bibinfo{author}{\bibfnamefont{G.}~\bibnamefont{Weidenspointner}}
  \bibnamefont{et~al.} (\bibinfo{year}{2007}), \eprint{astro-ph/0702623}.

\bibitem[{\citenamefont{Weidenspointner et~al.}(2008)}]{Weidenspointner:2008zz}
\bibinfo{author}{\bibfnamefont{G.}~\bibnamefont{Weidenspointner}}
  \bibnamefont{et~al.}, \bibinfo{journal}{Nature}
  \textbf{\bibinfo{volume}{451}}, \bibinfo{pages}{159} (\bibinfo{year}{2008}).

\bibitem[{\citenamefont{Beacom et~al.}(2005)\citenamefont{Beacom, Bell, and
  Bertone}}]{Beacom:2004pe}
\bibinfo{author}{\bibfnamefont{J.~F.} \bibnamefont{Beacom}},
  \bibinfo{author}{\bibfnamefont{N.~F.} \bibnamefont{Bell}}, \bibnamefont{and}
  \bibinfo{author}{\bibfnamefont{G.}~\bibnamefont{Bertone}},
  \bibinfo{journal}{Phys. Rev. Lett.} \textbf{\bibinfo{volume}{94}},
  \bibinfo{pages}{171301} (\bibinfo{year}{2005}), \eprint{astro-ph/0409403}.

\bibitem[{\citenamefont{Jean et~al.}(2006)}]{Jean:2005af}
\bibinfo{author}{\bibfnamefont{P.}~\bibnamefont{Jean}} \bibnamefont{et~al.},
  \bibinfo{journal}{Astron. Astrophys.} \textbf{\bibinfo{volume}{445}},
  \bibinfo{pages}{579} (\bibinfo{year}{2006}), \eprint{astro-ph/0509298}.

\bibitem[{\citenamefont{Beacom and Yuksel}(2006)}]{Beacom:2005qv}
\bibinfo{author}{\bibfnamefont{J.~F.} \bibnamefont{Beacom}} \bibnamefont{and}
  \bibinfo{author}{\bibfnamefont{H.}~\bibnamefont{Yuksel}},
  \bibinfo{journal}{Phys. Rev. Lett.} \textbf{\bibinfo{volume}{97}},
  \bibinfo{pages}{071102} (\bibinfo{year}{2006}), \eprint{astro-ph/0512411}.

\bibitem[{\citenamefont{Boehm and Fayet}(2004)}]{Boehm:2003hm}
\bibinfo{author}{\bibfnamefont{C.}~\bibnamefont{Boehm}} \bibnamefont{and}
  \bibinfo{author}{\bibfnamefont{P.}~\bibnamefont{Fayet}},
  \bibinfo{journal}{Nucl. Phys.} \textbf{\bibinfo{volume}{B683}},
  \bibinfo{pages}{219} (\bibinfo{year}{2004}), \eprint{hep-ph/0305261}.

\bibitem[{\citenamefont{Boehm et~al.}(2004)\citenamefont{Boehm, Hooper, Silk,
  Casse, and Paul}}]{Boehm:2003bt}
\bibinfo{author}{\bibfnamefont{C.}~\bibnamefont{Boehm}},
  \bibinfo{author}{\bibfnamefont{D.}~\bibnamefont{Hooper}},
  \bibinfo{author}{\bibfnamefont{J.}~\bibnamefont{Silk}},
  \bibinfo{author}{\bibfnamefont{M.}~\bibnamefont{Casse}}, \bibnamefont{and}
  \bibinfo{author}{\bibfnamefont{J.}~\bibnamefont{Paul}},
  \bibinfo{journal}{Phys. Rev. Lett.} \textbf{\bibinfo{volume}{92}},
  \bibinfo{pages}{101301} (\bibinfo{year}{2004}), \eprint{astro-ph/0309686}.

\bibitem[{\citenamefont{Hooper and Zurek}(2008)}]{Hooper:2008im}
\bibinfo{author}{\bibfnamefont{D.}~\bibnamefont{Hooper}} \bibnamefont{and}
  \bibinfo{author}{\bibfnamefont{K.~M.} \bibnamefont{Zurek}}
  (\bibinfo{year}{2008}), \eprint{arXiv:0801.3686 [hep-ph]}.

\bibitem[{\citenamefont{Hooper and Wang}(2004)}]{Hooper:2004qf}
\bibinfo{author}{\bibfnamefont{D.}~\bibnamefont{Hooper}} \bibnamefont{and}
  \bibinfo{author}{\bibfnamefont{L.-T.} \bibnamefont{Wang}},
  \bibinfo{journal}{Phys. Rev.} \textbf{\bibinfo{volume}{D70}},
  \bibinfo{pages}{063506} (\bibinfo{year}{2004}), \eprint{hep-ph/0402220}.

\bibitem[{\citenamefont{Pospelov and Ritz}(2007)}]{Pospelov:2007xh}
\bibinfo{author}{\bibfnamefont{M.}~\bibnamefont{Pospelov}} \bibnamefont{and}
  \bibinfo{author}{\bibfnamefont{A.}~\bibnamefont{Ritz}},
  \bibinfo{journal}{Phys. Lett.} \textbf{\bibinfo{volume}{B651}},
  \bibinfo{pages}{208} (\bibinfo{year}{2007}), \eprint{hep-ph/0703128}.

\bibitem[{\citenamefont{Finkbeiner and Weiner}(2007)}]{Finkbeiner:2007kk}
\bibinfo{author}{\bibfnamefont{D.~P.} \bibnamefont{Finkbeiner}}
  \bibnamefont{and} \bibinfo{author}{\bibfnamefont{N.}~\bibnamefont{Weiner}},
  \bibinfo{journal}{Phys. Rev.} \textbf{\bibinfo{volume}{D76}},
  \bibinfo{pages}{083519} (\bibinfo{year}{2007}), \eprint{astro-ph/0702587}.

\bibitem[{\citenamefont{Drees and Nojiri}(1993)}]{Drees:1992am}
\bibinfo{author}{\bibfnamefont{M.}~\bibnamefont{Drees}} \bibnamefont{and}
  \bibinfo{author}{\bibfnamefont{M.~M.} \bibnamefont{Nojiri}},
  \bibinfo{journal}{Phys. Rev.} \textbf{\bibinfo{volume}{D47}},
  \bibinfo{pages}{376} (\bibinfo{year}{1993}), \eprint{hep-ph/9207234}.

\bibitem[{\citenamefont{Gunion et~al.}(1990)\citenamefont{Gunion, Haber, Kane,
  and Dawson}}]{HiggsHG}
\bibinfo{author}{\bibfnamefont{J.~F.} \bibnamefont{Gunion}},
  \bibinfo{author}{\bibfnamefont{H.~E.} \bibnamefont{Haber}},
  \bibinfo{author}{\bibfnamefont{G.~L.} \bibnamefont{Kane}}, \bibnamefont{and}
  \bibinfo{author}{\bibfnamefont{S.}~\bibnamefont{Dawson}}
  (\bibinfo{year}{1990}), \bibinfo{note}{perseus Publishing, Cambridge, MA}.

\bibitem[{\citenamefont{Voloshin and Zakharov}(1980)}]{Voloshin:1980zf}
\bibinfo{author}{\bibfnamefont{M.~B.} \bibnamefont{Voloshin}} \bibnamefont{and}
  \bibinfo{author}{\bibfnamefont{V.~I.} \bibnamefont{Zakharov}},
  \bibinfo{journal}{Phys. Rev. Lett.} \textbf{\bibinfo{volume}{45}},
  \bibinfo{pages}{688} (\bibinfo{year}{1980}).

\bibitem[{\citenamefont{Voloshin}(1986)}]{Voloshin:1985tc}
\bibinfo{author}{\bibfnamefont{M.~B.} \bibnamefont{Voloshin}},
  \bibinfo{journal}{Sov. J. Nucl. Phys.} \textbf{\bibinfo{volume}{44}},
  \bibinfo{pages}{478} (\bibinfo{year}{1986}).

\bibitem[{\citenamefont{Grinstein et~al.}(1988)\citenamefont{Grinstein, Hall,
  and Randall}}]{Grinstein:1988yu}
\bibinfo{author}{\bibfnamefont{B.}~\bibnamefont{Grinstein}},
  \bibinfo{author}{\bibfnamefont{L.~J.} \bibnamefont{Hall}}, \bibnamefont{and}
  \bibinfo{author}{\bibfnamefont{L.}~\bibnamefont{Randall}},
  \bibinfo{journal}{Phys. Lett.} \textbf{\bibinfo{volume}{B211}},
  \bibinfo{pages}{363} (\bibinfo{year}{1988}).

\bibitem[{\citenamefont{Moskalenko and Strong}(1999)}]{Moskalenko:1999sb}
\bibinfo{author}{\bibfnamefont{I.~V.} \bibnamefont{Moskalenko}}
  \bibnamefont{and} \bibinfo{author}{\bibfnamefont{A.~W.}
  \bibnamefont{Strong}}, \bibinfo{journal}{Phys. Rev.}
  \textbf{\bibinfo{volume}{D60}}, \bibinfo{pages}{063003}
  (\bibinfo{year}{1999}), \eprint{astro-ph/9905283}.

\bibitem[{\citenamefont{Strong et~al.}(2007)\citenamefont{Strong, Moskalenko,
  and Ptuskin}}]{Strong:2007nh}
\bibinfo{author}{\bibfnamefont{A.~W.} \bibnamefont{Strong}},
  \bibinfo{author}{\bibfnamefont{I.~V.} \bibnamefont{Moskalenko}},
  \bibnamefont{and} \bibinfo{author}{\bibfnamefont{V.~S.}
  \bibnamefont{Ptuskin}}, \bibinfo{journal}{Ann. Rev. Nucl. Part. Sci.}
  \textbf{\bibinfo{volume}{57}}, \bibinfo{pages}{285} (\bibinfo{year}{2007}),
  \eprint{astro-ph/0701517}.

\bibitem[{\citenamefont{Strong and
  Moskalenko}(1999{\natexlab{a}})}]{Strong:1999sv}
\bibinfo{author}{\bibfnamefont{A.~W.} \bibnamefont{Strong}} \bibnamefont{and}
  \bibinfo{author}{\bibfnamefont{I.~V.} \bibnamefont{Moskalenko}}
  (\bibinfo{year}{1999}{\natexlab{a}}), \eprint{astro-ph/9906228}.

\bibitem[{\citenamefont{Strong et~al.}(2006)\citenamefont{Strong, Moskalenko,
  and Ptuskin}}]{Galprop1}
\bibinfo{author}{\bibfnamefont{A.~W.} \bibnamefont{Strong}},
  \bibinfo{author}{\bibfnamefont{I.~V.} \bibnamefont{Moskalenko}},
  \bibnamefont{and} \bibinfo{author}{\bibfnamefont{V.~S.}
  \bibnamefont{Ptuskin}}, \emph{\bibinfo{title}{GALPROP C++ v.50: Explanatory
  Supplement}} (\bibinfo{year}{2006}).

\bibitem[{\citenamefont{Abdo et~al.}(2007)}]{Abdo:2006fq}
\bibinfo{author}{\bibfnamefont{A.~A.} \bibnamefont{Abdo}} \bibnamefont{et~al.},
  \bibinfo{journal}{Astrophys. J.} \textbf{\bibinfo{volume}{658}},
  \bibinfo{pages}{L33} (\bibinfo{year}{2007}), \eprint{astro-ph/0611691}.

\bibitem[{\citenamefont{Strong et~al.}(2004)\citenamefont{Strong, Moskalenko,
  and Reimer}}]{Strong:2004de}
\bibinfo{author}{\bibfnamefont{A.~W.} \bibnamefont{Strong}},
  \bibinfo{author}{\bibfnamefont{I.~V.} \bibnamefont{Moskalenko}},
  \bibnamefont{and} \bibinfo{author}{\bibfnamefont{O.}~\bibnamefont{Reimer}},
  \bibinfo{journal}{Astrophys. J.} \textbf{\bibinfo{volume}{613}},
  \bibinfo{pages}{962} (\bibinfo{year}{2004}), \eprint{astro-ph/0406254}.

\bibitem[{\citenamefont{Navarro et~al.}(1996)\citenamefont{Navarro, Frenk, and
  White}}]{Navarro:1995iw}
\bibinfo{author}{\bibfnamefont{J.~F.} \bibnamefont{Navarro}},
  \bibinfo{author}{\bibfnamefont{C.~S.} \bibnamefont{Frenk}}, \bibnamefont{and}
  \bibinfo{author}{\bibfnamefont{S.~D.~M.} \bibnamefont{White}},
  \bibinfo{journal}{Astrophys. J.} \textbf{\bibinfo{volume}{462}},
  \bibinfo{pages}{563} (\bibinfo{year}{1996}), \eprint{astro-ph/9508025}.

\bibitem[{\citenamefont{Merritt et~al.}(2005)\citenamefont{Merritt, Navarro,
  Ludlow, and Jenkins}}]{Merritt:2005xc}
\bibinfo{author}{\bibfnamefont{D.}~\bibnamefont{Merritt}},
  \bibinfo{author}{\bibfnamefont{J.~F.} \bibnamefont{Navarro}},
  \bibinfo{author}{\bibfnamefont{A.}~\bibnamefont{Ludlow}}, \bibnamefont{and}
  \bibinfo{author}{\bibfnamefont{A.}~\bibnamefont{Jenkins}},
  \bibinfo{journal}{Astrophys. J.} \textbf{\bibinfo{volume}{624}},
  \bibinfo{pages}{L85} (\bibinfo{year}{2005}), \eprint{astro-ph/0502515}.

\bibitem[{\citenamefont{Kelner et~al.}(2006)\citenamefont{Kelner, Aharonian,
  and Bugayov}}]{Kelner:2006tc}
\bibinfo{author}{\bibfnamefont{S.~R.} \bibnamefont{Kelner}},
  \bibinfo{author}{\bibfnamefont{F.~A.} \bibnamefont{Aharonian}},
  \bibnamefont{and} \bibinfo{author}{\bibfnamefont{V.~V.}
  \bibnamefont{Bugayov}}, \bibinfo{journal}{Phys. Rev.}
  \textbf{\bibinfo{volume}{D74}}, \bibinfo{pages}{034018}
  (\bibinfo{year}{2006}), \eprint{astro-ph/0606058}.

\bibitem[{\citenamefont{Sanuki et~al.}(2000)}]{Sanuki:2000wh}
\bibinfo{author}{\bibfnamefont{T.}~\bibnamefont{Sanuki}} \bibnamefont{et~al.},
  \bibinfo{journal}{Astrophys. J.} \textbf{\bibinfo{volume}{545}},
  \bibinfo{pages}{1135} (\bibinfo{year}{2000}), \eprint{astro-ph/0002481}.

\bibitem[{\citenamefont{Menn et~al.}(2000)}]{Menn:2000}
\bibinfo{author}{\bibfnamefont{W.}~\bibnamefont{Menn}} \bibnamefont{et~al.},
  \bibinfo{journal}{Astrophys. J.} \textbf{\bibinfo{volume}{533}},
  \bibinfo{pages}{281} (\bibinfo{year}{2000}).

\bibitem[{\citenamefont{Alcaraz et~al.}(2000)}]{Alcaraz:2000vp}
\bibinfo{author}{\bibfnamefont{J.}~\bibnamefont{Alcaraz}} \bibnamefont{et~al.}
  (\bibinfo{collaboration}{AMS}), \bibinfo{journal}{Phys. Lett.}
  \textbf{\bibinfo{volume}{B490}}, \bibinfo{pages}{27} (\bibinfo{year}{2000}).

\bibitem[{\citenamefont{Diemand et~al.}(2007)\citenamefont{Diemand, Kuhlen, and
  Madau}}]{Diemand:2006ik}
\bibinfo{author}{\bibfnamefont{J.}~\bibnamefont{Diemand}},
  \bibinfo{author}{\bibfnamefont{M.}~\bibnamefont{Kuhlen}}, \bibnamefont{and}
  \bibinfo{author}{\bibfnamefont{P.}~\bibnamefont{Madau}},
  \bibinfo{journal}{Astrophys. J.} \textbf{\bibinfo{volume}{657}},
  \bibinfo{pages}{262} (\bibinfo{year}{2007}), \eprint{astro-ph/0611370}.

\bibitem[{\citenamefont{Hooper and Kribs}(2004)}]{Hooper:2004xn}
\bibinfo{author}{\bibfnamefont{D.}~\bibnamefont{Hooper}} \bibnamefont{and}
  \bibinfo{author}{\bibfnamefont{G.~D.} \bibnamefont{Kribs}},
  \bibinfo{journal}{Phys. Rev.} \textbf{\bibinfo{volume}{D70}},
  \bibinfo{pages}{115004} (\bibinfo{year}{2004}), \eprint{hep-ph/0406026}.

\bibitem[{\citenamefont{Grajek et~al.}(2008)\citenamefont{Grajek, Kane, Phalen,
  Pierce, and Watson}}]{Grajek:2008jb}
\bibinfo{author}{\bibfnamefont{P.}~\bibnamefont{Grajek}},
  \bibinfo{author}{\bibfnamefont{G.}~\bibnamefont{Kane}},
  \bibinfo{author}{\bibfnamefont{D.~J.} \bibnamefont{Phalen}},
  \bibinfo{author}{\bibfnamefont{A.}~\bibnamefont{Pierce}}, \bibnamefont{and}
  \bibinfo{author}{\bibfnamefont{S.}~\bibnamefont{Watson}}
  (\bibinfo{year}{2008}), \eprint{0807.1508}.

\bibitem[{\citenamefont{Nagai and Nakayama}(2008)}]{Nagai:2008se}
\bibinfo{author}{\bibfnamefont{M.}~\bibnamefont{Nagai}} \bibnamefont{and}
  \bibinfo{author}{\bibfnamefont{K.}~\bibnamefont{Nakayama}}
  (\bibinfo{year}{2008}), \eprint{0807.1634}.

\bibitem[{\citenamefont{Gates et~al.}(1995)\citenamefont{Gates, Gyuk, and
  Turner}}]{Gates:1995dw}
\bibinfo{author}{\bibfnamefont{E.~I.} \bibnamefont{Gates}},
  \bibinfo{author}{\bibfnamefont{G.}~\bibnamefont{Gyuk}}, \bibnamefont{and}
  \bibinfo{author}{\bibfnamefont{M.~S.} \bibnamefont{Turner}},
  \bibinfo{journal}{Astrophys. J.} \textbf{\bibinfo{volume}{449}},
  \bibinfo{pages}{L123} (\bibinfo{year}{1995}), \eprint{astro-ph/9505039}.

\bibitem[{\citenamefont{Bergstrom et~al.}(1998)\citenamefont{Bergstrom, Ullio,
  and Buckley}}]{Bergstrom:1997fj}
\bibinfo{author}{\bibfnamefont{L.}~\bibnamefont{Bergstrom}},
  \bibinfo{author}{\bibfnamefont{P.}~\bibnamefont{Ullio}}, \bibnamefont{and}
  \bibinfo{author}{\bibfnamefont{J.~H.} \bibnamefont{Buckley}},
  \bibinfo{journal}{Astropart. Phys.} \textbf{\bibinfo{volume}{9}},
  \bibinfo{pages}{137} (\bibinfo{year}{1998}), \eprint{astro-ph/9712318}.

\bibitem[{\citenamefont{Bottino et~al.}(2008)\citenamefont{Bottino, Donato,
  Fornengo, and Scopel}}]{Bottino:2008mf}
\bibinfo{author}{\bibfnamefont{A.}~\bibnamefont{Bottino}},
  \bibinfo{author}{\bibfnamefont{F.}~\bibnamefont{Donato}},
  \bibinfo{author}{\bibfnamefont{N.}~\bibnamefont{Fornengo}}, \bibnamefont{and}
  \bibinfo{author}{\bibfnamefont{S.}~\bibnamefont{Scopel}}
  (\bibinfo{year}{2008}), \eprint{0806.4099}.

\bibitem[{\citenamefont{Torii et~al.}(2001)}]{Torii:2001aw}
\bibinfo{author}{\bibfnamefont{S.}~\bibnamefont{Torii}} \bibnamefont{et~al.},
  \bibinfo{journal}{Astrophys. J.} \textbf{\bibinfo{volume}{559}},
  \bibinfo{pages}{973} (\bibinfo{year}{2001}).

\bibitem[{\citenamefont{DuVernois et~al.}(2001)}]{DuVernois:2001bb}
\bibinfo{author}{\bibfnamefont{M.~A.} \bibnamefont{DuVernois}}
  \bibnamefont{et~al.}, \bibinfo{journal}{Astrophys. J.}
  \textbf{\bibinfo{volume}{559}}, \bibinfo{pages}{296} (\bibinfo{year}{2001}).

\bibitem[{\citenamefont{Boezio et~al.}(2000)}]{Boezio:2000}
\bibinfo{author}{\bibfnamefont{M.}~\bibnamefont{Boezio}} \bibnamefont{et~al.},
  \bibinfo{journal}{Astrophys. J} \textbf{\bibinfo{volume}{532}},
  \bibinfo{pages}{653} (\bibinfo{year}{2000}).

\bibitem[{\citenamefont{Grimani et~al.}(2002)}]{Grimani:2002yz}
\bibinfo{author}{\bibfnamefont{C.}~\bibnamefont{Grimani}} \bibnamefont{et~al.},
  \bibinfo{journal}{Astron. Astrophys.} \textbf{\bibinfo{volume}{392}},
  \bibinfo{pages}{287} (\bibinfo{year}{2002}).

\bibitem[{\citenamefont{Strong and
  Moskalenko}(1999{\natexlab{b}})}]{Strong:1999su}
\bibinfo{author}{\bibfnamefont{A.~W.} \bibnamefont{Strong}} \bibnamefont{and}
  \bibinfo{author}{\bibfnamefont{I.~V.} \bibnamefont{Moskalenko}}
  (\bibinfo{year}{1999}{\natexlab{b}}), \eprint{astro-ph/9903370}.

\bibitem[{\citenamefont{{Moskalenko}}(2007)}]{Moskemail}
\bibinfo{author}{\bibfnamefont{I.~V.} \bibnamefont{{Moskalenko}}},
  \emph{\bibinfo{title}{Private communication}} (\bibinfo{year}{2007}).

\bibitem[{\citenamefont{Schlickeiser}(2002)}]{Schlickeiser}
\bibinfo{author}{\bibfnamefont{R.}~\bibnamefont{Schlickeiser}},
  \emph{\bibinfo{title}{Cosmic Ray Astrophysics}}
  (\bibinfo{publisher}{Springer}, \bibinfo{year}{2002}).

\bibitem[{\citenamefont{Hooper and Silk}(2005)}]{Hooper:2004bq}
\bibinfo{author}{\bibfnamefont{D.}~\bibnamefont{Hooper}} \bibnamefont{and}
  \bibinfo{author}{\bibfnamefont{J.}~\bibnamefont{Silk}},
  \bibinfo{journal}{Phys. Rev.} \textbf{\bibinfo{volume}{D71}},
  \bibinfo{pages}{083503} (\bibinfo{year}{2005}), \eprint{hep-ph/0409104}.

\bibitem[{\citenamefont{Strong and Moskalenko}(1998)}]{Strong:1998pw}
\bibinfo{author}{\bibfnamefont{A.~W.} \bibnamefont{Strong}} \bibnamefont{and}
  \bibinfo{author}{\bibfnamefont{I.~V.} \bibnamefont{Moskalenko}},
  \bibinfo{journal}{Astrophys. J.} \textbf{\bibinfo{volume}{509}},
  \bibinfo{pages}{212} (\bibinfo{year}{1998}), \eprint{astro-ph/9807150}.

\bibitem[{\citenamefont{Dobler and Finkbeiner}(2007)}]{Dobler:2007wv}
\bibinfo{author}{\bibfnamefont{G.}~\bibnamefont{Dobler}} \bibnamefont{and}
  \bibinfo{author}{\bibfnamefont{D.~P.} \bibnamefont{Finkbeiner}}
  (\bibinfo{year}{2007}), \eprint{arXiv:0712.1038 [astro-ph]}.

\bibitem[{\citenamefont{de~Blois~(France)}()}]{pamprelim}
\bibinfo{author}{\bibfnamefont{C.}~\bibnamefont{de~Blois~(France)}},
  \emph{\bibinfo{title}{Le xxth rencontres de blois - challenges in particle
  astrophysics}}.

\bibitem[{\citenamefont{Pospelov et~al.}(2007)\citenamefont{Pospelov, Ritz, and
  Voloshin}}]{Pospelov:2007mp}
\bibinfo{author}{\bibfnamefont{M.}~\bibnamefont{Pospelov}},
  \bibinfo{author}{\bibfnamefont{A.}~\bibnamefont{Ritz}}, \bibnamefont{and}
  \bibinfo{author}{\bibfnamefont{M.~B.} \bibnamefont{Voloshin}}
  (\bibinfo{year}{2007}), \eprint{arXiv:0711.4866 [hep-ph]}.

\bibitem[{\citenamefont{Michel}(1950)}]{Michel:1949qe}
\bibinfo{author}{\bibfnamefont{L.}~\bibnamefont{Michel}},
  \bibinfo{journal}{Proc. Phys. Soc.} \textbf{\bibinfo{volume}{A63}},
  \bibinfo{pages}{514} (\bibinfo{year}{1950}).

\end{thebibliography}
\bibliographystyle{apsrev}

\end{document}